\documentclass[12pt,onecolumn]{article}
\usepackage{color}
\usepackage{amsmath}
\usepackage{amssymb}
\usepackage{amsfonts}
\usepackage{geometry}
\usepackage{setspace}
\usepackage{amsmath,bm}
\usepackage{graphicx}
\usepackage[section]{placeins}
\usepackage{subfigure}
\usepackage[authoryear]{natbib}

\author{Shuwei Zhou$^{3,4}$, Xiaoying Zhuang$^{4,5}$, Hehua Zhu$^{4}$, Timon Rabczuk$^{1,2*}$}
\title {Phase field modelling of crack propagation, branching and coalescence in rocks}
\begin{document}
\bibliographystyle{unsrtnat}
\setcitestyle{round,aysep={},yysep={,}}
\date{}
\maketitle

\spacing{1.5}

\noindent
1 Division of Computational Mechanics, Ton Duc Thang University, Ho Chi Minh City, Viet Nam \\
2 Faculty of Civil Engineering, Ton Duc Thang University, Ho Chi Minh City, Viet Nam \\
3 Institute of Structural Mechanics, Bauhaus-University Weimar, Weimar 99423, Germany\\
4 Department of Geotechnical Engineering, College of Civil Engineering, Tongji University, Shanghai 200092, P.R. China\\
5 Institute of Continuum Mechanics, Leibniz University Hannover, Hannover 30167, Germany\\
* Corresponding author: timon.rabczuk@tdt.edu.vn\\

\begin{abstract}
\noindent We present a phase field model (PFM)  for simulating complex crack patterns including crack propagation, branching and coalescence in rock. The phase field model is implemented in COMSOL and is based on the strain decomposition for the elastic energy, which drives the evolution of the phase field. Then, numerical simulations of notched semi-circular bend (NSCB) tests and Brazil splitting tests are performed. Subsequently, crack propagation and coalescence in rock plates with multiple echelon flaws and twenty parallel flaws are studied. Finally, complex crack patterns are presented for a plate subjected to increasing internal pressure, the (3D) Pertersson beam and a 3D NSCB test. All results are in good agreement with previous experimental and numerical results.
\end{abstract}


\section {Introduction}\label{Introduction}

Fracture-induced failure has gained extensive concern in engineering because of the huge threat to engineering safety \citep{anderson2005fracture}.  The prediction of fracture in rock is challenging. Rock masses have many pre-existing flaws, such as micro cracks, voids and soft minerals. Many efforts have been made to study crack propagation in rock, see for instance the contributions in \citet{bobet1998fracture}, \citet{wong2001analysis}, \citet{sagong2002coalescence}, \citet{wong2009crack}, \citet{park2009crack}, \citet{park2010crack}, \citet{lee2011experimental}, and \citet{zhou2014experimental}.  However, many studies focus on  uniaxial compressive loads since tensile loads or more complicated load cases, which are more difficult to perform in practical tests.

Numerical methods are a good alternative to study  fracture problems. They are less expensive than experimental tests and can provide physical insight difficult to gain through 'pure' experimental testing. Computational methods for fracture can be classified in discrete and continuous approaches. Efficient remeshing techniques \citep{Areias201727,Areias2013113,Areias20131099}, multiscale method \citep{budarapu2014adaptive, budarapu2014efficient, yang2015meshless}, strain-softening element \citep{Areias201450}, the extended finite element method \citep{nanthakumar2014detection,moes2002extended}, the phantom node method \citep{rabczuk2008new,chau2012phantom,vu2013phantom} and specific meshfree methods \citep{rabczuk2007simplified,rabczuk2007meshfree, rabczuk2007three,rabczuk2008new,rabczuk2008discontinuous,rabczuk2008geometrically,Amiri201445} are classical representatitves of the first class. The cracking particles method (CPM) \citep{Rabczuk20042316,Rabczuk20072777,rabczuk2010simple}, Peridynamics \citep{rabczuk2017peridynamics} and dual-horizon peridynamics \citep{ren2016dual,ren2017dual} are also discrete crack approaches but they share the simplicity of continuous approaches to fracture as they also do not require any explicit representation of the crack surface and any crack tracking algorithms. Element-erosion \citep{belytschko1987three,johnson1987eroding} directly sets the stresses of the elements to zero when the elements fulfill the fracture criterion. However, the element-erosion method cannot simulate crack branching correctly \citep{song2008comparative}. Gradient models \citep{thai2016higher}, non-local models \citep{pijaudier2004non}, models based on the screend-poisson equation \citep{Areias2016116} and also phase field models are typical continuous approaches to fracture.

In this paper, we pursue the phase field model (PFM) \citep{bourdin2008variational,miehe2010phase,miehe2010thermodynamically,hesch2014thermodynamically,borden2012phase} to model crack propagation, branching and coalescence in rock. The origins of the PFM can be traced back to \citet{bourdin2008variational}, but a thermodynamic consistent framework  was first presented by \citet{miehe2010phase}. Considerable attention has been paid to PFMs due to their ease in implementation and applicability to multi-physics problems. The PFM does not treat the crack as a physical discontinuity but uses a scalar field (the phase field) to smoothly transit the intact material to the broken one. Thus, the sharp crack is represented by a 'damage-like' zone. The shape of the crack is controlled by a length scale parameter and  propagation of the crack is obtained through the solution of a differential equation. Thus, the PFM does not require any external criterion for fracture and additional work to track the fracture surface algorithmically \citep{borden2012phase}. It is believed that for this reason, the phase field is therefore has some advantage over other approaches in modeling branching and merging of multiple cracks.

Phase field models have been discretized in the context of the finite element method \citep{Areias2016322}, meshfree methods \citep{Amiri2014102} and isogeometric analysis  \citep{borden2012phase}; the latter two approaches use a fourth-order differential equation for the phase field exploiting the higher continuity of the meshfree and isogeometric approximation. The PFM for brittle cracks has also been implemented in commercial software such as ABAQUS  \citep{msekh2015abaqus,liu2016abaqus}. However, the extension of the implementation in ABAQUS to problems with more fields --  as hydraulic fracturing -- is difficult. Hence, we present an implementation of the phase field model in COMSOL Multiphysics, a software particularly dedicated to multi-field modeling. 

This paper is organized as follows. The phase field model for brittle fractures  is presented in Section 2. Subsequently, the numerical implementation of the phase field model in COMSOL is described in Section 3. Then, simulations of  initiation, propagation, branching, and coalescence of cracks in rock are shown in Section 4 before Section 5 concludes our manuscript.

\section {Theory of phase field modeling}\label{Theory of phase field modeling}
\subsection {Theory of brittle fracture}\label{Theory of brittle fracture}

\begin{figure}[htbp]
	\centering
	\includegraphics[width = 6cm]{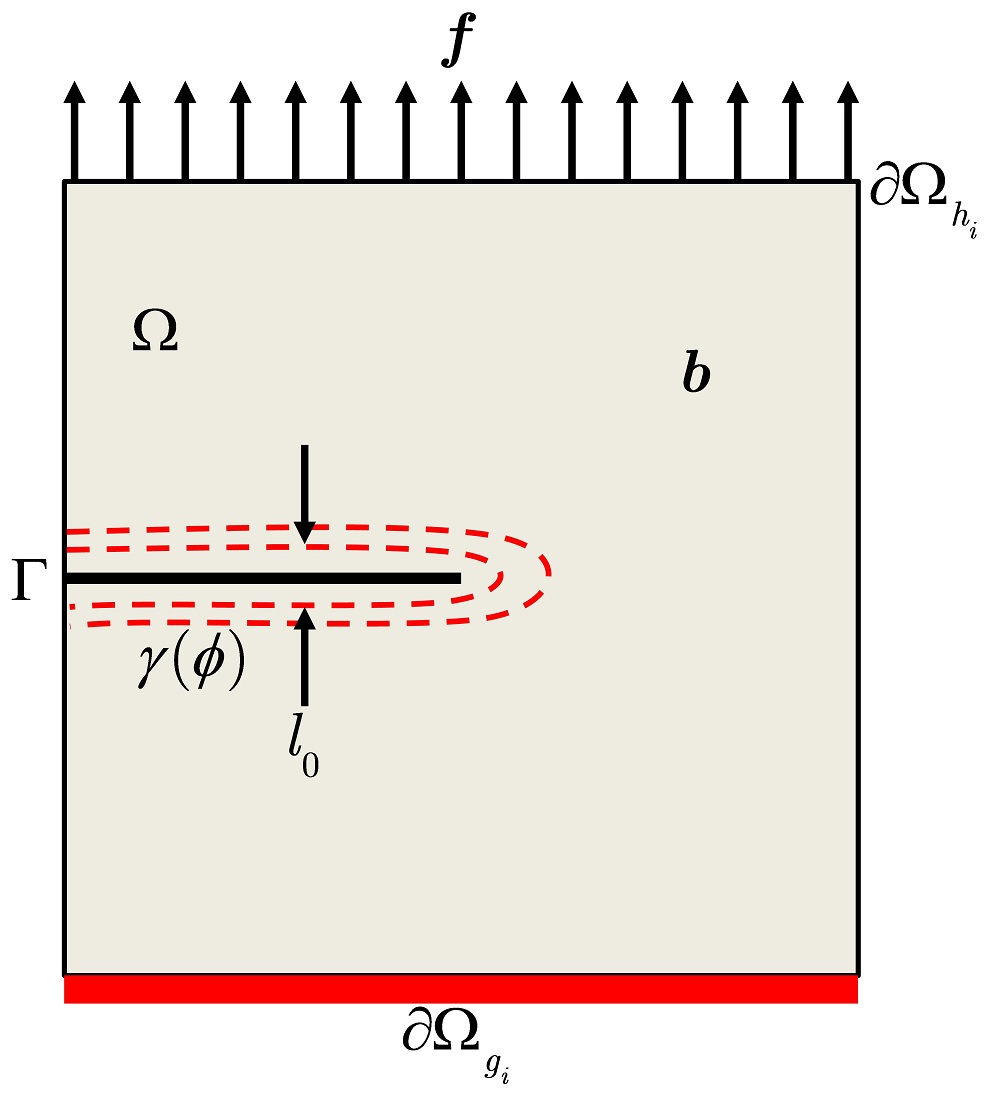}
	\caption{Phase field approximation of the crack surface}
	\label{Phase field approximation of the crack surface}
	\end{figure}

Consider an elastic body $\Omega\subset \mathbb R^d$ ($d\in \{1,2,3\} $) as shown in Figure \ref {Phase field approximation of the crack surface}, whose external boundary and internal discontinuity boundary are denoted as $\partial \Omega$ and $\Gamma $, respectively; $\bm x $ is the position vector and $\bm u(\bm x,t)\subset \mathbb R^d$ the displacement vector at time $t$. In Fig. \ref {Phase field approximation of the crack surface}, the body $\Omega $ satisfies the time-dependent Dirichlet boundary conditions ($u_i(\bm x,t)=g_i(\bm x,t)$ on $\partial \Omega_{g_i} \in \Omega$), and also the time-dependent Neumann conditions on $\partial \Omega_{h_i} \in \Omega$; $\bm b(\bm x,t)\subset \mathbb R^d$ is the body force and $\bm f(\bm x,t)$ the traction  on  boundary $\partial \Omega_{h_i}$.

Given that the stored elastic energy can be transformed into dissipative forms of energy, the classical Griffith's theory \citep{anderson2005fracture} for brittle fracture states that the crack starts to propagate when the stored energy is sufficient to overcome the fracture resistance of the material. Therefore, the crack propagation is regarded as a process to minimize a free energy $L$ that consists of the kinetic energy $\Psi_{kin}(\bm{\dot u})$, elastic energy $\Psi_{\varepsilon}$, fracture energy $\Psi_f$ and external work $W_{ext}$:
	\begin{equation}
	L = 	\Psi_{kin}(\bm{\dot u})-\underbrace{\int_{\Omega}\psi_{\varepsilon}(\bm \varepsilon) \mathrm{d}{\Omega}}_{\Psi_{\varepsilon}}- \underbrace{\int_{\Gamma}G_c \mathrm{d}S}_{\Psi_f}+\underbrace{\int_{\Omega} \bm b\cdot{\bm u}\mathrm{d}{\Omega} + \int_{\partial\Omega_{h_i}} \bm f\cdot{\bm u}\mathrm{d}S}_{W_{ext}}
	\label{energy functional L}
	\end{equation} 

\noindent where $\bm{\dot u}=\frac {\partial {\bm u}}{\partial t}$, $\psi_{\varepsilon}$ is the elastic energy density, and $G_c$ is the critical energy release rate. The linear strain tensor $\bm\varepsilon = \bm\varepsilon(\bm u)$ is given by
	\begin{equation}
	\bm\varepsilon=\frac 1 2 \left[\nabla \bm u+(\nabla \bm u)^\mathrm{ T }\right]
	\end{equation}

The kinetic energy  is given by
	\begin{equation}
	\Psi_{kin}(\bm{\dot u})=\frac 1 2\int_{\Omega}\rho\bm{\dot u}^2 \mathrm{d}{\Omega}
	\label{kinetic energy}
	\end{equation}

\noindent where $\rho$ indicates the density.

\subsection{Phase filed approximation for the fracture energy}\label{Phase filed approximation for the fracture energy}

The phase field method  \citep{miehe2010phase, miehe2010thermodynamically,borden2012phase} uses a scalar field, i.e. the phase field, to smear out the crack surface (see Fig. \ref{Phase field approximation of the crack surface}) over the domain $\Omega$. The phase field $\phi(\bm x,t)\in[0,1]$ has to satisfy the following conditions:
	\begin{equation}
	\phi = 
		\begin{cases}
		0,\hspace{1cm}\text{if material is intact}\\1,\hspace{1cm}\text{if material is cracked}
		\end{cases}
	\end{equation}

A typical one dimensional phase field approximated by the exponential function is given by \citep{miehe2010phase} 
	\begin{equation}
	\phi(x)= e^{-|x|/l_0}
	\end{equation}

\begin{figure}[htbp]
	\centering
	\includegraphics[width = 8cm]{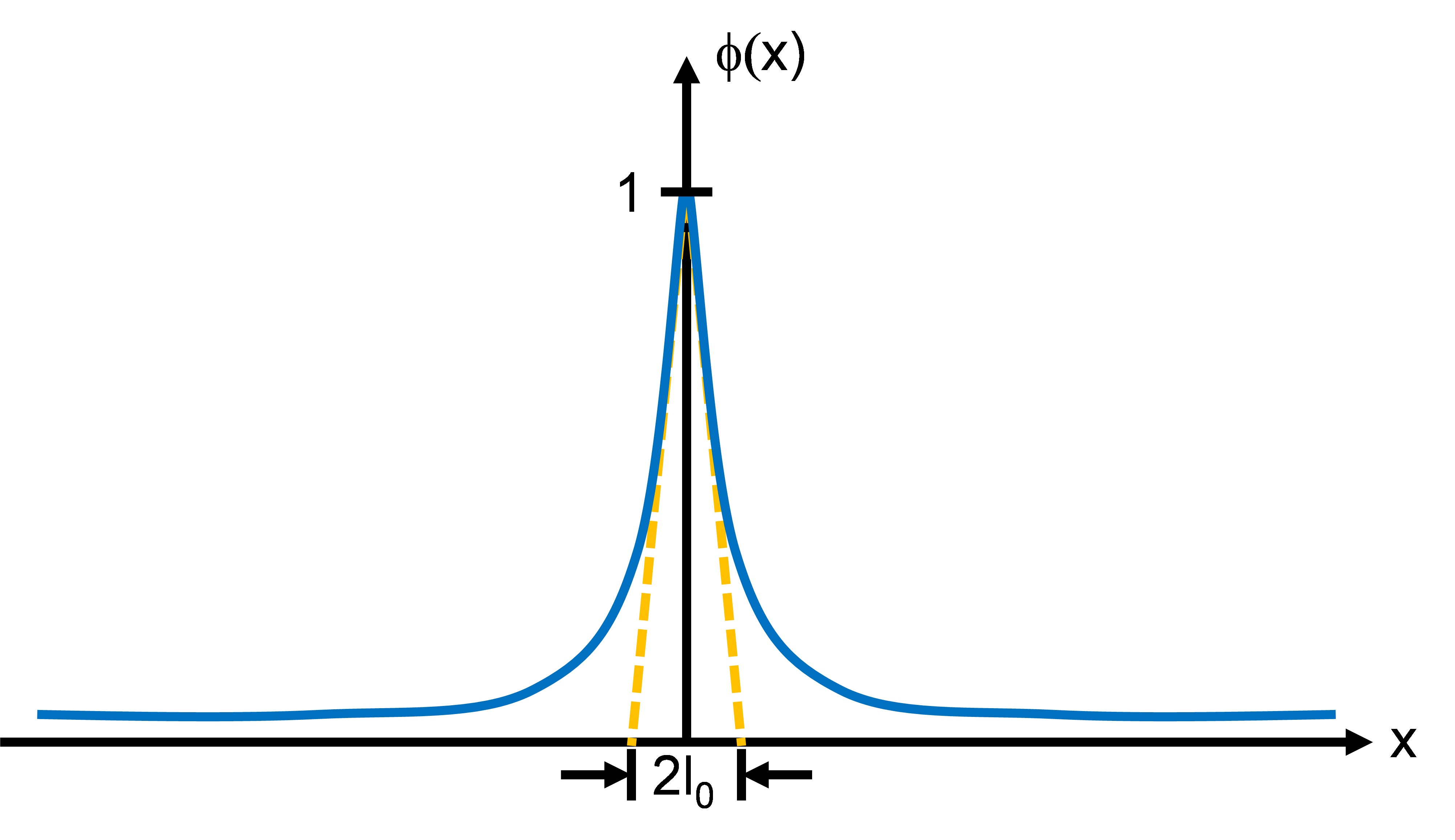}
	\caption{Distribution of the one dimensional phase field across a crack}
	\label{Distribution of the one dimensional phase field across a crack}
	\end{figure}

\noindent  $l_0$ denoting the length scale parameter, which controls the transition region of the phase field and thereby reflects the width of the crack. The distribution of the one dimensional phase field is shown in Fig. \ref{Distribution of the one dimensional phase field across a crack}. The crack region will have a larger width as $l_0$ increases and the phase field will represent a sharp crack when $l_0$ tends to zero. 

It can be shown that the crack surface density per unit volume of the solid is given by \citep{miehe2010phase}
	\begin{equation}
	\gamma(\phi,\bigtriangledown\phi)=\frac{\phi^2}{2l_0}+\frac{l_0}2|\nabla\phi|^2
	\label{the crack surface density per unit volume of the solid}
	\end{equation}

Thus, the fracture energy is approximated by
	\begin{equation}
	\int_{\Gamma}G_c \mathrm{d}S=\int_{\Omega}G_c\left(\frac{\phi^2}{2l_0}+\frac{l_0}2|\nabla\phi|^2 \right)\mathrm{d}{\Omega}
	\label{fracture energy}
	\end{equation}

The variational approach \citep{bourdin2000numerical} states that the crack surface energy is transformed from the elastic energy, which drives the evolution of the phase field. To capture cracks only under tension,  the elastic energy is decomposed into tensile and compressive parts \citep{miehe2010thermodynamically}:
	\begin{equation}
	\bm\varepsilon_{\pm}=\sum_{a=1}^d \langle\varepsilon_a\rangle_{\pm}\bm n_a\otimes\bm n_a
	\end{equation}

\noindent where $\bm\varepsilon_+$  and $\bm\varepsilon_-$  are the tensile and compressive strain tensors, respectively. In addition, $\varepsilon_a$ is the principal strain and  $\bm n_a$  is the direction of the principal strain. The operators $\langle\centerdot\rangle_{\pm}$  are defined as : $\langle\centerdot\rangle_{\pm}=(\centerdot \pm |\centerdot|)/ 2$. Consequently, the positive and negative elastic energy densities are expressed as
	\begin{equation}
	\psi_{\varepsilon}^{\pm}(\bm \varepsilon) = \frac{\lambda}{2}\langle \mathrm{tr}(\bm\varepsilon)\rangle_{\pm}^2+\mu \mathrm{tr} \left(\bm\varepsilon_{\pm}^2\right) 
	\end{equation}

\noindent where $\lambda>0$ and $\mu>0$ are the Lam\'e constants. The Lam\'e constants are related to the Young's modulus $E$ and Poisson's ratio $\nu$ of the solid through the well known relation:
	\begin{equation}
		  \left\{
	   \begin{aligned}
	\lambda&=\frac{E\nu}{(1+\nu)(1-2\nu)}
	\\ 	\mu&=\frac{E}{2(1+\nu)}
	   \end{aligned}\right.
	\end{equation}

The phase field is assumed to affect only the positive elastic energy density, which introduces a stiffness reduction as \citep{borden2012phase}
	\begin{equation}
	\psi_{\varepsilon}(\bm\varepsilon)=\left[(1-k)(1-\phi)^2+k\right]\psi_{\varepsilon}^+(\bm \varepsilon)+\psi_{\varepsilon}^-(\bm \varepsilon)
	\label{elastic energy}
	\end{equation}

\noindent where $0<k\ll1$ is a stability parameter for avoiding numerical singularities because the positive elastic energy density disappears as the phase field $\phi$ tends to 1. 

\subsection{Governing equations}\label{Governing equations}

By substituting Eqs. \eqref{kinetic energy}, \eqref{fracture energy}, and \eqref{elastic energy} into Eq. \eqref{energy functional L}, the energy functional $L$ is rewritten as
	\begin{multline}
	L=\frac 1 2\int_{\Omega}\rho\bm{\dot u}^2 \mathrm{d}{\Omega}-\int_{\Omega}\left\{\left[(1-k)(1-\phi)^2+k\right]\psi_{\varepsilon}^+(\bm \varepsilon)+\psi_{\varepsilon}^-(\bm \varepsilon)\right\}\mathrm{d}{\Omega}-
\\ \int_{\Omega}G_c\left(\frac{\phi^2}{2l_0}+\frac{l_0}2|\nabla \phi|^2 \right)\mathrm{d}{\Omega}+\int_{\Omega} \bm b \cdot \bm u \mathrm{d}{\Omega}+ \int_{\partial\Omega_{h_i}} \bm f \cdot \bm u \mathrm{d}S
	\end{multline}

Employ the first variation of the functional $\delta L=0$, it can be shown that the strong form of the governing equations are given by \citep{borden2012phase}	\begin{equation}
	  \left\{
	   \begin{aligned}
	\text{Div}(\bm\sigma)+\bm b=\rho \ddot{\bm u}
	\\ \left[\frac{2l_0(1-k)\psi_{\varepsilon}^+}{G_c}+1\right]\phi-l_0^2\nabla^2\phi=\frac{2l_0(1-k)\psi_{\varepsilon}^+}{G_c}
	   \end{aligned}\right.
	\label{governing equation1}
	\end{equation}

\noindent where $\ddot {\bm u}=\frac {\partial^2 {\bm u}}{\partial t^2}$  and  $\bm\sigma$ is Cauchy stress tensor given by
	\begin{equation}
		\begin{aligned}
		\bm\sigma&=\partial_{\bm\varepsilon} \psi_{\varepsilon}\\
		&=\left [(1-k)(1-\phi)^2+k \right]\left[\lambda \langle tr(\bm\varepsilon)\rangle_+ \bm I+ 2\mu \bm\varepsilon_+ \right]+\lambda \langle tr(\bm\varepsilon)\rangle_- \bm I+ 2\mu \bm\varepsilon_-
		\end{aligned}
	\end{equation}

\noindent with unit tensor $\bm I$ $\in \mathbb R^{d\times d}$.

The phase field requires the irreversibility condition $\Gamma(\bm x,s)\in\Gamma(\bm x,t)(s<t)$ during compression or unloading, i.e. the crack cannot be healed. Therefore, we introduce a strain-history field $H(\bm x,t)$ \citep{miehe2010phase,miehe2010thermodynamically} to ensure a monotonically increasing phase field:
	\begin{equation}
	H(\bm x,t) = \max \limits_{x\in[0,t]}\psi_\varepsilon^+\left(\bm\varepsilon(\bm x,s)\right)
	\end{equation}

The history field $H$ satisfies the Kuhn-Tucker conditions for loading and unloading \citep{miehe2010phase}:
	\begin{equation}
	\psi_\varepsilon^+-H\le0,\hspace{0.5cm}\dot{H}\ge0,\hspace{0.5cm}\dot{H}(\psi_\varepsilon^+-H)=0
	\end{equation}
By replacing $\psi_\varepsilon^+$  by  $H(\bm x,t)$  in Eq. \eqref{governing equation1}, the strong form is rewritten as
	\begin{equation}
	  \left\{
	   \begin{aligned}
	\text{Div}(\bm\sigma)+\bm b=\rho{\ddot {\bm u}}
	\\ \left[\frac{2l_0(1-k)H}{G_c}+1\right]\phi-l_0^2\nabla^2\phi=\frac{2l_0(1-k)H}{G_c}
	   \end{aligned}\right.
	\label{governing equation2}
	\end{equation}

We denote $\bm m$ as the outward-pointing normal vector to the boundaries, and the governing equations are subjected to the Dirichlet and Neumann boundary conditions
	\begin{equation}
	  \left\{
	   \begin{aligned}
	&\bm u = \bm g \hspace{2cm} &\mathrm{on}\hspace{0.5cm} \partial\Omega_{g_i}\\
	&\bm\sigma \cdot \bm m = \bm f &\mathrm{on}\hspace{0.5cm} \partial\Omega_{h_i}
	\\ &\nabla \phi \cdot \bm m = 0  &\mathrm{on}\hspace{0.5cm} \partial\Omega
	\end{aligned}\right.\label{boundary conditions}
	\end{equation}

\noindent along with the initial conditions
	\begin{equation}
	  \left\{
	   \begin{aligned}
	&\bm u(\bm x,0)=\bm u_0(\bm x)\hspace{2cm} &\bm x\in\Omega
	\\ &\bm v(\bm x,0)=\bm v_0(\bm x)\hspace{2cm} &\bm x\in\Omega
	\\&\phi(\bm x,0)=\phi_0(\bm x)\hspace{2cm} &\bm x\in\Omega
	\end{aligned}\right.\label{initial conditions}
	\end{equation}

\subsection{The choice of $l_0$}\label{The choice of l_0}

\citet{borden2012phase} and \citet{zhang2017numerical} proposed an analytical solution for the critical tensile stress $\sigma_{cr}$ that a one-dimensional bar can sustain:
	\begin{equation}
	\sigma_{cr}=\frac 9 {16} \sqrt{\frac{EG_c}{3l_0}}
	\label{critical stress}
	\end{equation}

There is an apparent singularity when $l_0$ tends to zero, i.e. in case of a sharp crack, which is phyiscally meaningless.  However, assuming all other parameters except of $l_0$ are known, eq. (\ref{critical stress}) can be solved for $l_0$:
	\begin{equation}
	l_0=\frac {27EG_c} {256\sigma_{cr}^2}
	\label{l0}
	\end{equation}

In Eq. \eqref{l0}, the critical energy release rate $G_c$ and Young's modulus $E$ can be obtained by conducting regular experimental tests, while the critical stress $\sigma_{cr}$ can be approximated by the tensile strength $\sigma_t$ through the standard tensile test. Hence, Eq. \eqref{l0} can estimate the length scale. Note that the accuracy is unknown for more complex cases.

\section {Numerical implementation}
\subsection{Finite element method}\label{Finite element method}
We use the finite element method to solve the governing equations \eqref{governing equation2} given in weak form by
	\begin{equation}
	\int_{\Omega}\left(-\rho \bm{\ddot u} \cdot \delta \bm u -\bm\sigma:\delta \bm {\varepsilon}\right) \mathrm{d}\Omega +\int_{\Omega}\bm b \cdot \delta \bm u  \mathrm{d}\Omega +\int_{\Omega_{h_i}}\bm f \cdot \delta \bm u  \mathrm{d}S=0
	\label{weak form 1}
	\end{equation}

\noindent and
	\begin{equation}
	\int_{\Omega}-2(1-k)H(1-\phi)\delta\phi\mathrm{d}\Omega+\int_{\Omega}G_c\left(l_0\nabla\phi\cdot\nabla\delta\phi+\frac{1}{l_0}\phi\delta\phi\right)\mathrm{d}\Omega=0
	\label{weak form 2}
	\end{equation}

We use the standard vector-matrix notation and denote the nodal values of the displacement and phase field as $\bm u_i$ and $\phi_i$. The discretization is thereby given by 
	\begin{equation}
	\bm u = \bm N_u \bm d,\hspace{0.5cm} \phi = \bm N_{\phi} \hat{\bm\phi}
	\end{equation}

\noindent where $\bm d$ and $\hat{\bm\phi}$ are the vectors consisting of node values $\bm u_i$ and $\phi_i$. $\bm N_u$ and $\bm N_{\phi}$ are shape function matrices given by
	\begin{equation}		
			\bm N_u = \left[ \begin{array}{ccccccc}
			N_{1}&0&0&\dots&N_{n}&0&0\\
			0&N_{1}&0&\dots&0&N_{n}&0\\
			0&0&N_{1}&\dots&0&0&N_{n}
			\end{array}\right], \hspace{0.5cm}
			\bm N_\phi = \left[ \begin{array}{cccc}
			N_{1}&N_{2}&\dots&N_{n}
			\end{array}\right]
	\end{equation}

\noindent where $n$ is the node number in one element and $N_i$ is the shape function of node $i$. The same discretization is applied to the test functions and we obtain
	\begin{equation}
	\delta \bm u = \bm N_u \delta \bm d,\hspace{0.5cm} \delta \phi = \bm N_{\phi} \delta \hat{\bm\phi}
	\end{equation}

\noindent where $\delta \bm d$ and $\delta \hat{\bm\phi}$ are the vectors consisting of node values of the test functions.

The gradients are thereby calculated by
	\begin{equation}
	\bm \varepsilon =  \bm B_u \bm d,\hspace{0.5cm} \nabla\phi = \bm B_\phi \hat{\bm\phi}, \hspace{0.5cm}\bm \delta \varepsilon =  \bm B_u \delta \bm d,\hspace{0.5cm} \nabla\phi = \bm B_\phi \delta \hat{\bm\phi}
	\end{equation}

\noindent where $\bm B_u$ and $\bm B_\phi$ are the derivatives of the shape functions defined by
	\begin{equation}
	\bm B_u=\left[
		\begin{array}{ccccccc}
		N_{1,x}&0&0&\dots&N_{n,x}&0&0\\
		0&N_{1,y}&0&\dots&0&N_{n,y}&0\\
		0&0&N_{1,z}&\dots&0&0&N_{n,z}\\
		N_{1,y}&N_{1,x}&0&\dots&N_{n,y}&N_{n,x}&0\\
		0&N_{1,z}&N_{1,y}&\dots&0&N_{n,z}&N_{n,y}\\
		N_{1,z}&0&N_{1,x}&\dots&N_{n,z}&0&N_{n,x}
		\end{array}\right],\hspace{0.2cm}
		\bm B_\phi=\left[
		\begin{array}{cccc}
		N_{1,x}&N_{2,x}&\dots&N_{n,x}\\
		N_{1,y}&N_{2,y}&\dots&N_{n,y}\\
		N_{1,z}&N_{2,z}&\dots&N_{n,z}
		\end{array}\right]
		\label{BiBu}
	\end{equation}

By applying the finite element approximation, the equations of weak form \eqref{weak form 1} and \eqref{weak form 2} are then written as
	\begin{equation}
	-(\delta\bm d)^\mathrm{T} \left[\int_{\Omega}\rho\bm N_u^\mathrm{T}\bm N_u \mathrm{d}\Omega \ddot{\bm d}+\int_{\Omega} \bm B_u^\mathrm{T} \bm D_e \bm B_u \mathrm{d}\Omega \bm d \right]+ 
	(\delta\bm d)^\mathrm{T} \left[\int_{\Omega}\bm N_u^\mathrm{T}\bm b \mathrm{d}\Omega+\int_{\Omega_{h_i}} \bm N_u^\mathrm{T} \bm f\mathrm{d}S \right]=0
	\label{discrete equation 1}
	\end{equation}
	\begin{equation}
	-(\delta\hat{\bm \phi})^{\mathrm{T}} \int_{\Omega}\left\{\bm B_\phi^{\mathrm{T}} G_c l_0 \bm B_\phi +\bm N_\phi^{\mathrm{T}} \left [ \frac{G_c}{l_0} + 2(1-k)H \right ]  \bm N_\phi \right \} \mathrm{d}\Omega \hat{\bm \phi}+ (\delta\hat{\bm \phi})^\mathrm{T} \int_{\Omega}2(1-k)H\bm N_\phi^{\mathrm{T}} \mathrm{d}\Omega = 0
	\label{discrete equation 2}      
	\end{equation}

\noindent where $\bm D_e$ is the degraded stiffness matrix. $\bm D_e$ can be calculated from the fourth order elasticity tensor $\bm D$:
	\begin{equation}
		\begin{aligned}
		\bm D &= \frac {\partial \bm\sigma}{\partial \bm\varepsilon}\\
	&=\lambda\left\{ \left[(1-k)(1-\phi)^2+k \right]H_\varepsilon(tr(\bm\varepsilon))+H_\varepsilon(-tr(\bm\varepsilon))\right\}\bm J +2\mu\left\{\left[(1-k)(1-\phi)^2+k \right]\bm P^++\bm P^- \right \}
		\end{aligned}
	\end{equation}

\noindent where $J_{ijkl}=\delta_{ij}\delta_{kl}$, $\delta_{ij}$ being the Kronecker and $P_{ijkl}^\pm= \sum_{a=1}^3\sum_{b=1}^3 H_\varepsilon(\varepsilon_a)\delta_{ab}n_{ai}n_{aj}n_{bk}n_{bl}+\sum_{a=1}^3\sum_{b\neq a}^3 \frac 1 2 \frac {\langle \varepsilon_a\rangle_\pm - \langle \varepsilon_b\rangle_\pm}{\varepsilon_a-\varepsilon_b}n_{ai}n_{bj}(n_{ak}n_{bl}+n_{bk}n_{al})$ with $n_{ai}$ the $i$-th component of vector $\bm n_a$; $H_\varepsilon \langle x \rangle$  is the Heaviside function:
	\begin{equation}
	H_\varepsilon \langle x \rangle = \left\{
		\begin{aligned}
		1, \hspace{0.5cm} x>0\\
		0, \hspace{0.5cm} x\leq0
		\end{aligned}	
	\right.
	\end{equation}

Since $P_{ijkl}^\pm$ cannot be computed when $\varepsilon_a=\varepsilon_b$, we apply a ``perturbation'' technology for the principal strains \citep{miehe1993computation} with an unchanged $\varepsilon_2$:
	\begin{equation}
	  \left\{
	   \begin{aligned}
	&\varepsilon_1 = \varepsilon_1(1+\delta)\hspace{0.5cm} &\mathrm{if}\hspace{0.1cm}\varepsilon_1 = \varepsilon_2
	\\ &\varepsilon_3 = \varepsilon_3(1-\delta)\hspace{0.5cm} &\mathrm{if}\hspace{0.1cm}\varepsilon_2 = \varepsilon_3
	\end{aligned}\right.
	\end{equation}

\noindent with the perturbation $\delta=1\times 10^{-9}$. 

For admissible arbitrary test functions, Eqs. \eqref{discrete equation 1} and \eqref{discrete equation 2} always hold, thereby producing the discretized weak form as
	\begin{equation}
	-\underbrace{\int_{\Omega}\rho\bm N_u^\mathrm{T}\bm N \mathrm{d}\Omega \ddot{\bm d}}_{\bm F_u^{ine}=\bm M \ddot{\bm d}}-\underbrace{\int_{\Omega} \bm B_u^\mathrm{T} \bm D_e \bm B_u \mathrm{d}\Omega \bm d}_{\bm F_u^{int}=\bm K_u \bm d} + \underbrace{\int_{\Omega}\bm N_u^\mathrm{T}\bm b \mathrm{d}\Omega+\int_{\Omega_{h_i}} \bm N_u^\mathrm{T} \bm f\mathrm{d}S}_{\bm F_u^{ext}}=0
	\end{equation}
	\begin{equation}
	-\underbrace{ \int_{\Omega}\left\{\bm B_\phi^{\mathrm{T}} G_c l_0 \bm B_\phi +\bm N_\phi^{\mathrm{T}} \left [ \frac{G_c}{l_0} + 2(1-k)H \right ] \bm N_\phi \right \} \mathrm{d}\Omega \hat{\bm \phi}}_{\bm F_\phi^{int}=\bm K_\phi \hat{\bm \phi}}+ \underbrace{\int_{\Omega}2(1-k)H\bm N_\phi^{\mathrm{T}} \mathrm{d}\Omega}_{\bm F_\phi^{ext}} = 0 
	\end{equation}

\noindent where $\bm F_u^{ine}$, $\bm F_u^{int}$, and $\bm F_u^{ext}$ are the inertial, internal, and external forces for the displacement field and $\bm F_\phi^{int}$ and $\bm F_\phi^{ext}$ are the internal and external force terms of the phase field. In addition, the mass and stiffness matrices follow
	\begin{equation}
	\left\{\begin{aligned}\bm M &= \int_{\Omega}\rho\bm N_u^\mathrm{T}\bm N \mathrm{d}\Omega\\
\bm K_u &= \int_{\Omega} \bm B_u^\mathrm{T} \bm D_e \bm B_u \mathrm{d}\Omega\\
	\bm K_\phi &= \int_{\Omega}\left\{\bm B_\phi^{\mathrm{T}} G_c l_0 \bm B_\phi +\bm N_\phi^{\mathrm{T}} \left [ \frac{G_c}{l_0} + 2(1-k)H \right ] \bm N_\phi \right \} \mathrm{d}\Omega
	\end{aligned}\right .
	\end{equation}

In this paper, we use a staggered scheme to solve the displacement and phase fields. Thus, the Newton-Raphson approach is adopted to obtain the residual of the discrete equations $\bm R_{u}=\bm F_u^{ext} - \bm F_u^{ine} - \bm F_u^{int} = 0 $ and $ \bm R_{\phi}=\bm F_\phi^{ext} - \bm F_\phi^{int}=0$, respectively. 

\subsection{COMSOL implementation}
We implemented the phase field approach into COMSOL Multiphysics, which can simulate mathematical and physical problems easily by adding application-specific modules. Therefore, it is suitable for multi-field modeling. In this paper, we construct three main modules, namely, the Solid Mechanics, History-strain and Phase Field Modules, which employ the standard finite element discretization in space as described in Subsection \ref{Finite element method}. In addition, a pre-set Storage Module is employed to evaluate and store the intermediate field variables in a time step, such as the positive elastic energy and principal strains.

Based on a linear elastic material library, the Solid Mechanics Module is used for the displacement $\bm u$. The boundary and initial conditions shown in Section \ref{Theory of phase field modeling} are added to the Solid Mechanics Module, while a non-linear stress-strain relationship is considered. The stiffness matrix $D_e$ is modified in a time step as follows
	\begin{equation}
	\bm D_e= \left [
	\begin{array}{cccccc}
	D_{1111} & D_{1122} & D_{1133} & D_{1112} & D_{1123} & D_{1113}\\
	D_{2211} & D_{2222} & D_{2233} & D_{2212} & D_{2223} & D_{2213}\\
	D_{3311} & D_{3322} & D_{3333} & D_{3312} & D_{3323} & D_{3313}\\
	D_{1211} & D_{1222} & D_{1233} & D_{1212} & D_{1223} & D_{1213}\\
	D_{2311} & D_{2322} & D_{2333} & D_{2312} & D_{2323} & D_{2313}\\
	D_{1311} & D_{1322} & D_{1333} & D_{1312} & D_{1323} & D_{1313}
	\end{array}
	\right ]
	\end{equation}

The governing equation \eqref{governing equation2} is presented for dynamic crack problems. However, for a quasi-static problem, the inertia term must be neglected in the Solid Mechanics Module. The Phase Field Module is established for the phase field $\phi$ by revising a pre-defined module, which is governed by the Helmholtz equation. The boundary condition Eq. \eqref{boundary conditions} and initial condition \eqref{initial conditions} are also implemented in this module. For the history strain field $H$, the Distributed ODEs and DAEs Interfaces are used to construct the History-strain Module, where the history strain field is not solved directly. We use a ``previous solution" solver to record  the results in the previous time steps and obtain the field $H$ by applying the following format in COMSOL:
	\begin{equation}
	\left\{
	\begin{aligned}
	&H-\psi_\varepsilon^+=0,\hspace{1cm}&\text{if} \hspace{0.2cm}\psi_\varepsilon^+>H\\
	&H-H=0,\hspace{1cm}&\text{if}  \hspace{0.2cm}\psi_\varepsilon^+\le H
	\end{aligned}\right.
	\end{equation}

\noindent Additionally, the initial condition $H_0(\bm x)=0$ is used for the History-strain Module. 

\begin{figure}[htbp]
	\centering
	\includegraphics[width = 10cm]{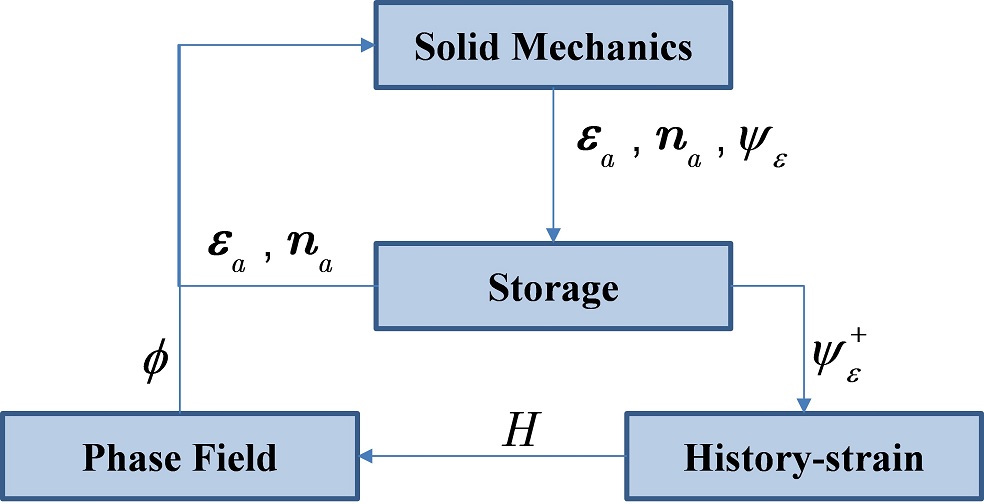}
	\caption{Relationship between all the modules established}
	\label{Relationship between all the modules established}
	\end{figure}

Figure \ref{Relationship between all the modules established} shows the relationship among all the established modules. The mechanical responses, such as the principal strains, the directions of principal strain, and the elastic energy, are naturally exported from the Solid Mechanics Module and stored in the Storage Module. The History-strain Module then call the positive elastic energy and update the local history strain field $H$. The Phase Field Module employs the updated $H$ to solve the phase field. In a time step, the updated phase field and the stored principal strains as well as the directions are used to modify the stiffness matrix of the Solid Mechanics Module and subsequently to obtain the mechanical responses. 

\begin{figure}[htbp]
	\centering
	\includegraphics[width = 15cm]{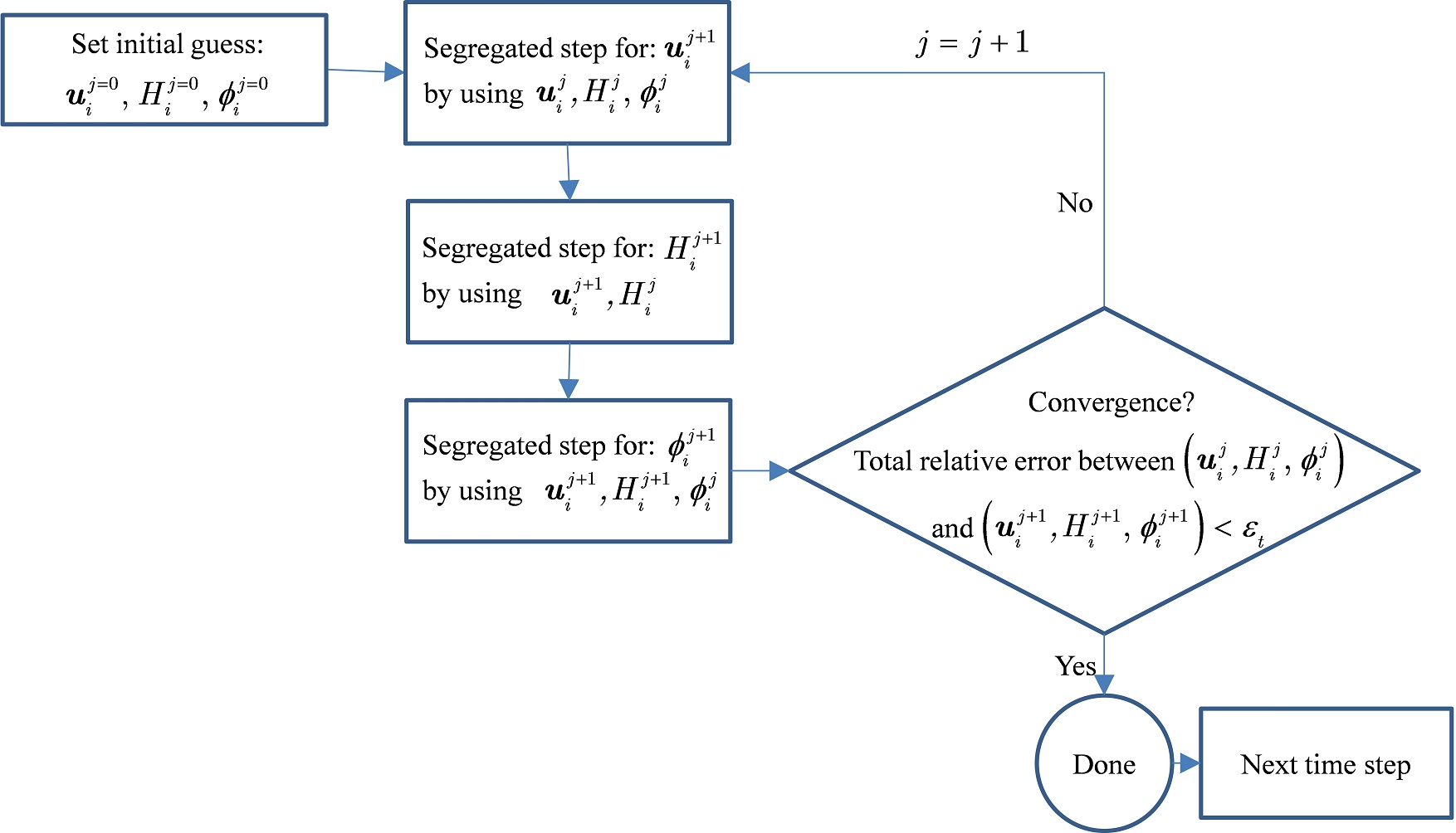}
	\caption{Segregated scheme for the coupled calculation in phase field modeling}
	\label{Segregated scheme for the coupled calculation in phase field modeling}
	\end{figure}

\begin{figure}[htbp]
	\centering
	\includegraphics[width = 10cm]{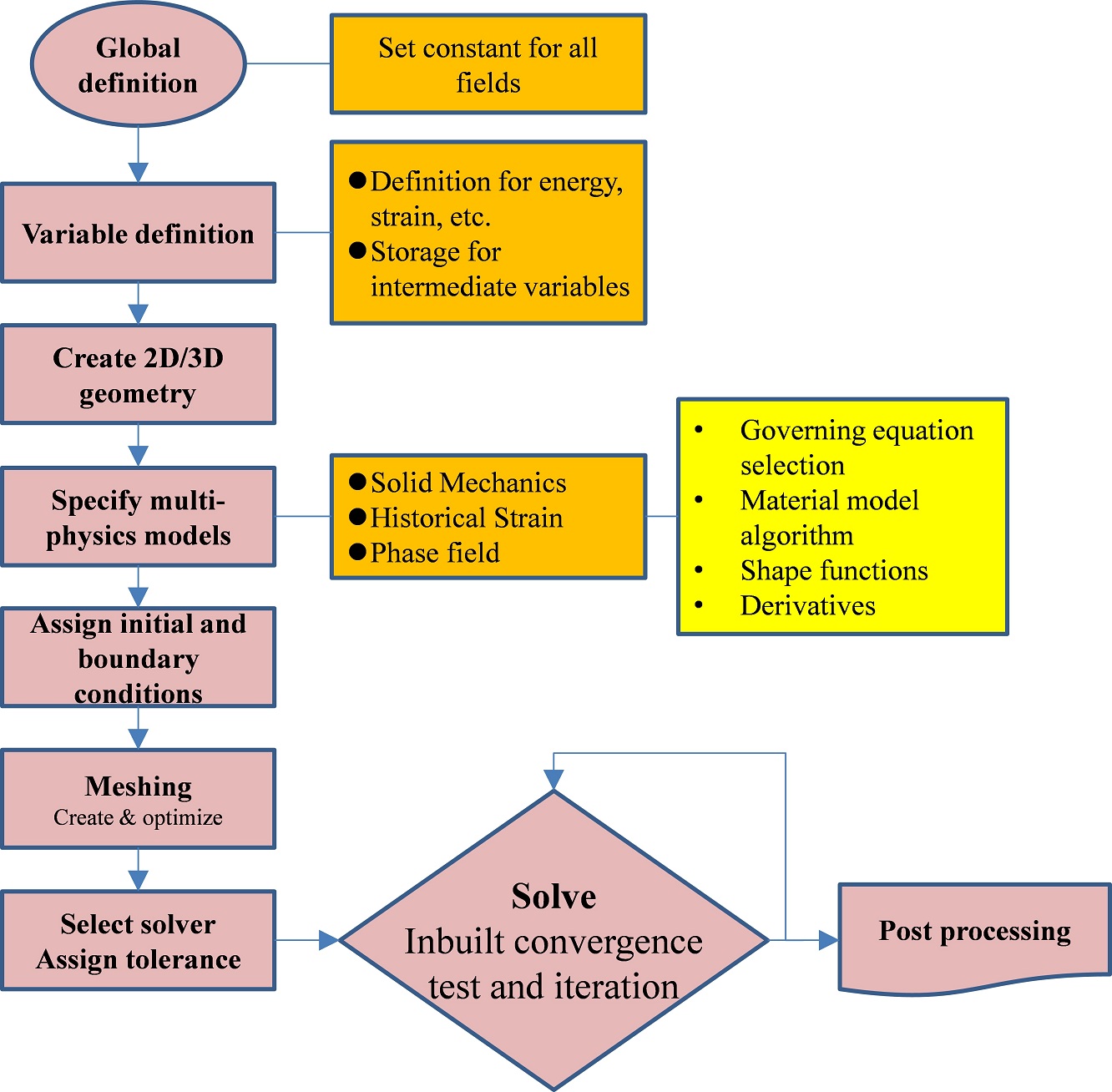}
	\caption{COMSOL implementation of the phase field modeling}
	\label{COMSOL implementation of the phase field modeling}
	\end{figure}

The detailed procedure of the staggered (segregated) scheme is depicted in Fig. \ref{Segregated scheme for the coupled calculation in phase field modeling}. The equations of displacement, history strain and phase-field are solved independently. An implicit Generalized-$\alpha$ method \citep{borden2012phase} is used for time integration. The Generalized-$\alpha$ method is unconditionally stable and requires a prediction step and a correction step. When a new time step starts, an initial guess from the linear extrapolation of the previous solution is used in the prediction stage. Newton-Raphson iterations are used to solve the residuals for each module in the correction stage. That is, in the iteration step $j+1$ of a given time step $i$, the displacement $\bm u_i^{j+1}$ is first solved by using the results ($\bm u_i^j$, $H_i^j$ and $\phi_i^j$) of the previous iteration step $j$. Following the updated displacements $\bm u_i^{j+1}$, the history strain is updated. Subsequently, another Newton-Raphson iteration is applied to solve the phase-field $\phi_i^{j+1}$. The total relative error is estimated and the iterations continue until the tolerance requirement is met. The maximum number of iteration in one time step is set as 50 in our simulations. 
 We accelerate the convergence by using the Anderson acceleration \citep{toth2015convergence} where the dimension of the iteration space field is chosen as more than 50. A flow chart of our implementation is shown in Fig. \ref{COMSOL implementation of the phase field modeling}. The source code can be found in ``https://sourceforge.net/projects/phasefieldmodelingcomsol/".
	
\section {Numerical examples of crack propagation, branching and coalescence in rocks}
\subsection {Simulation of notched semi-circular bend (NSCB) tests}

\begin{figure}[htbp]
	\centering
	\includegraphics[width = 8cm]{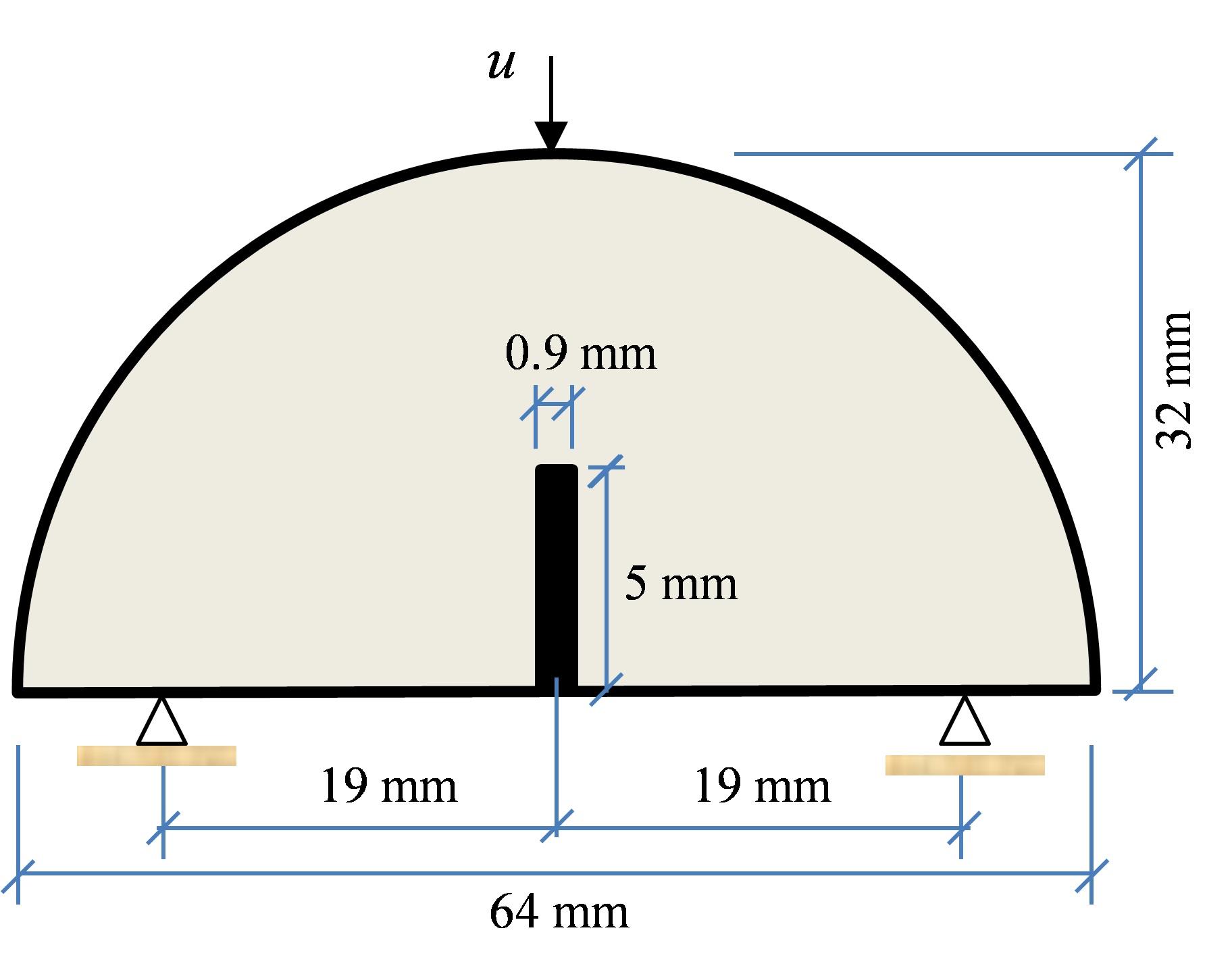}
	\caption{Geometry and boundary condition of the notched semi-circular bend (NSCB) tests}
	\label{Geometry and boundary condition of the notched semi-circular bend (NSCB) tests}
	\end{figure}

Let us consider the notched semi-circular bend (NSCB) test first. Geometry and boundary condition of the rock specimen are shown in Fig.\ref{Geometry and boundary condition of the notched semi-circular bend (NSCB) tests}. The mechanical properties of the rock are taken from the Laurentian granite (LG) from Grenville province of Canada \citep {gao2015application}. The rock density $\rho$ is 2630 kg/m$^3$, while the Young's modulus $E$ and Poisson's ratio $\nu$ are 92 GPa and 0.21, respectively. We follow the 1D solution of \citet{borden2012phase} for the critical stress of material softening, and choose $G_c$ = 7.6 J/m$^2$ and the length scale $l_0 = 4.5\times10^{-4}$ m. This produces a critical stress close to the quasi-static tensile strength of the specimen (12.8 MPa) \citep{gao2015application}.

The phase field modeling is performed by using 93587 6-node quadratic triangular elements and the maximum element size is $h =2.25\times10^{-4}$ m. We apply a vertical displacement on the top of the specimen to drive crack propagation from the tip of the notch. During the simulation, we apply the displacement increment $\Delta u  = 5\times10^{-7}$ mm in each time step.

\begin{figure}[htbp]
	\centering
	\subfigure[]{\includegraphics[width = 6cm]{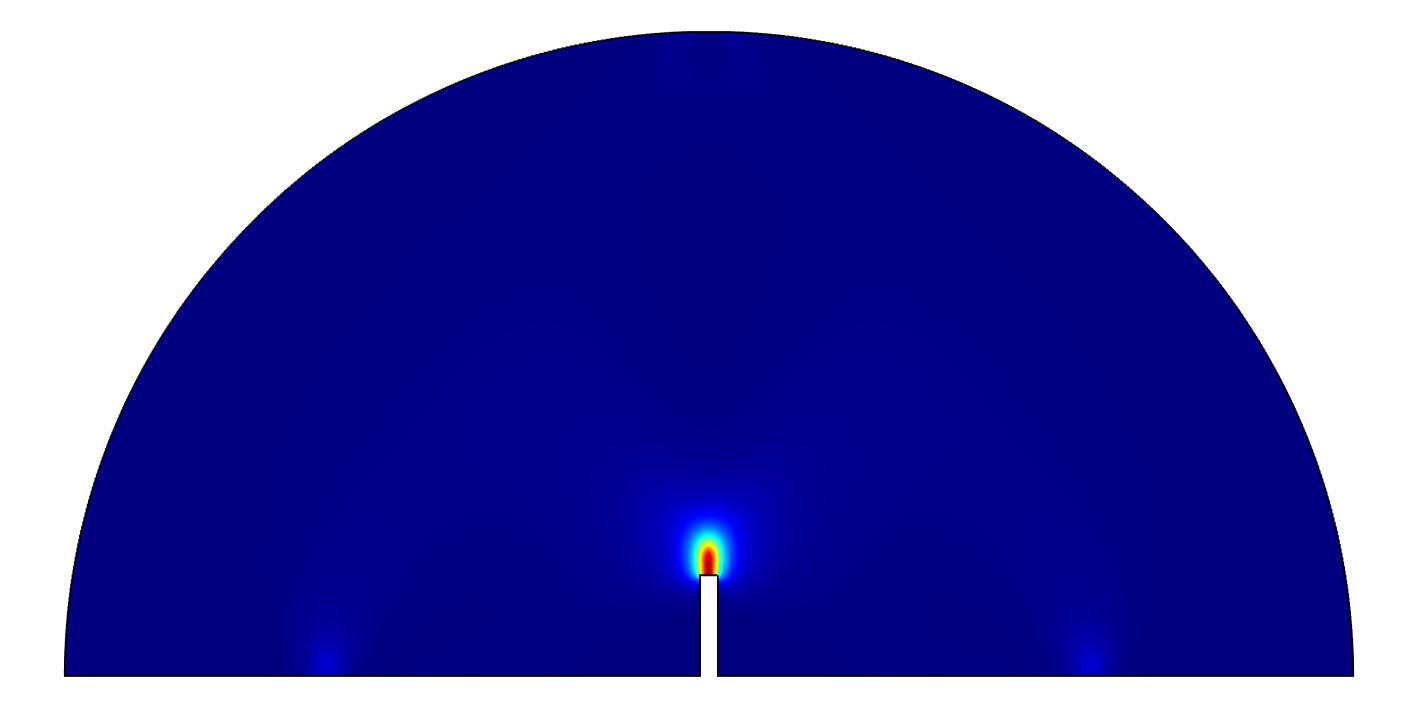}}
	\subfigure[]{\includegraphics[width = 6cm]{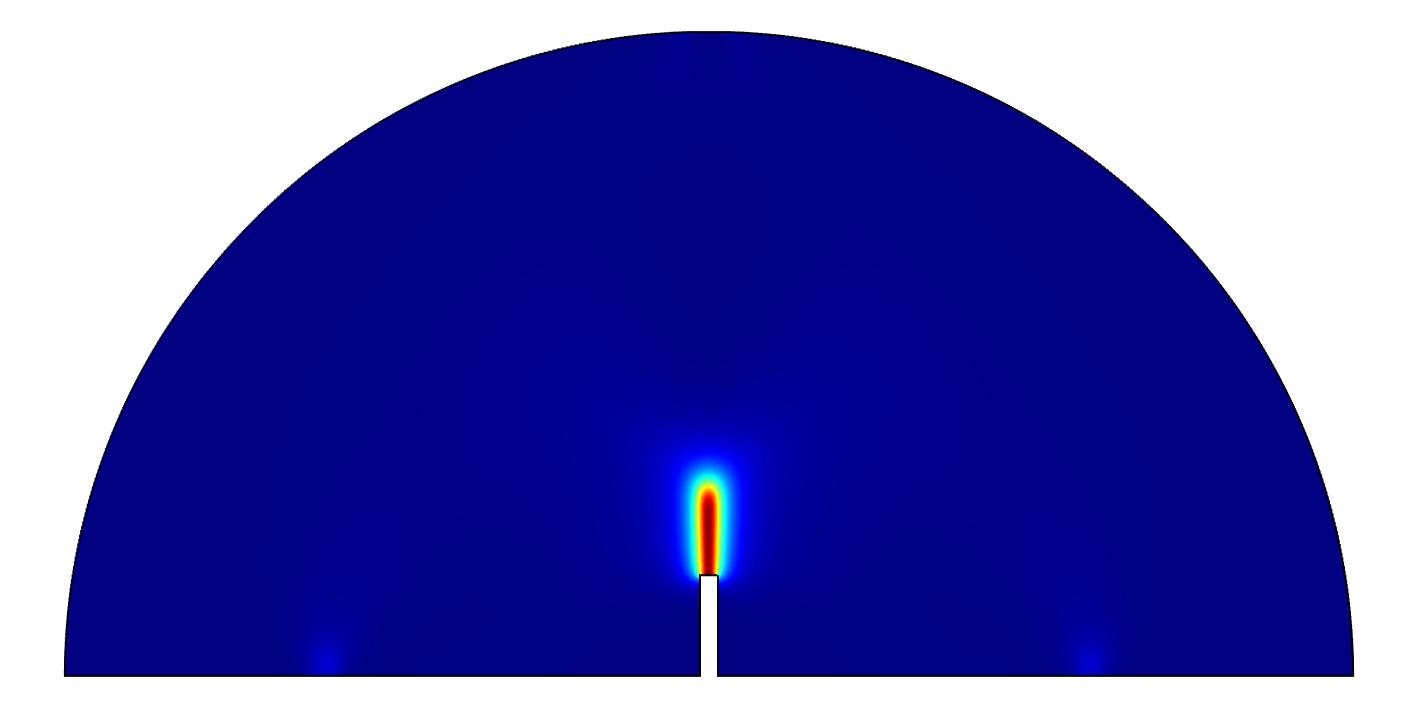}}
	
	\subfigure[]{\includegraphics[width = 6cm]{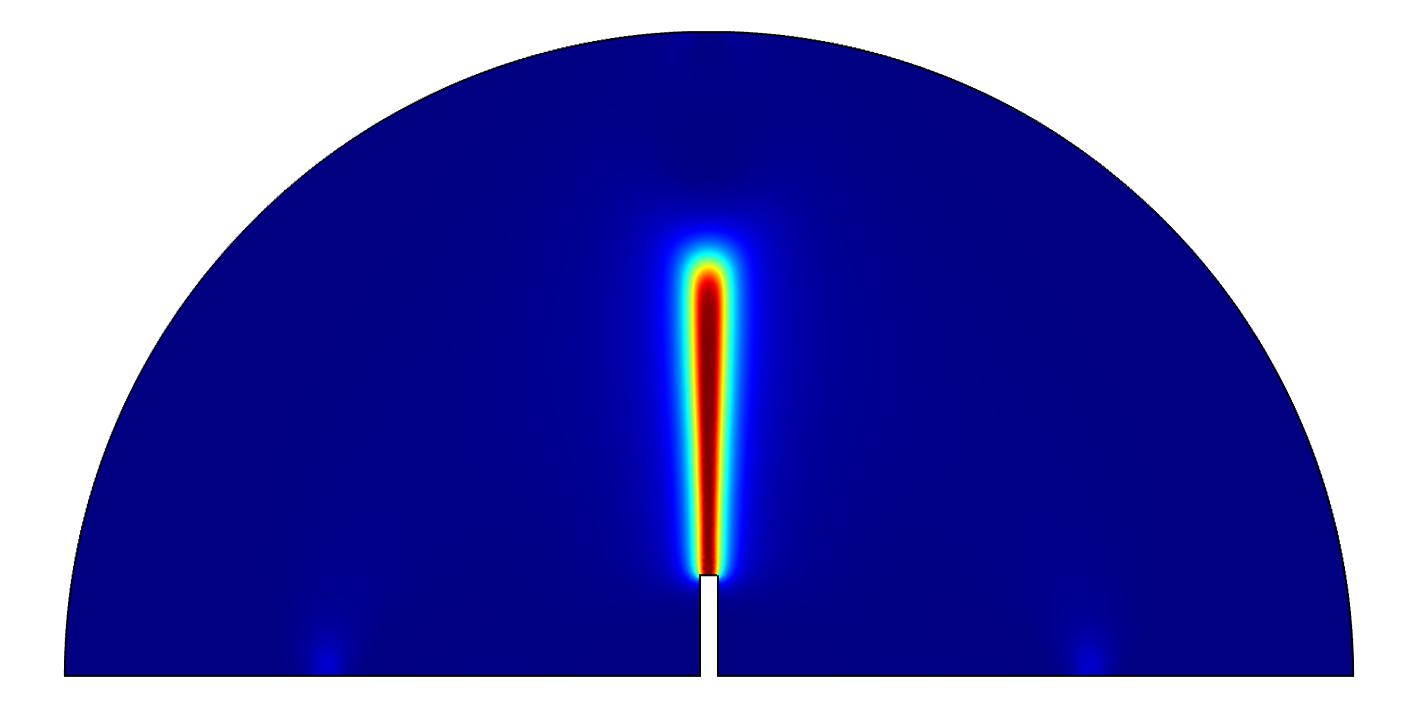}}
	\subfigure[]{\includegraphics[width = 6cm]{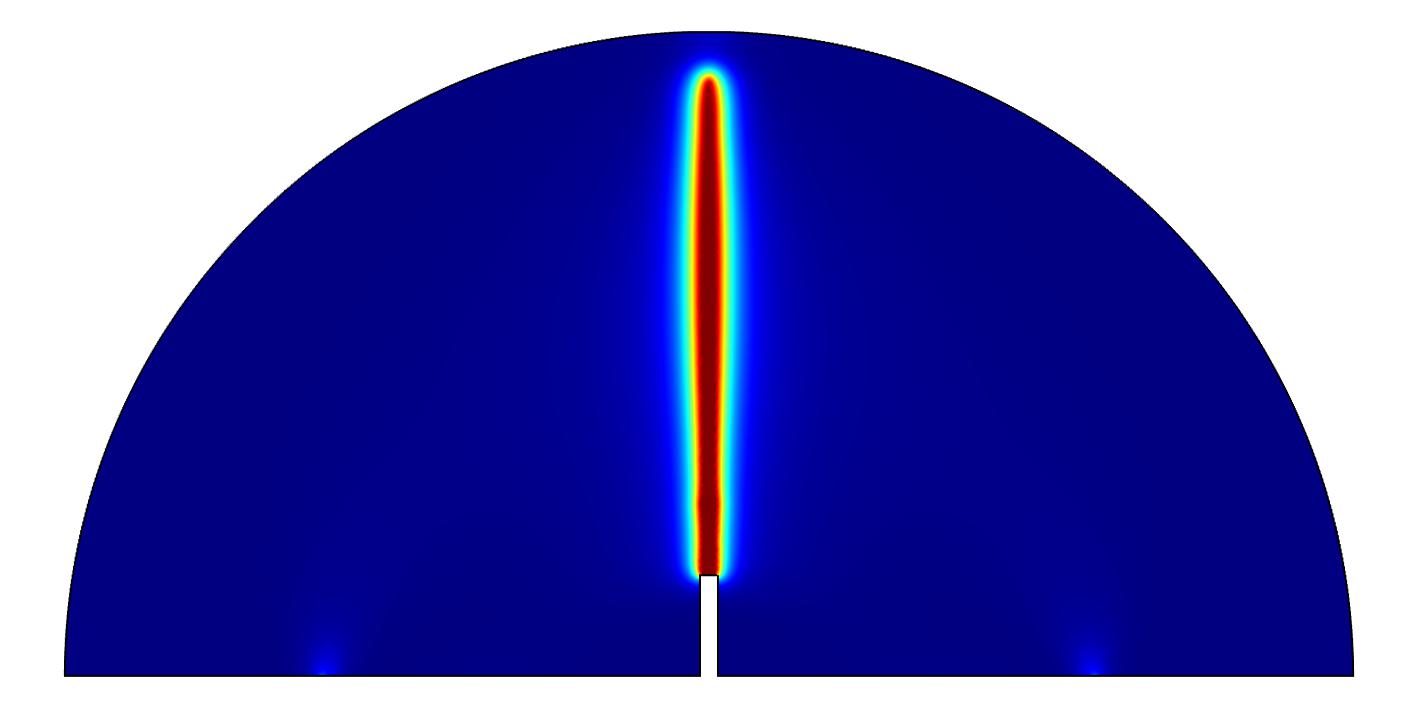}}
	\caption{Crack propagation of the notched semi-circular bend (NSCB) tests at a displacement of (a) $u = 6.72\times10^{-3}$ mm, (b) $u = 6.77\times10^{-3}$ mm, (c) $u = 6.74\times10^{-3}$ mm, and (d) $u=6.86\times10^{-3}$ mm for $G_c=7.6$ J/m$^2$}
	\label{Crack propagation of the notched semi-circular bend (NSCB) tests}
	\end{figure}

Figure \ref{Crack propagation of the notched semi-circular bend (NSCB) tests} shows the crack initiation and propagation in the rock specimen by using the phase field model.  When the applied displacement $u$ reaches to $6.72\times10^{-3}$ mm, the crack initiates from the tip of the notch. Subsequently, the crack propagates straightly in the vertical direction when $u$ accumulates to  $6.74\times10^{-3}$ mm and $6.77\times10^{-3}$ mm. When the displacement reaches to $6.86\times10^{-3}$ mm, the tip of the propagating crack is close to the upper boundary of the specimen and failure of the semi-circular rock specimen occurs. The crack patterns obtained by the phase field simulation are in good agreement with the experimental results in \citet{gao2015application}.

\begin{figure}[htbp]
	\centering
	\includegraphics[width = 10cm]{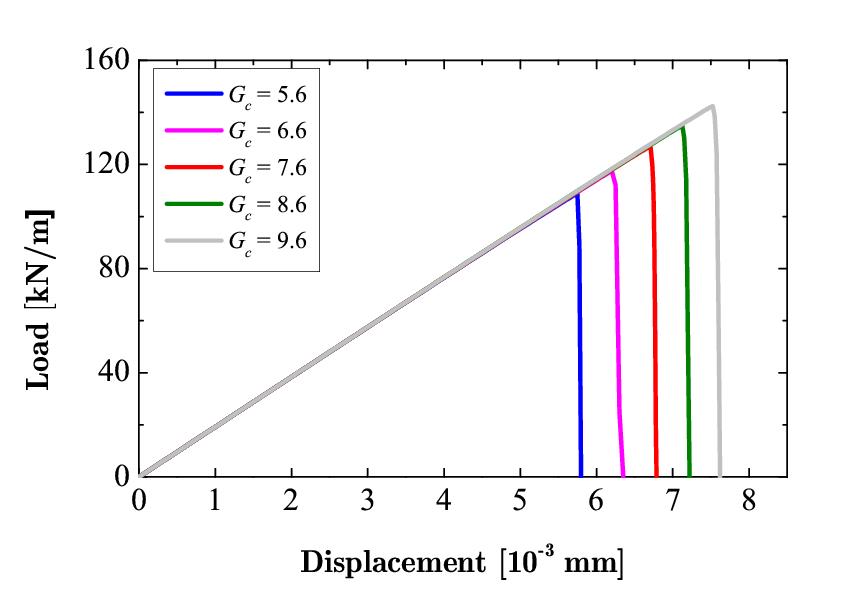}
	\caption{Load-displacement curves of the notched semi-circular bend (NSCB) tests}
	\label{Load-displacement curves of the notched semi-circular bend (NSCB) tests}
	\end{figure}

We test the crack patterns for $G_c = 5.6$, 6.6, 8.6, and 9.6 J/m$^2$. The results show that $G_c$ has little influence on the crack patterns. Different $G_c$ have the same crack paths. Figure \ref{Load-displacement curves of the notched semi-circular bend (NSCB) tests} gives the curves of the reaction force on the upper boundary of the specimen versus the displacement for different $G_c$. A sudden drop of the load is observed after the crack initiation. In addition, all the curves have the same slope before the crack initiation under different $G_c$, while the maximum load increases as $G_c$ increases .

\begin{figure}[htbp]
	\centering
	\includegraphics[width = 10cm]{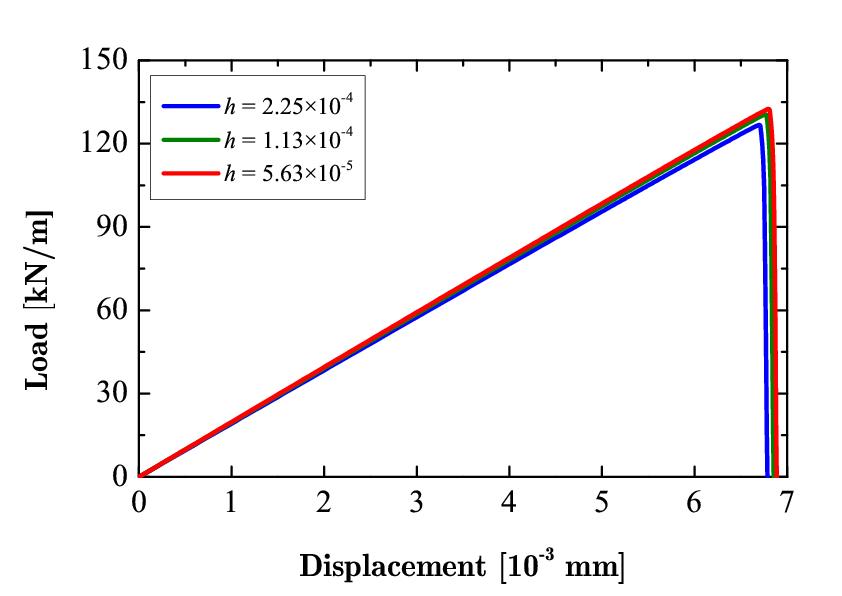}
	\caption{Load-displacement curves of the NSCB tests under different mesh size}
	\label{Load-displacement curves of the NSCB tests under different mesh size}
	\end{figure}

\begin{figure}[htbp]
	\centering
	\includegraphics[width = 10cm]{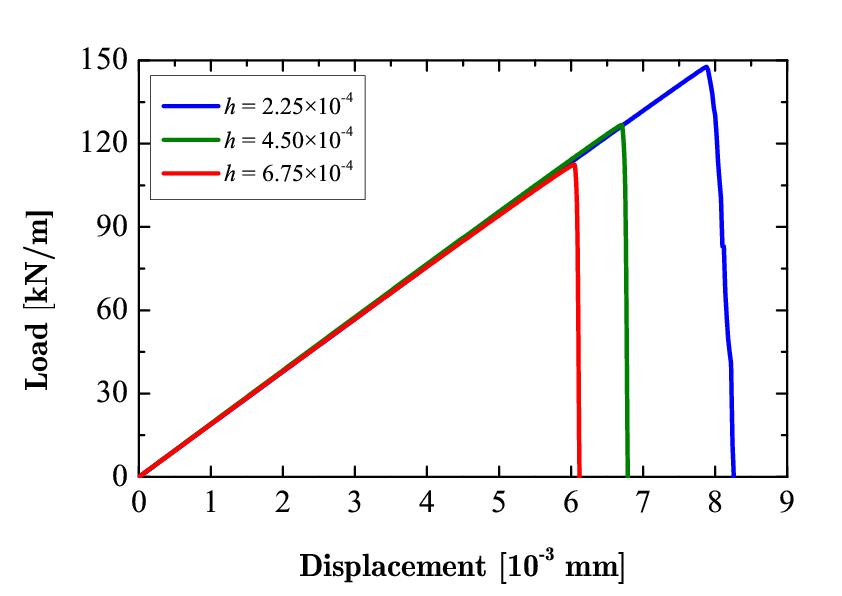}
	\caption{Load-displacement curves of the NSCB tests under different length scale $l_0$}
	\label{Load-displacement curves of the NSCB tests under different length scale}
	\end{figure}

We also test the influence of the maximum mesh size $h$ under a fixed $G_c$ = 7.6 J/m$^2$ and the length scale $l_0 = 4.5\times10^{-4}$ m. We choose $h =2.25\times10^{-4}$ m, $1.13\times10^{-4}$ m, and $5.63\times10^{-5}$ m in the tests. The resulting load-displacement curve is shown in Fig. \ref{Load-displacement curves of the NSCB tests under different mesh size}. As expected, the load-displacement curve converges with the mesh refinement. To test the influence of the length scale $l_0$, we fix the mesh size $h =2.25\times10^{-4}$ m and $G_c$ = 7.6 J/m$^2$, and then present the load-displacement curve for different $l_0$ in Fig. \ref{Load-displacement curves of the NSCB tests under different length scale}. Figure \ref{Load-displacement curves of the NSCB tests under different length scale} shows a decreasing peak load and displacement range when the length scale increases.

\subsection{Simulation of Brazil splitting tests}

\begin{figure}[htbp]
	\centering
	\includegraphics[width = 6cm]{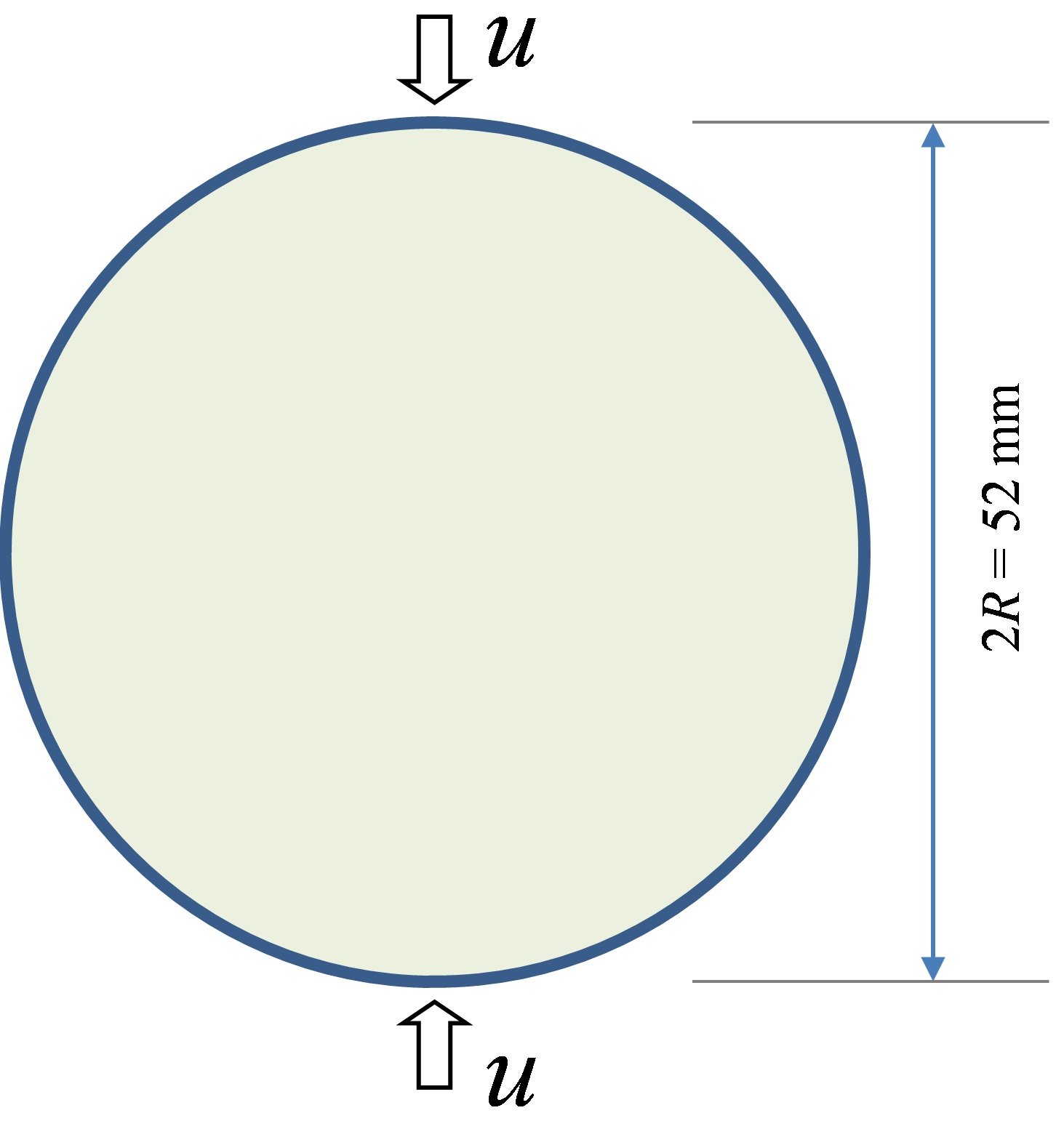}
	\caption{Geometry and boundary condition of the Brazil splitting tests}
	\label{Geometry and boundary condition of the Brazil splitting tests}
	\end{figure}

Brazil splitting tests are commonly used to obtain the tensile strength of rocks and many researchers simulated crack propagation in a Brazilian disc under compression, such as \citet{cai2013fracture} and \citet{zhou2016numerical}. Figure \ref{Geometry and boundary condition of the Brazil splitting tests} presents the geometry of the Brazilian disc along with the boundary conditions.

These material parameters are adopted: $\rho = 2630$ kg/m$^3$, $E$ = 31.5 GPa, and $\nu$ = 0.25. The length scale parameter $l_0$ is fixed to 1 mm. 26700 linear triangular elements (base mesh) are used to discretize the disc with the maximum element size $h=0.5$ mm, $k$ is set to $1 \times 10^{-9}$. Finally, we conduct the simulation by using five different $G_c$: 50, 75, 100, 125, and 150 J/m$^2$.

\begin{figure}[htbp]
	\centering
	\subfigure[]{\includegraphics[width = 6cm]{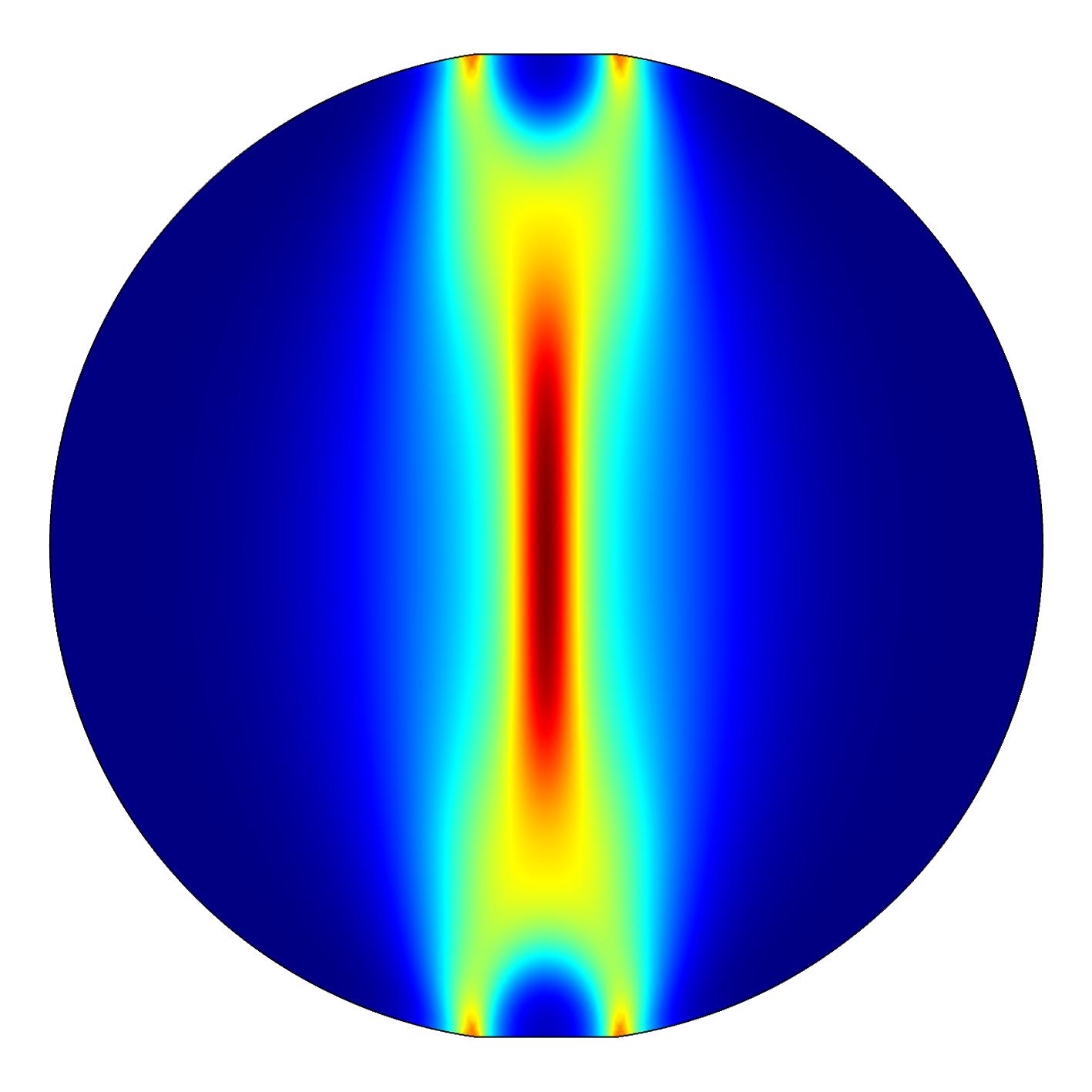}}
	\subfigure[]{\includegraphics[width = 6cm]{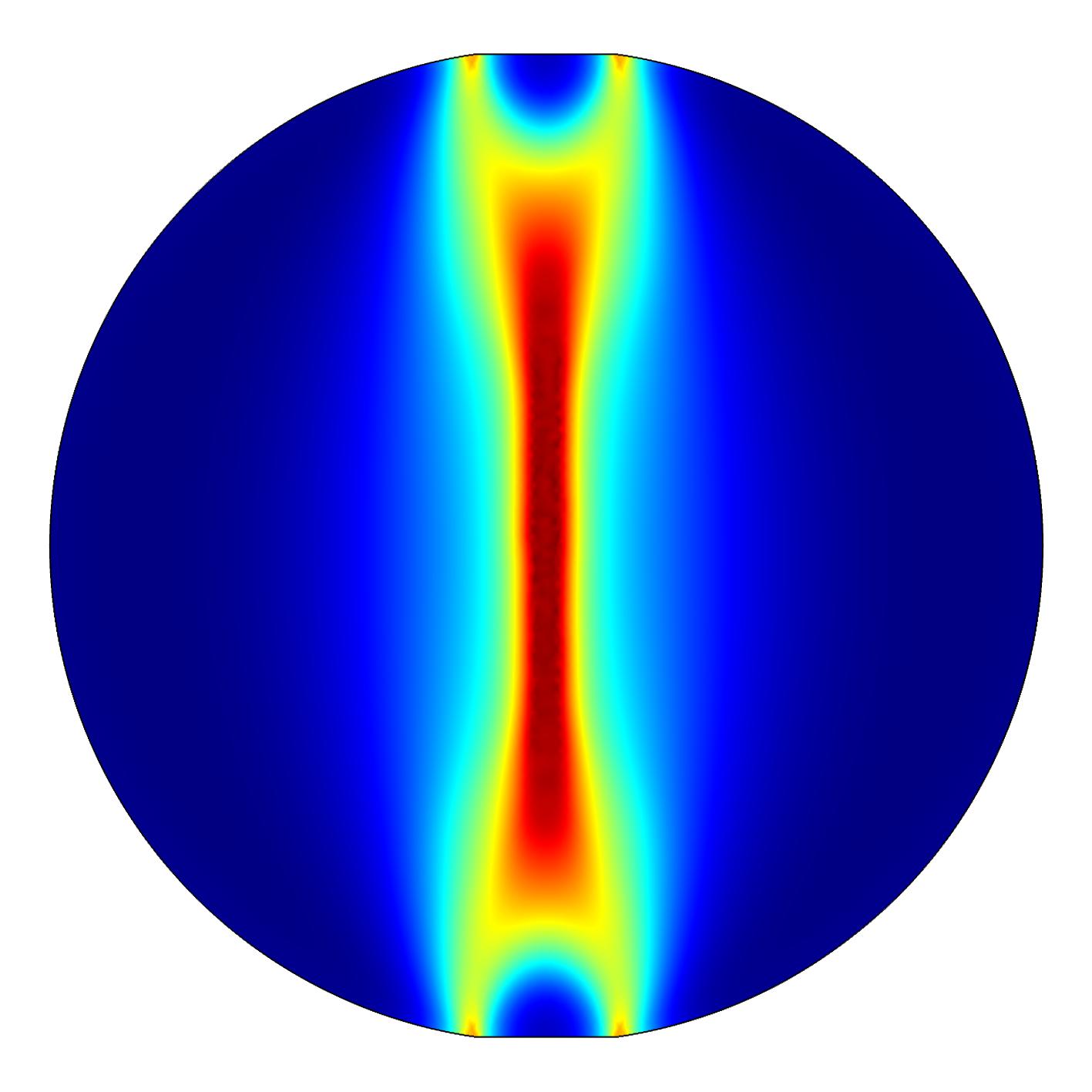}}
	
	\subfigure[]{\includegraphics[width = 6cm]{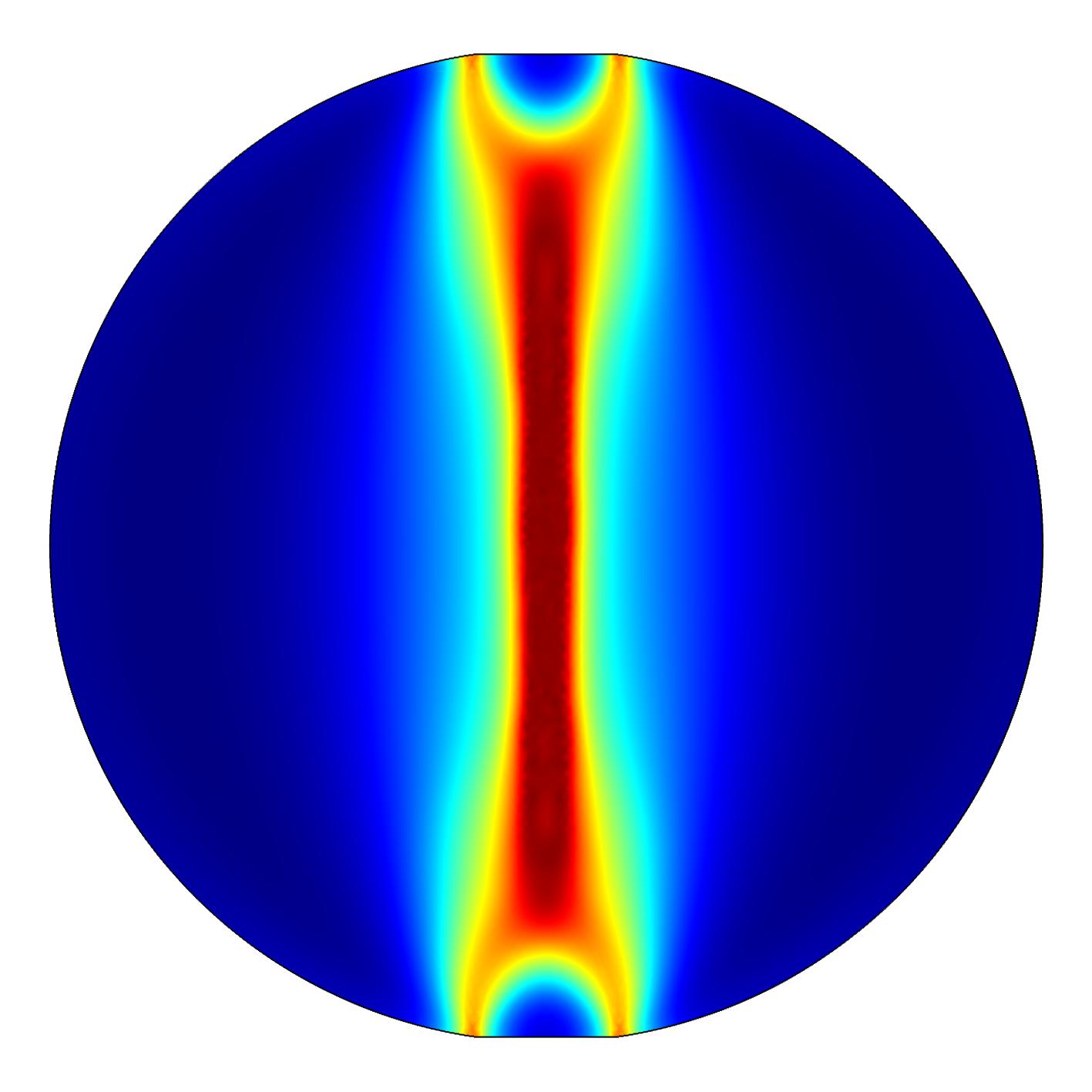}}
	\subfigure[]{\includegraphics[width = 6cm]{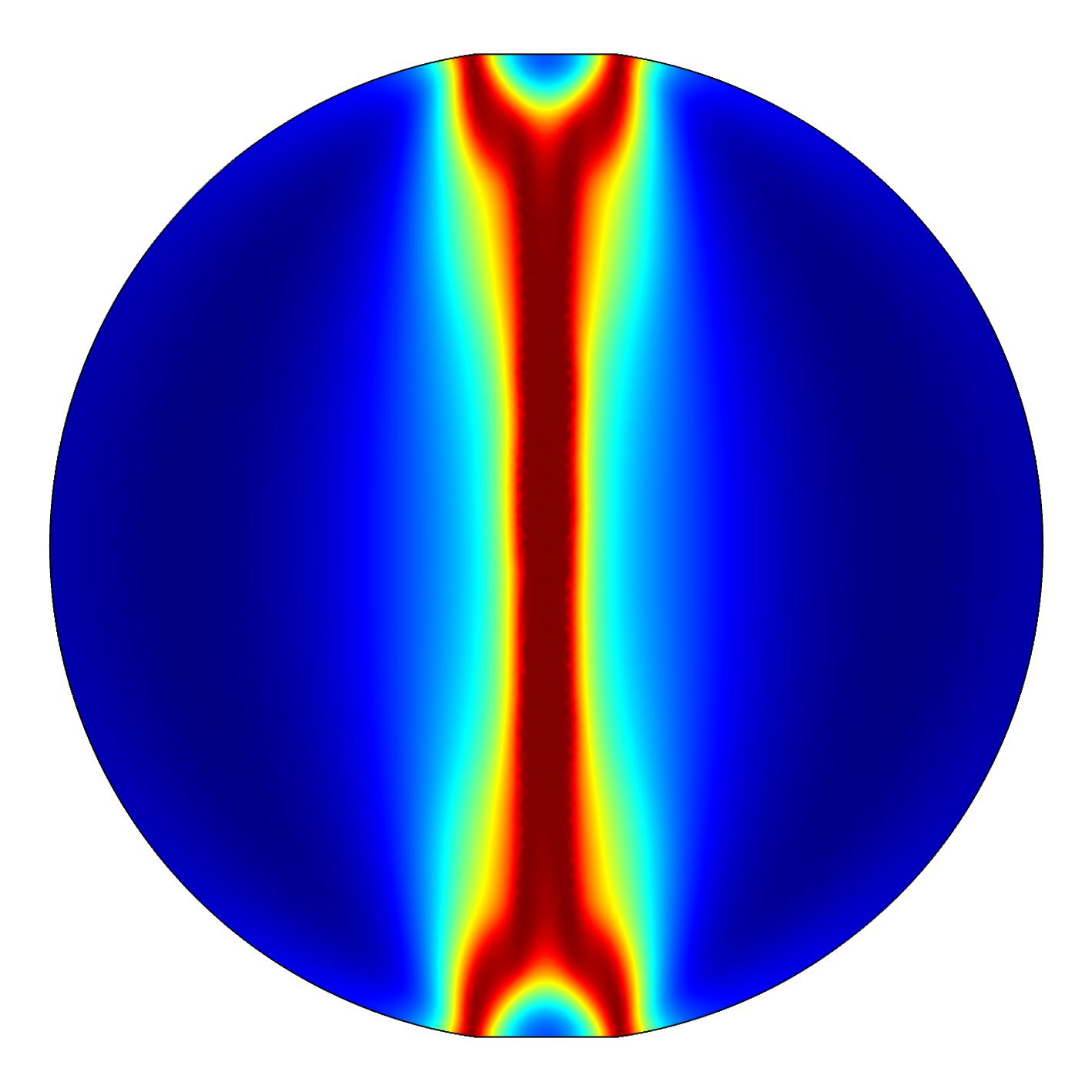}}
	\caption{Crack propagation of the Brazil splitting tests at a displacement of (a) $u = 0.476$ mm, (b) $u = 0.477$ mm, (c) $u = 0.478$ mm, and (d) $u=0.480$ mm for $G_c=100$ J/m$^2$}
	\label{Crack propagation of the Brazil splitting tests}
	\end{figure}

Figure \ref{Crack propagation of the Brazil splitting tests} shows the crack initiation and propagation in the Brazil splitting tests for $G_c$ = 100 J/m$^2$. When the displacement $u$ approaches a value of 0.476 mm, the crack occurs in the center of the disc where the maximum tensile stress occurs which is in good agreement with the experimental and analytical results \citep{atkinson1982combined,entacher2015rock}.  When $u$ reaches a value of 0.477 mm, the crack propagates with an increasing width. Then, the crack continues to propagate and the crack tips move close to the upper and bottom ends of the disc when $u$ reaches to 0.478 mm. The crack branching is observed when $u$ is close to 0.480 mm. In addition, the crack cannot penetrate deeply into the ends of the disc because of the locally compressed area around both ends.

\begin{figure}[htbp]
	\centering
	\includegraphics[width = 10cm]{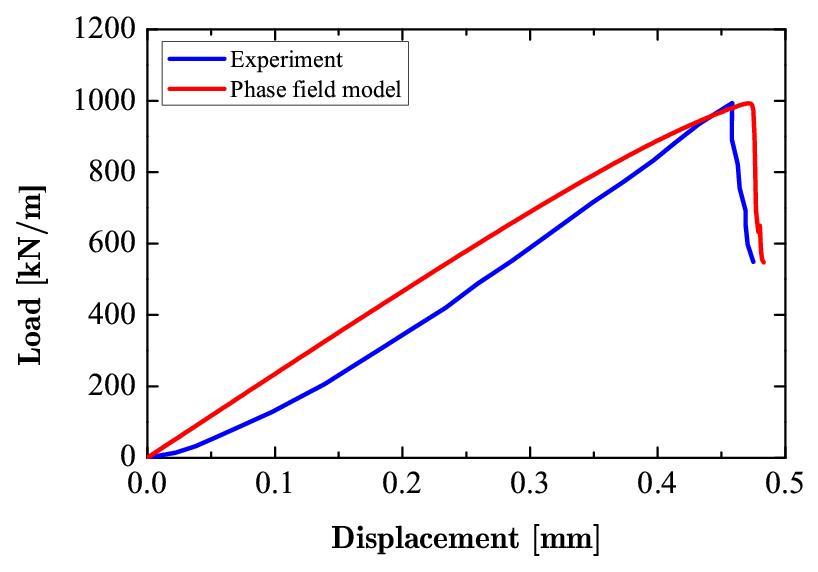}
	\caption{Comparison of the load-displacement curves obtained by the experimental test and phase field modeling}
	\label{Comparison of the load-displacement curves obtained by the experimental test and phase field modeling}
	\end{figure}

We compare the curve of the reaction force on the upper end of the Brazilian disc versus the displacement $u$ with the experimental result in Fig. \ref{Comparison of the load-displacement curves obtained by the experimental test and phase field modeling}. The experimental curve is originated from the work of \citet{erarslan2012investigating}. Figure \ref{Comparison of the load-displacement curves obtained by the experimental test and phase field modeling} shows that the phase field model can reproduce the experimental results well. The main reason for the difference in Fig. \ref{Comparison of the load-displacement curves obtained by the experimental test and phase field modeling} is that there are contact issues in the actual Brazilian tests and a compaction stage is commonly observed before the elastic stage.

\begin{figure}[htbp]
	\centering
	\includegraphics[width = 10cm]{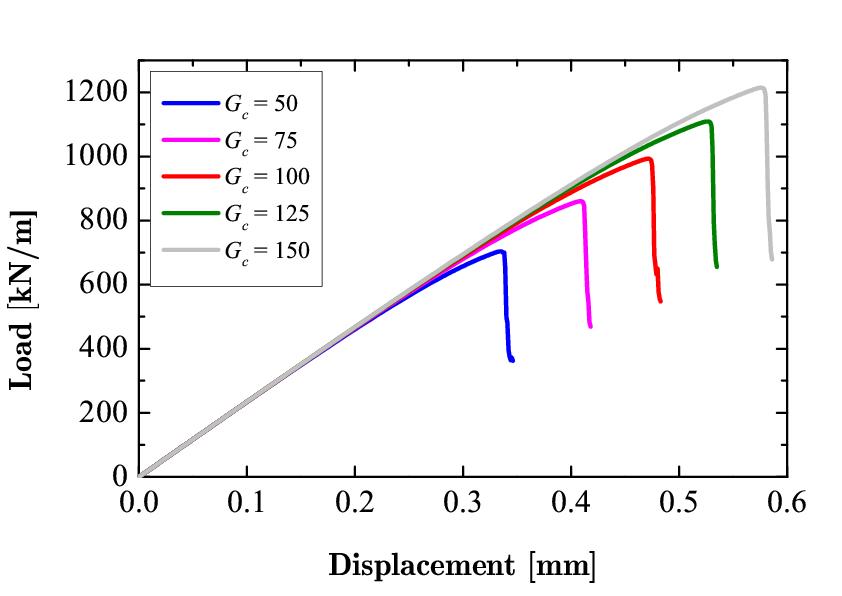}
	\caption{Load-displacement curves for the Brazil splitting tests for different $G_c$}
	\label{Load-displacement curves for the Brazil splitting tests for different Gc}
	\end{figure}

\begin{figure}[htbp]
	\centering
	\includegraphics[width = 10cm]{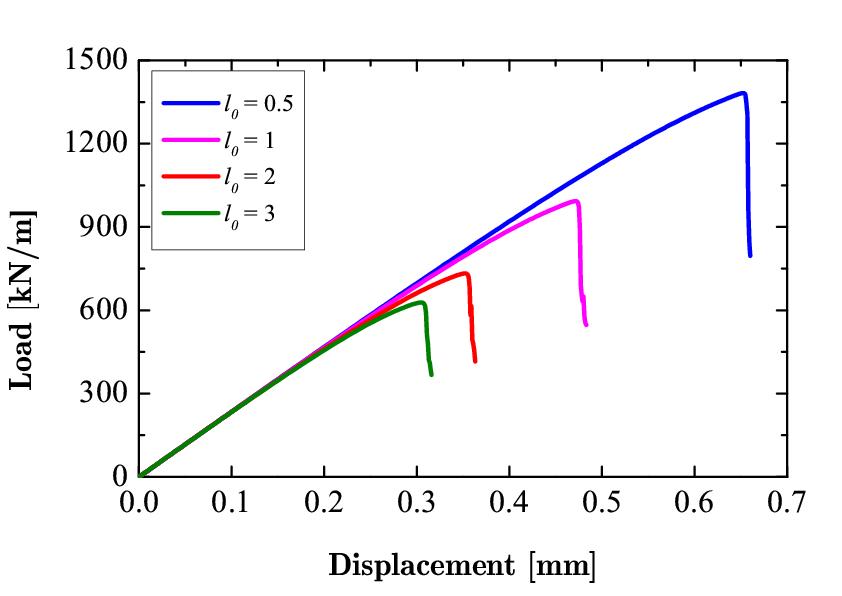}
	\caption{Load-displacement curves for the Brazil splitting tests for different $l_0$}
	\label{Load-displacement curves for the Brazil splitting tests for different l0}
	\end{figure}

Figure \ref{Load-displacement curves for the Brazil splitting tests for different Gc} presents the curves of the reaction force on the upper end of the Brazilian disc versus the displacement $u$ for different $G_c$. Similar to the NSCB tests, a sudden drop of the load is observed after the phase field increases to 1. In addition, the peak load increases with the increase in $G_c$. We then test the influence of the length scale $l_0$ under a fixed $G_c=100$ J/m$^2$ and mesh size $h=0.5$ mm. The tested length scale parameters are $l_0=$ 0.5, 1, 2, and 3 mm, respectively. The load-displacement curves under different length scale $l_0$ are shown in Fig. \ref{Load-displacement curves for the Brazil splitting tests for different l0}.  As shown in Fig. \ref{Load-displacement curves for the Brazil splitting tests for different l0}, the peak load of the specimen decreases with the increase in the length scale $l_0$.

\begin{figure}[htbp]
	\centering
	\includegraphics[width = 10cm]{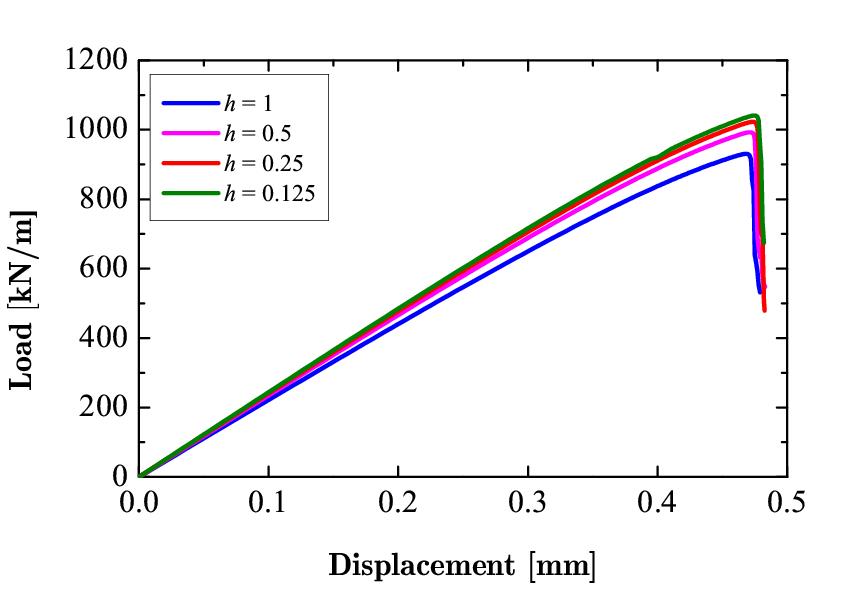}
	\caption{Load-displacement curves for the Brazil splitting tests for different $h$}
	\label{Load-displacement curves for the Brazil splitting tests for different h}
	\end{figure}

\begin{figure}[htbp]
	\centering
	\includegraphics[width = 10cm]{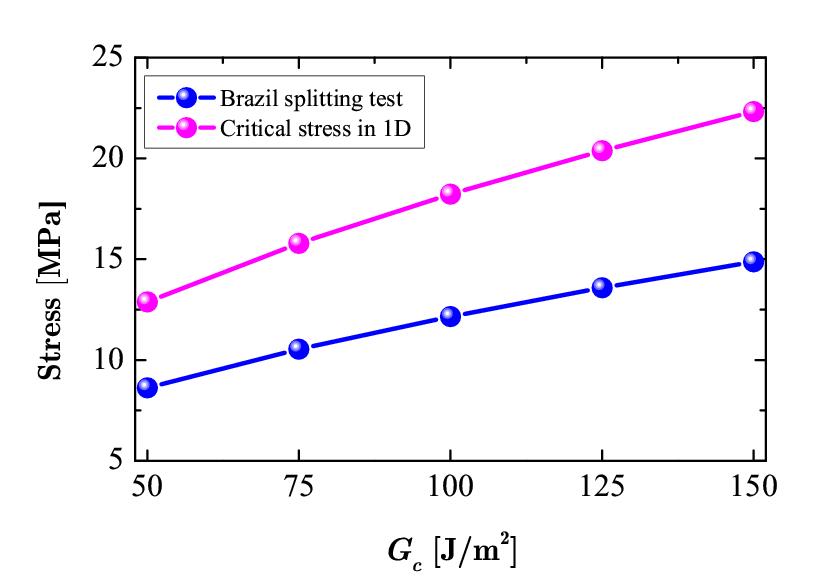}
	\caption{Comparison of tensile strength by Brazil splitting tests and critical stress for 1D}
	\label{Comparison of tensile strength by Brazil splitting tests and critical stress for 1D}
	\end{figure}

We also test the influence of mesh size $h$ under a fixed  $G_c=100$ J/m$^2$ and $l_0$ = 1 mm. The maximum mesh size $h$ is set as 1, 0.5, 0.25, and 0.125 m, respectively. The resulting load-displacement curve is shown in Fig. \ref{Load-displacement curves for the Brazil splitting tests for different h}. The load-displacement curve converges with mesh refinement as expected. In addition, in the Brazil splitting test, the tensile strength $\sigma_t$ of the rock specimen is given by
	\begin{equation}
	\sigma_t=\frac{2P_{peak}}{\pi D L}
	\end{equation}
 
\noindent where $P_{peak}$ is the peak load, and $D$ and $L$ are the diameter and length of the rock specimen. Thus, we compare the tensile strength by the phase field simulation and the critical stress for 1D by \citet{borden2012phase} in Fig. \ref{Comparison of tensile strength by Brazil splitting tests and critical stress for 1D}. The tensile strength increases at a decreasing rate with the increase in $G_c$. The critical stress has the same trend as the tensile strength. However, the critical stress is far larger than the tensile strength. This observation indicates that the critical stress for 1D analysis cannot be applied directly to 2D or 3D.

\subsection{Propagation of multiple echelon flaws}
\begin{figure}
	\centering
	\includegraphics[width = 6cm]{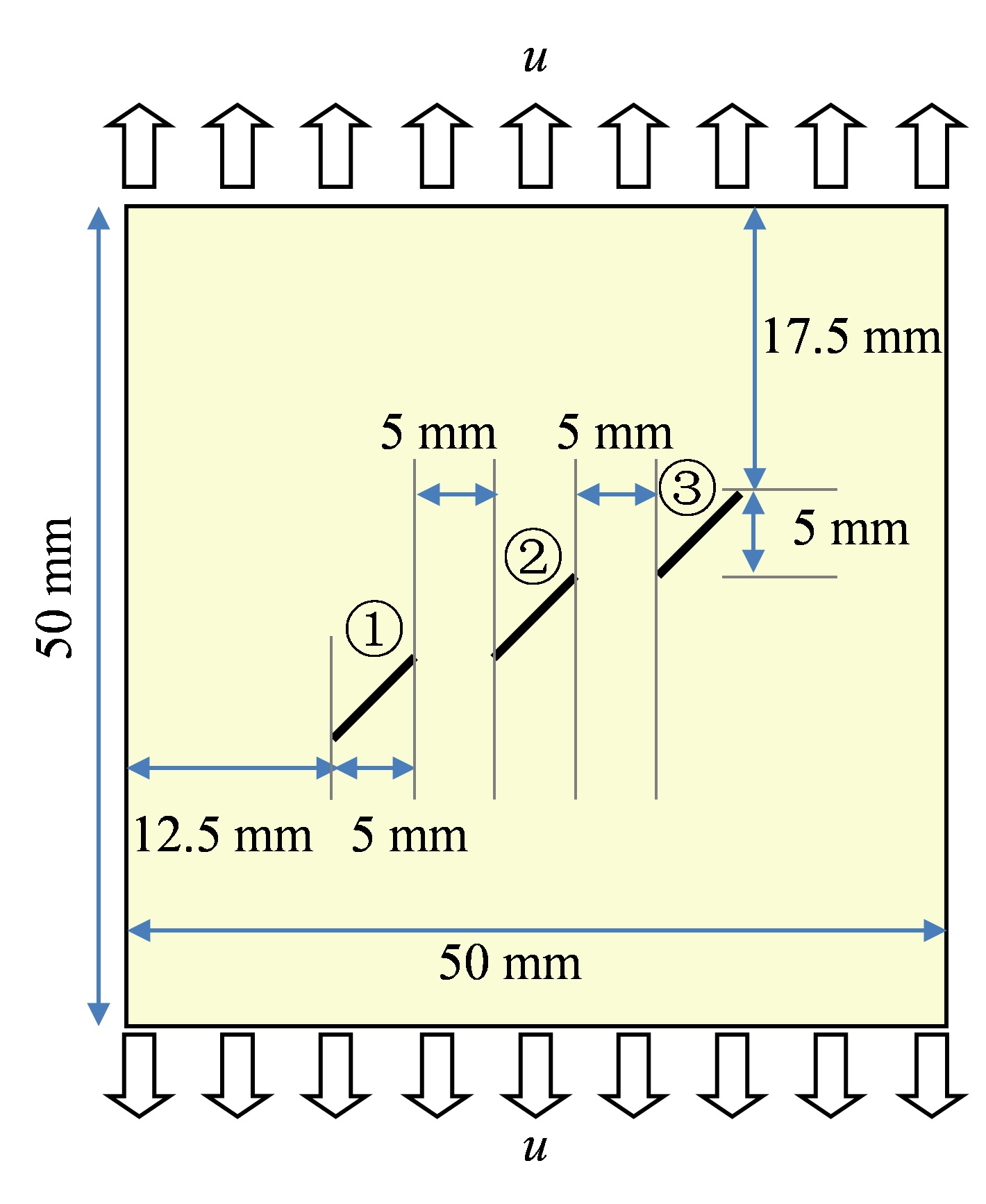}
	\caption{Geometry and boundary condition of three pre-existing flaws}
	\label{Geometry and boundary condition of the 3 flaws}
	\end{figure}

We consider a 50 mm $\times$ 50 mm square rock sample subjected to tension. The rock sample has three pre-existing flaws, whose position and geometry are shown in Fig. \ref{Geometry and boundary condition of the 3 flaws}.  All the flaws have the same length, spacing, and inclination angle of $45 ^\circ$. These parameters are adopted: the rock density $\rho=2500$ kg/m$^3$, the Young's modulus $E=30$ GPa, the Poisson's ratio $\nu = 0.333$, $G_c = 3$ J/m$^2$, $k=1\times10^{-5}$, and the length scale $l_0=0.25$ mm. Vertical displacements are applied on the top and bottom boundaries of the rock sample as shown in Fig. \ref{Geometry and boundary condition of the 3 flaws}. The simulation is performed by using 106852 6-node quadratic triangular elements where the maximum element size $h$ is 0.25 mm. In each time step, the displacement increment is $\Delta u  = 5\times10^{-7}$ mm.

\begin{figure}[htbp]
	\centering
	\subfigure[]{\includegraphics[width = 6cm]{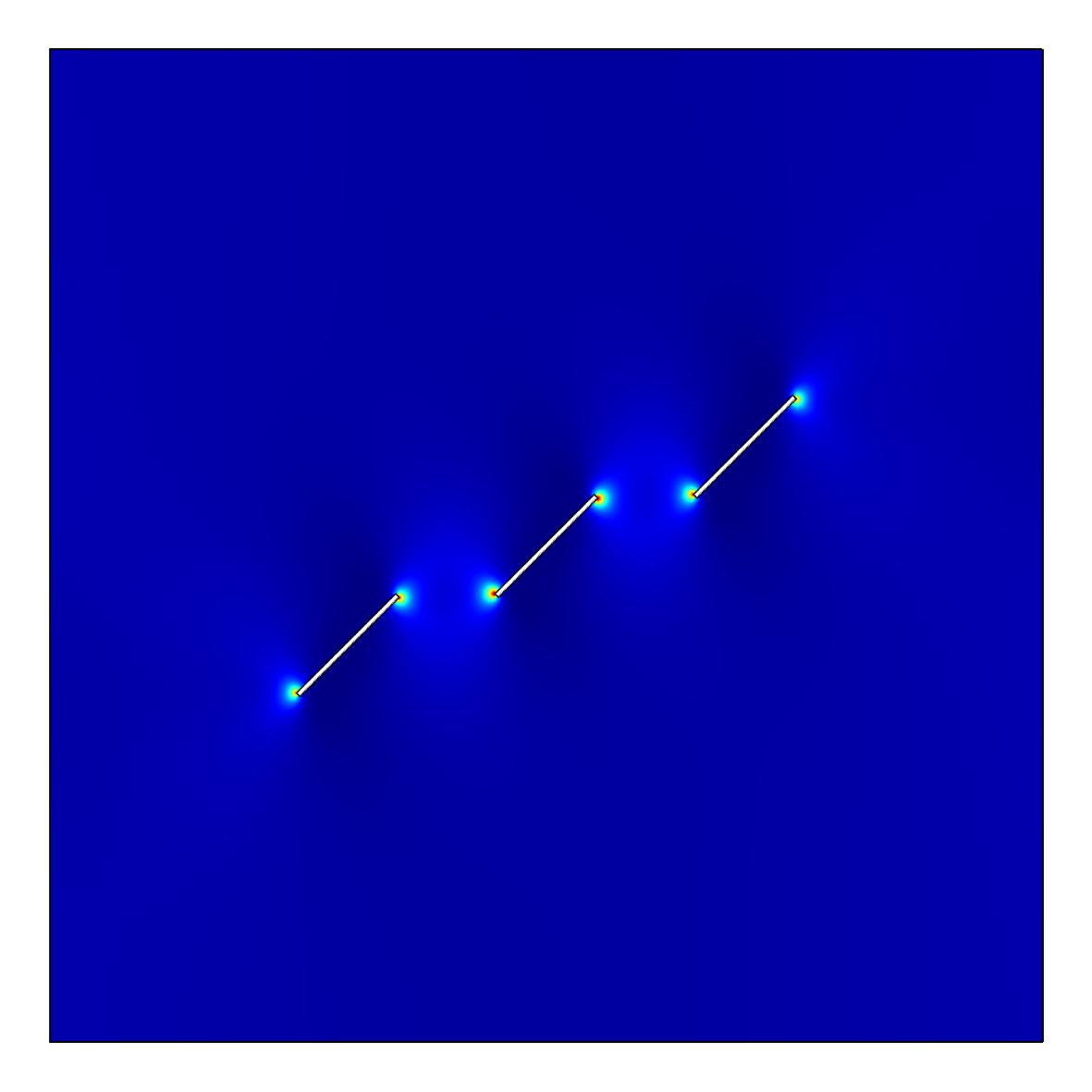}}
	\subfigure[]{\includegraphics[width = 6cm]{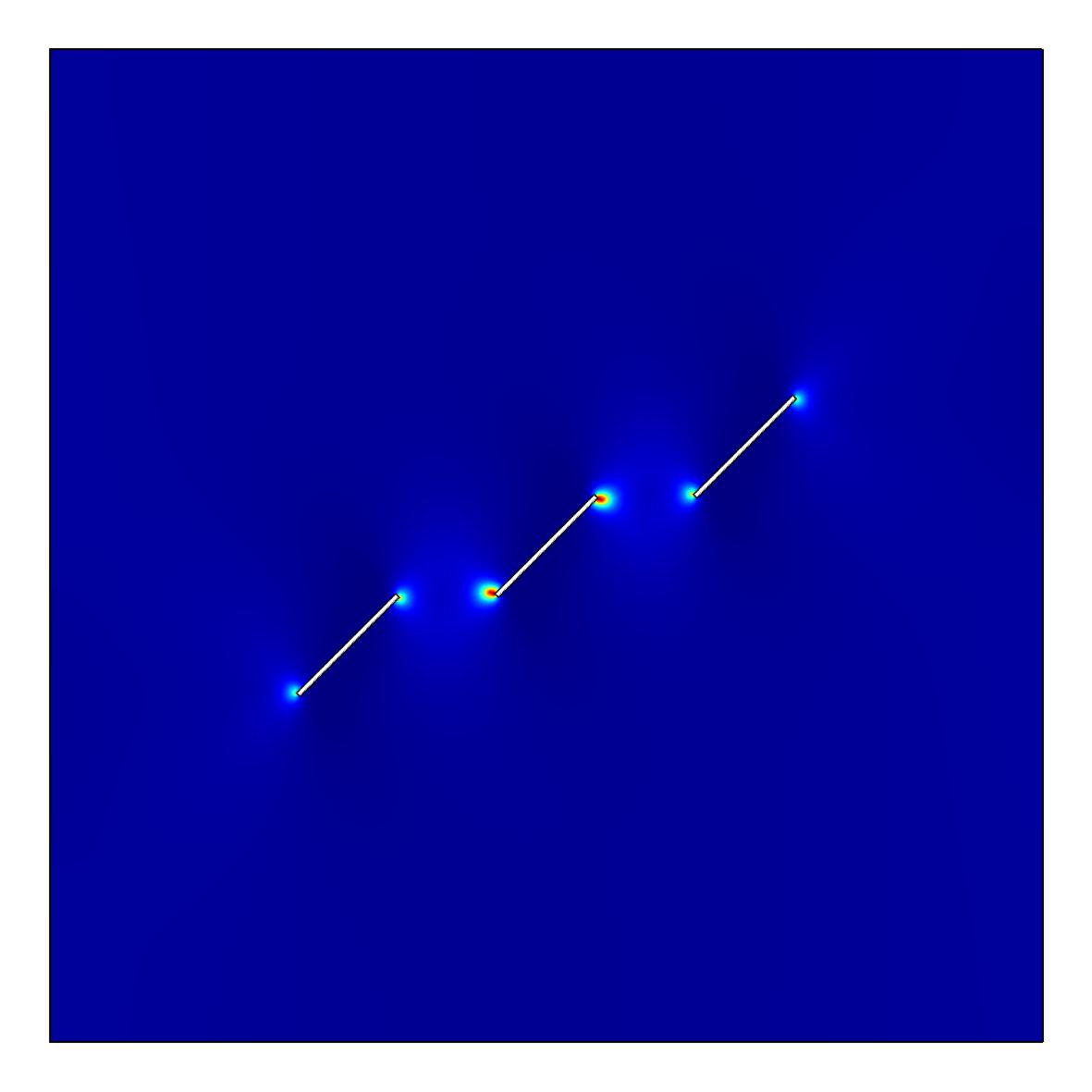}}
	
	\subfigure[]{\includegraphics[width = 6cm]{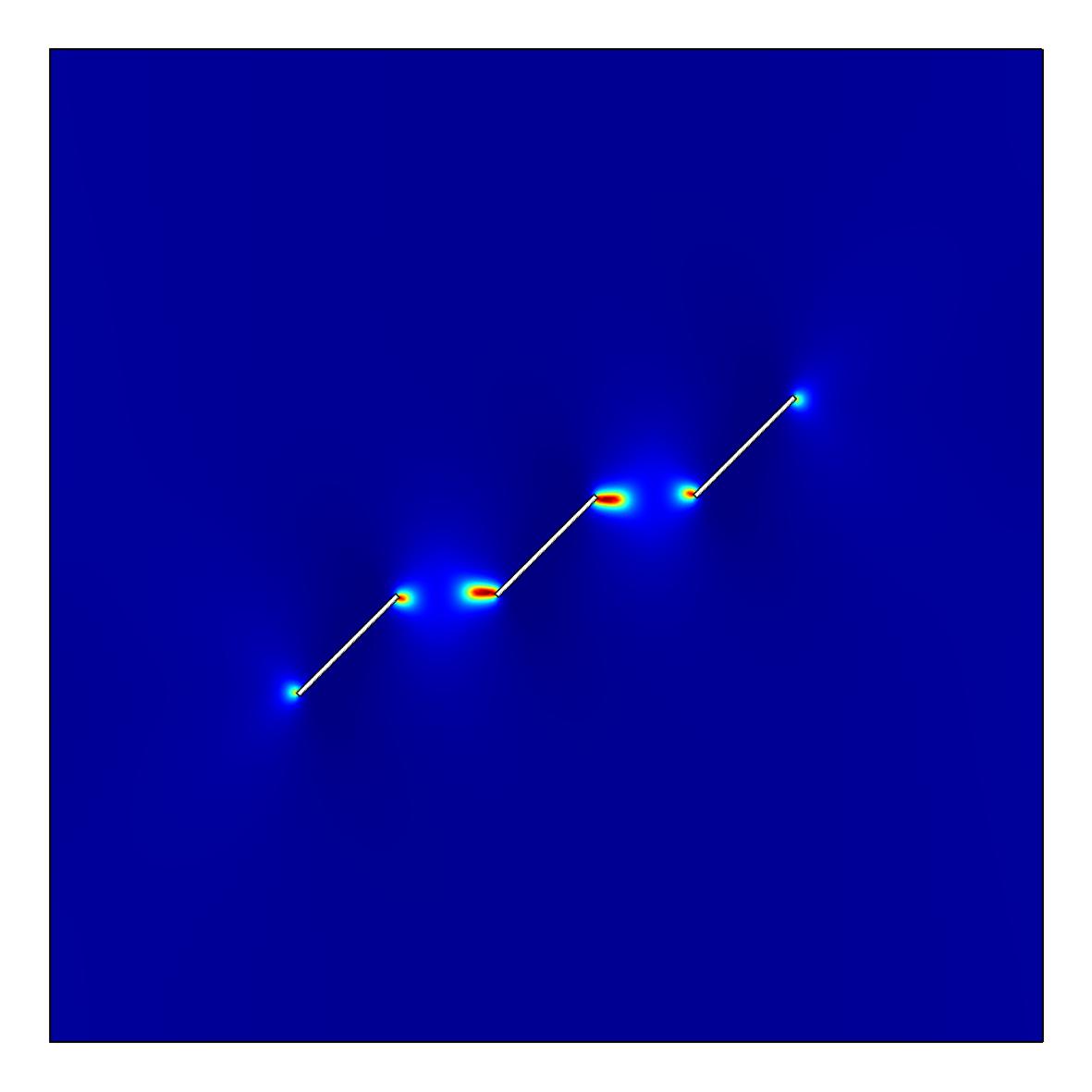}}
	\subfigure[]{\includegraphics[width = 6cm]{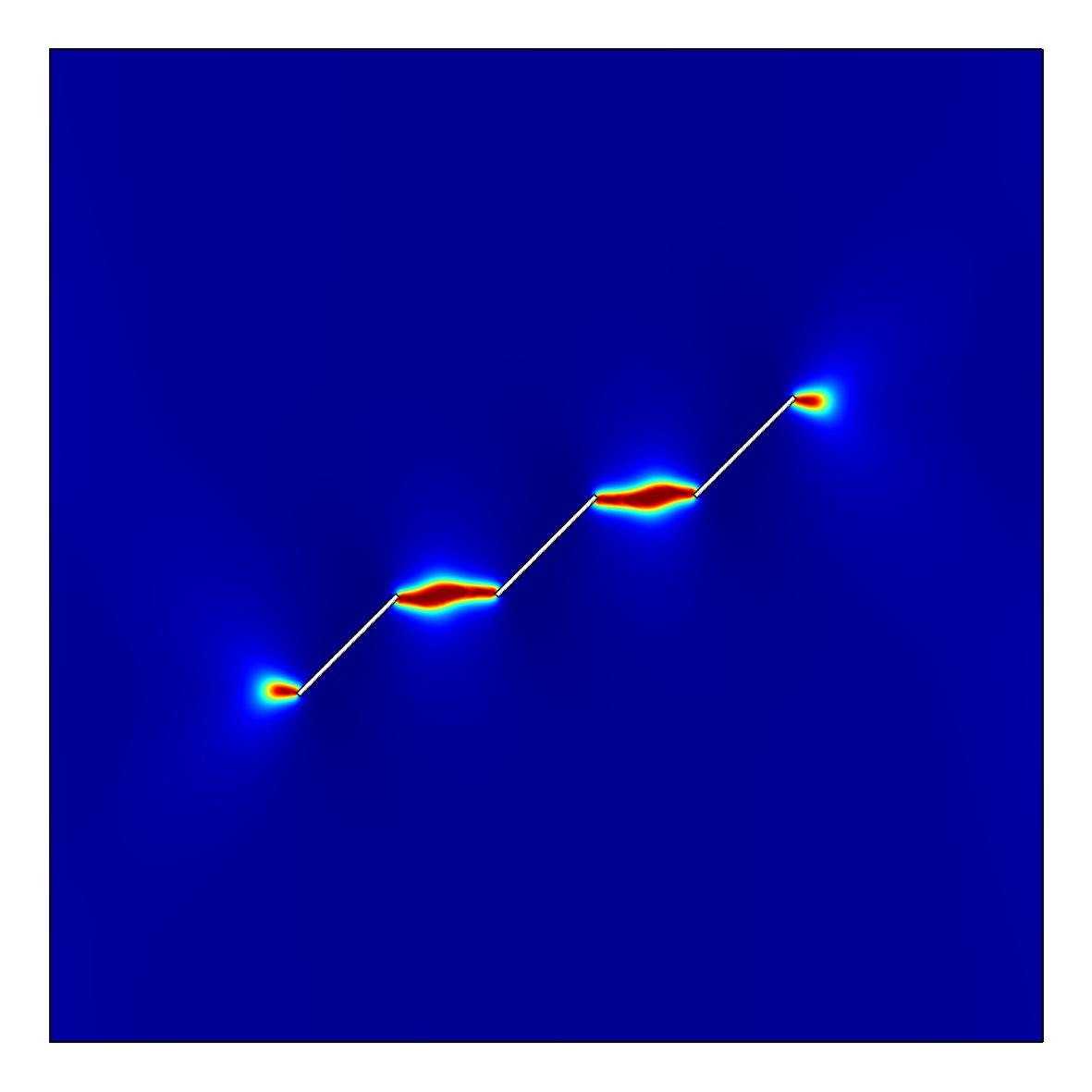}}

	\subfigure[]{\includegraphics[width = 6cm]{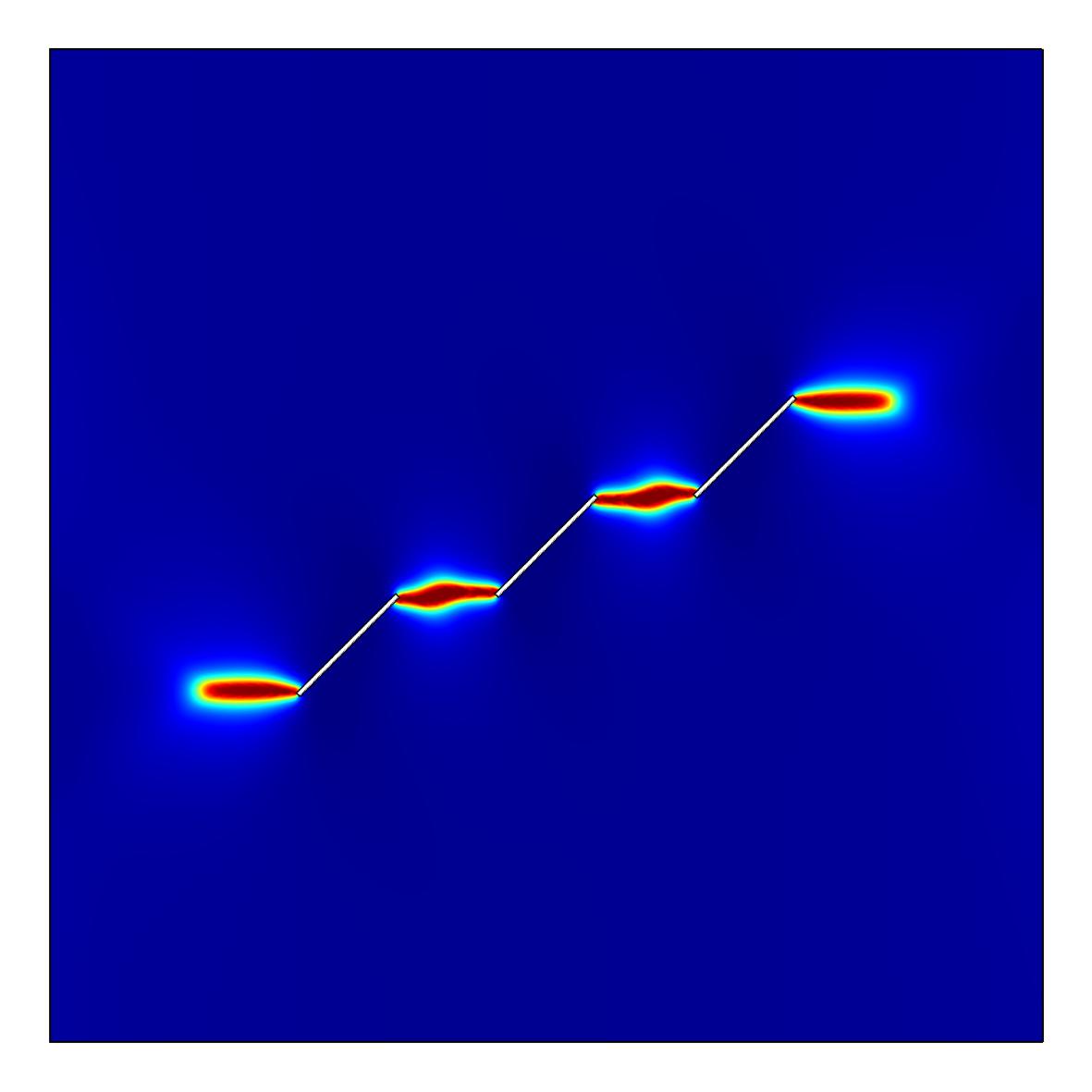}}
	\subfigure[]{\includegraphics[width = 6cm]{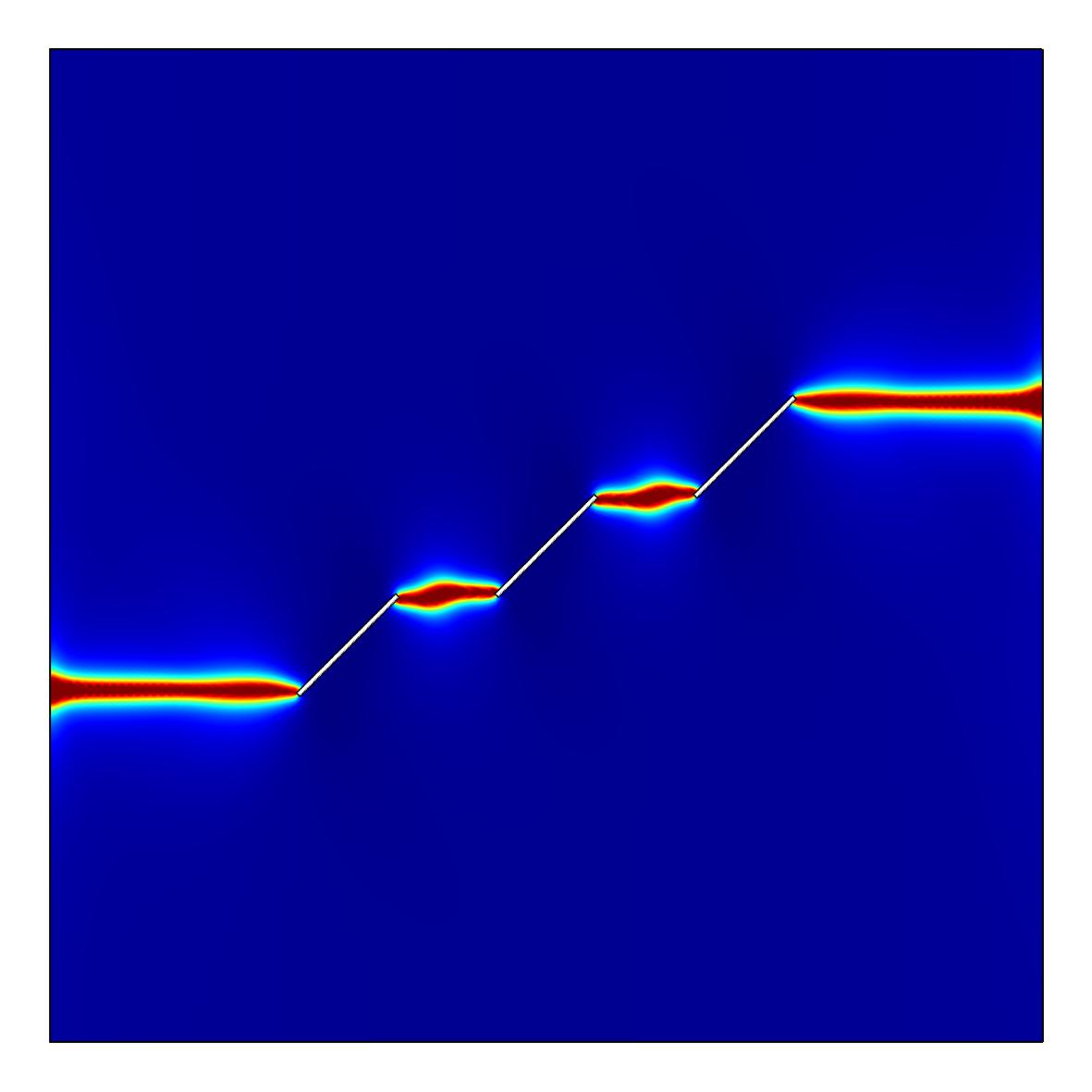}}
	\caption{Propagation and coalescence of the three pre-existing flaws in a square rock sample at a displacement of (a) $u = 2.6\times10^{-3}$ mm, (b) $u = 2.63\times10^{-3}$ mm, (c) $u = 2.64\times10^{-3}$ mm,  (d) $u=2.65\times10^{-3}$ mm, (e) $u=2.66\times10^{-3}$ mm, and (f) $u=2.69\times10^{-3}$ mm}
	\label{Propagation and coalescence of the three pre-existing flaws in a square rock sample}
	\end{figure}

\begin{figure}[htbp]
	\centering
	\includegraphics[width = 10cm]{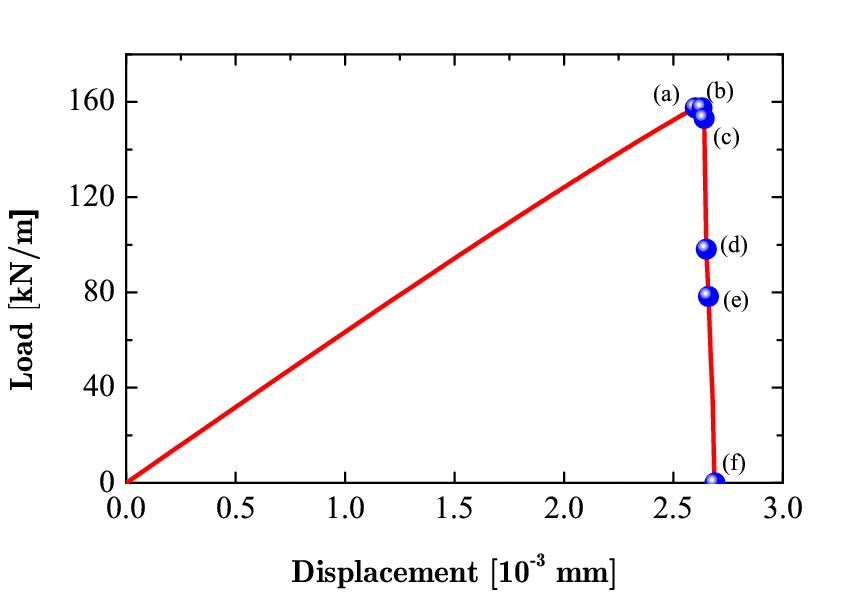}
	\caption{Load-displacement curves for the square rock sample with three pre-existing flaws}
	\label{Load-displacement curves for the square rock sample with three pre-existing flaws}
	\end{figure}

Figure \ref{Propagation and coalescence of the three pre-existing flaws in a square rock sample}(a)-(f) shows the propagation and coalescence of the three pre-existing flaws. We also calculate the reaction force on the upper boundary of the rock sample, and present the load-displacement curves in Fig. \ref{Load-displacement curves for the square rock sample with three pre-existing flaws}. As the displacement $u$ increases, both the load and the phase field around the tips of the flaws increase. When $u$ reaches to $2.6\times10^{-3}$ mm, the phase field increases close to 1 and the load approaches to the maximum value. When $u$ reaches to $2.63\times10^{-3}$ mm, the first tensile cracks occurs around the left and right tips of the flaw $\textcircled{2}$.  These two cracks are perpendicular to the direction of the applied displacement. In addition, the load reaches to the maximum as shown in Fig. \ref{Load-displacement curves for the square rock sample with three pre-existing flaws}. When the displacement $u$ reaches to $2.64\times10^{-3}$ mm, the cracks from the tips of the flaw $\textcircled{2}$ propagates perpendicular to $u$ while the load starts to drop after the peak load. Additionally, new cracks occur from the right tip of the flaw $\textcircled{1}$ and the left tip of the flaw $\textcircled{3}$. 

Figure \ref{Propagation and coalescence of the three pre-existing flaws in a square rock sample}(d) shows the coalescences of the cracks from the tips of the flaws $\textcircled{1}$, $\textcircled{2}$, and $\textcircled{3}$ when $u=2.65\times10^{-3}$ mm. Thus, the flaw $\textcircled{2}$ links the flaws $\textcircled{1}$ and $\textcircled{3}$, which is accompanied by a sudden drop of the load after the maximum value in Fig. \ref{Load-displacement curves for the square rock sample with three pre-existing flaws}. At the same time, new cracks from the left tip of the flaw $\textcircled{1}$ and the right tip of the flaw $\textcircled{3}$ are observed. When $u=2.66\times10^{-3}$ mm, the cracks from the flaws $\textcircled{1}$ and $\textcircled{3}$ continue to propagate and the load decreases. When the displacement increases to $2.69\times10^{-3}$ mm, the cracks initiating from the left tip of the flaw $\textcircled{1}$ and the right tip of the flaw $\textcircled{3}$ reach the boundaries of the rock sample. This indicates the rock loses its load-bearing capacity, which can also be verified by Fig. \ref{Load-displacement curves for the square rock sample with three pre-existing flaws}.

\begin{figure}
	\centering
	\includegraphics[width = 6cm]{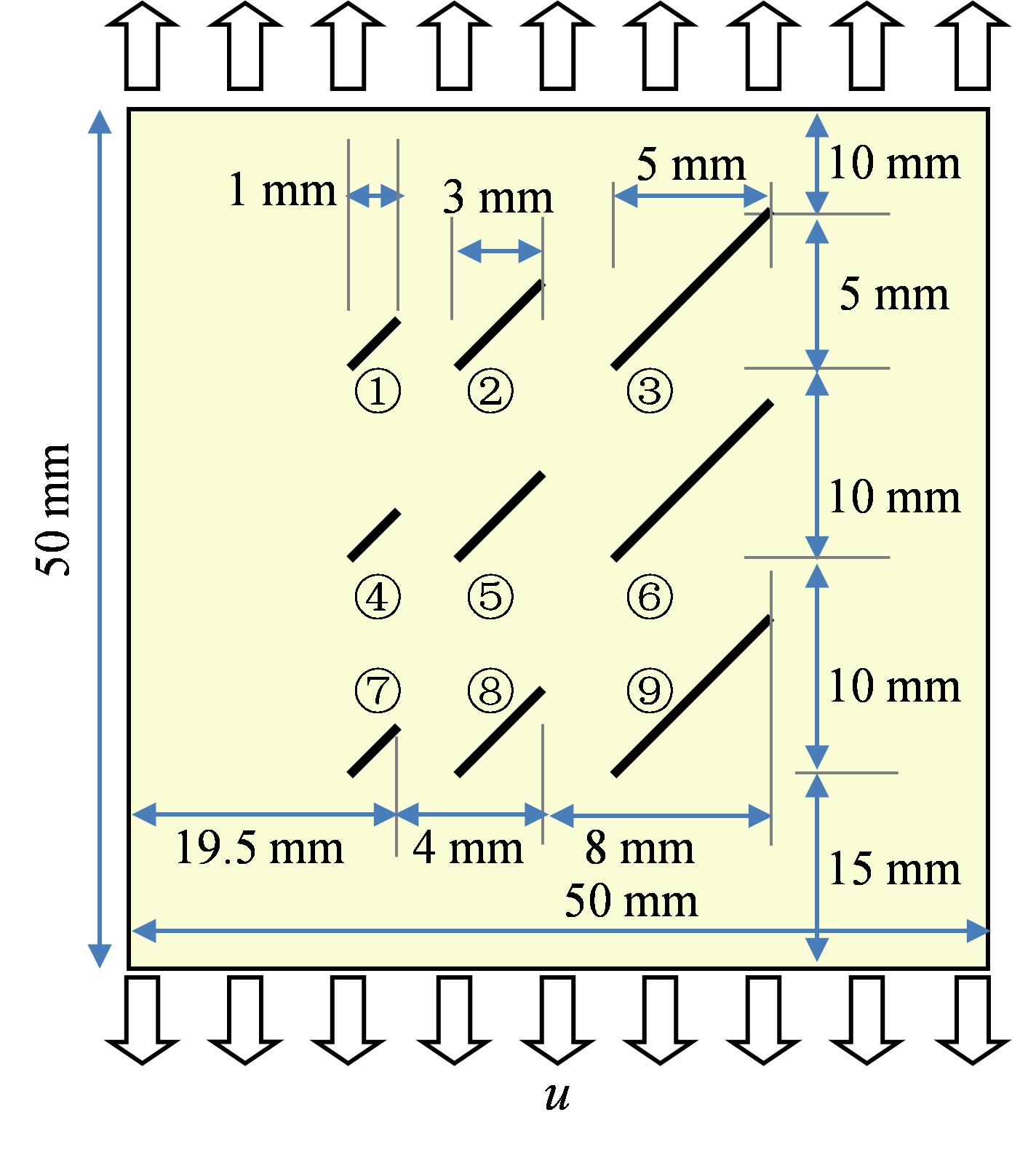}
	\caption{Geometry and boundary condition of nine pre-existing flaws}
	\label{Geometry and boundary condition of the 9 flaws}
	\end{figure}

We now consider the same square rock sample subjected to tension with nine pre-existing flaws in Fig.  \ref{Geometry and boundary condition of the 9 flaws}. These flaws have the same inclination angle of $45 ^\circ$, while the lengths and spacing are not fixed. The position and geometry of the flaws are  shown in Fig. \ref{Geometry and boundary condition of the 9 flaws}. The parameters are the same as those in the example of three flaws. A total of 107982 6-node quadratic elements are used to discretize the rock sample and the maximum element size $h$ is 0.25 mm. The displacement increment $\Delta u  = 1\times10^{-6}$ mm is applied in each time step.

\begin{figure}[htbp]
	\centering
	\subfigure[]{\includegraphics[width = 6cm]{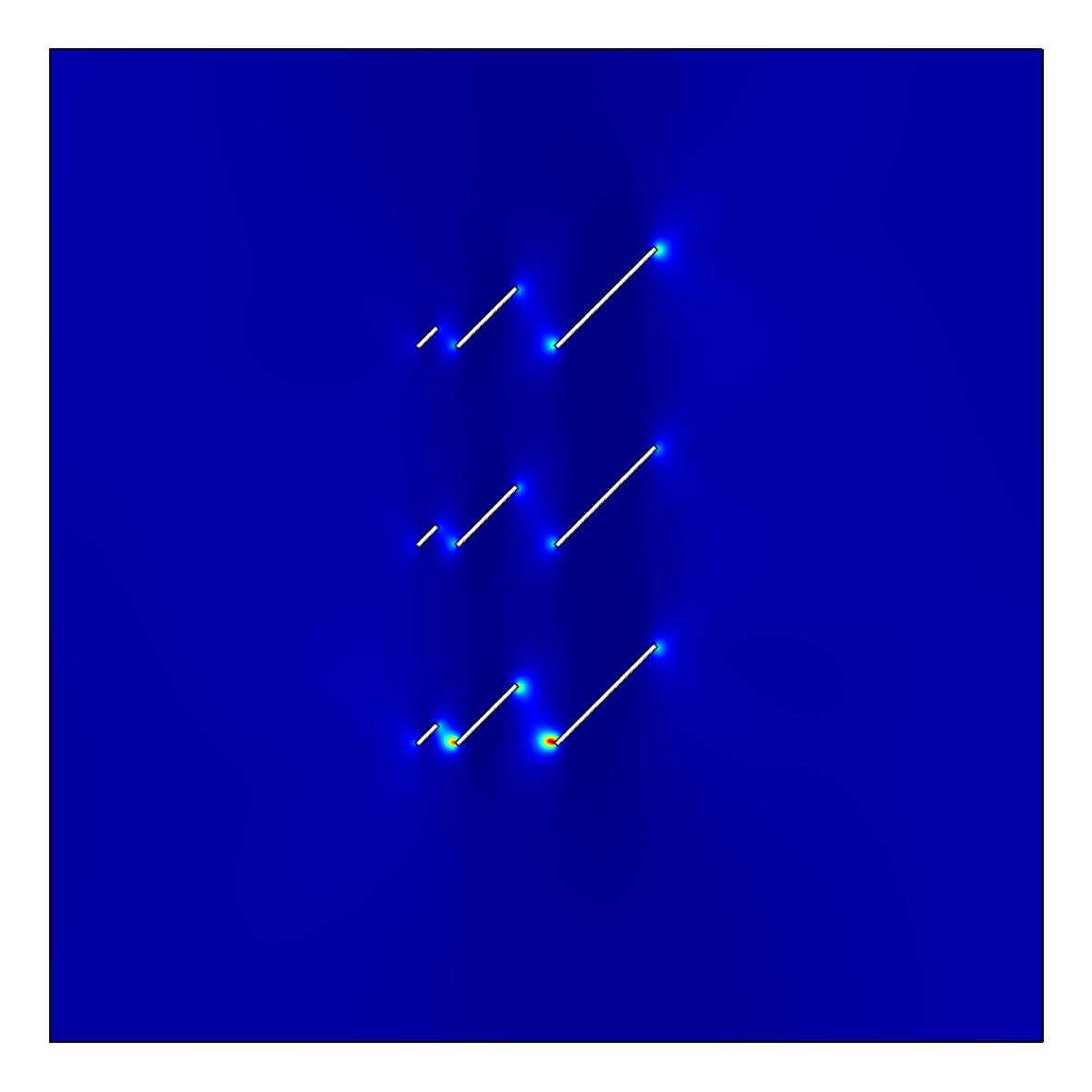}}
	\subfigure[]{\includegraphics[width = 6cm]{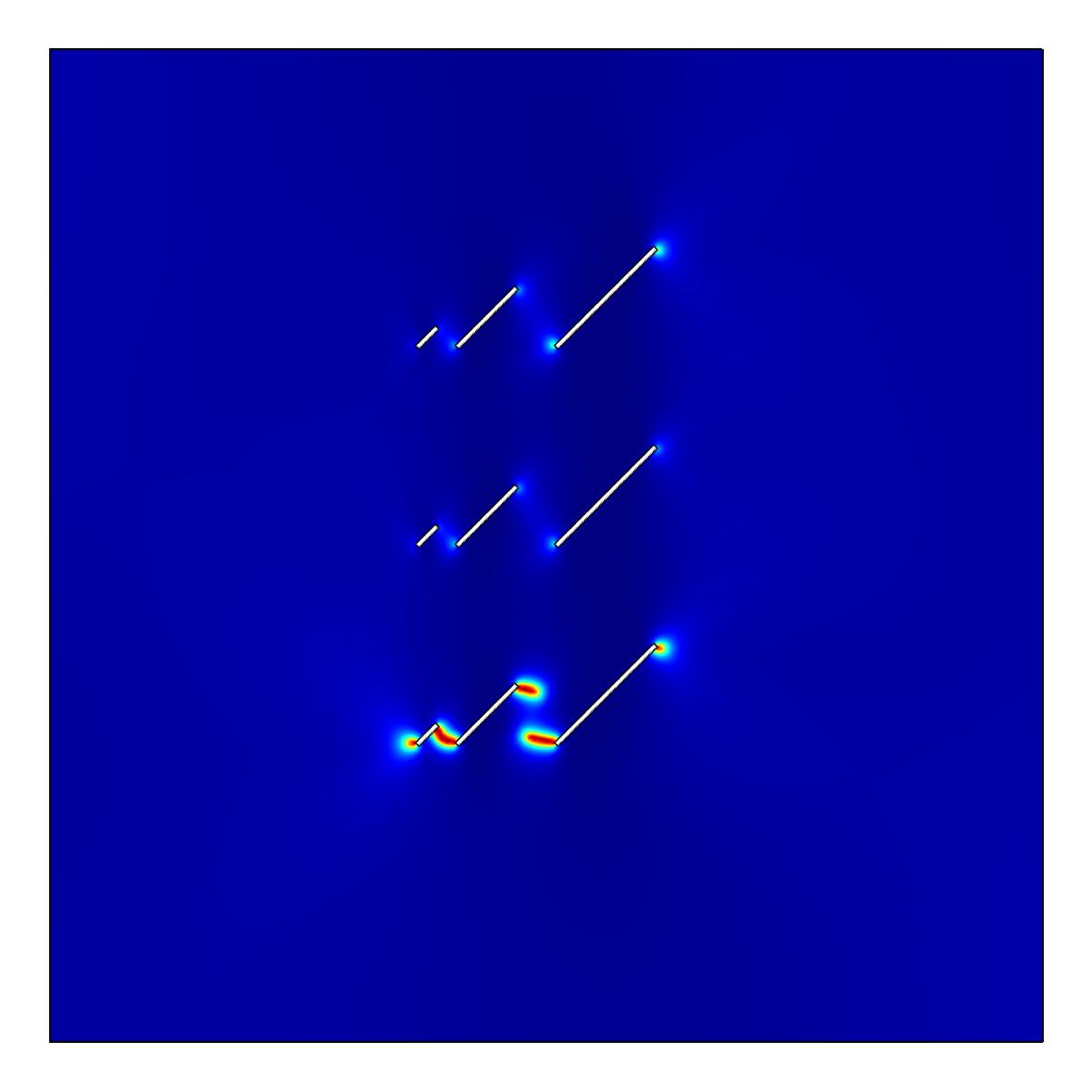}}
	
	\subfigure[]{\includegraphics[width = 6cm]{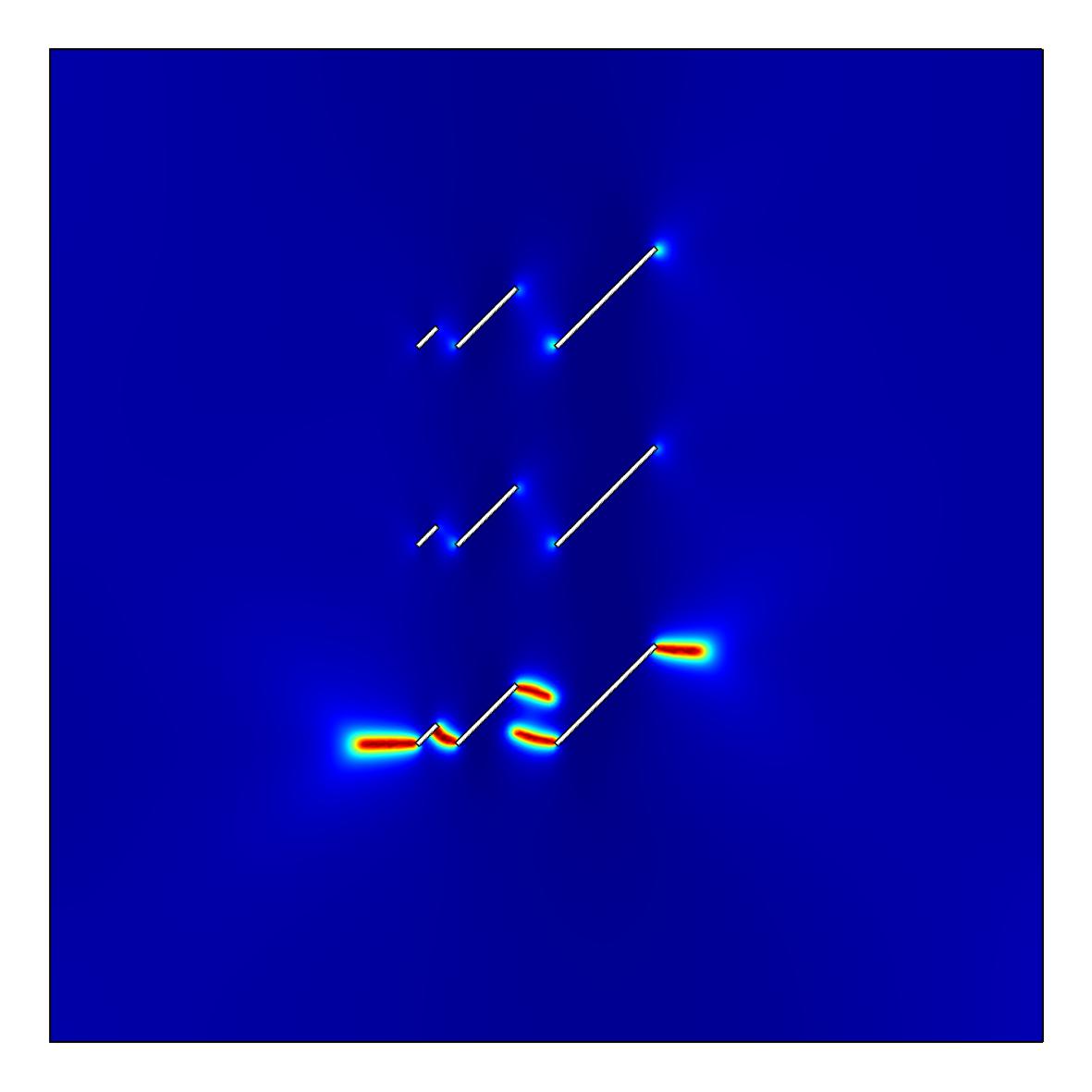}}
	\subfigure[]{\includegraphics[width = 6cm]{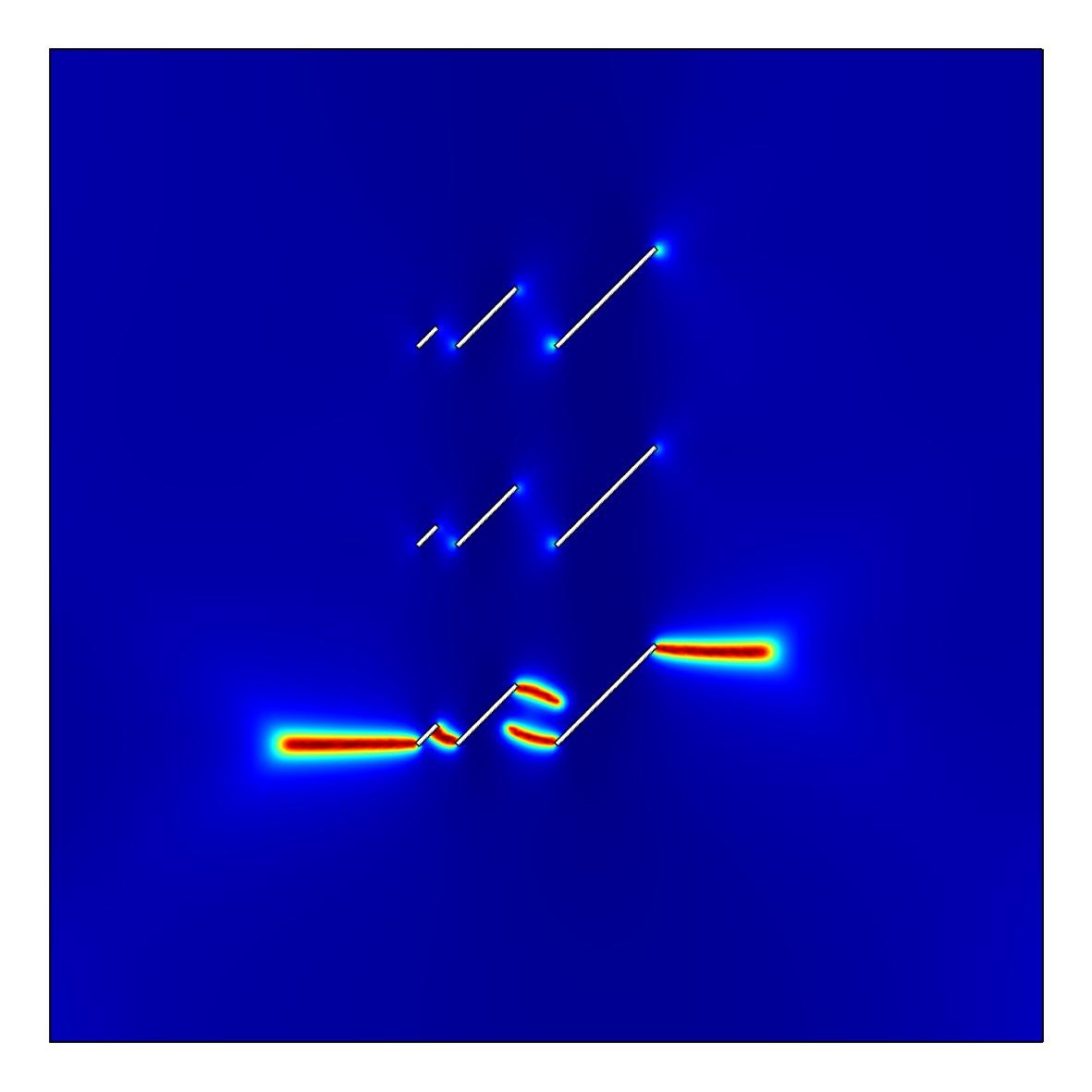}}

	\subfigure[]{\includegraphics[width = 6cm]{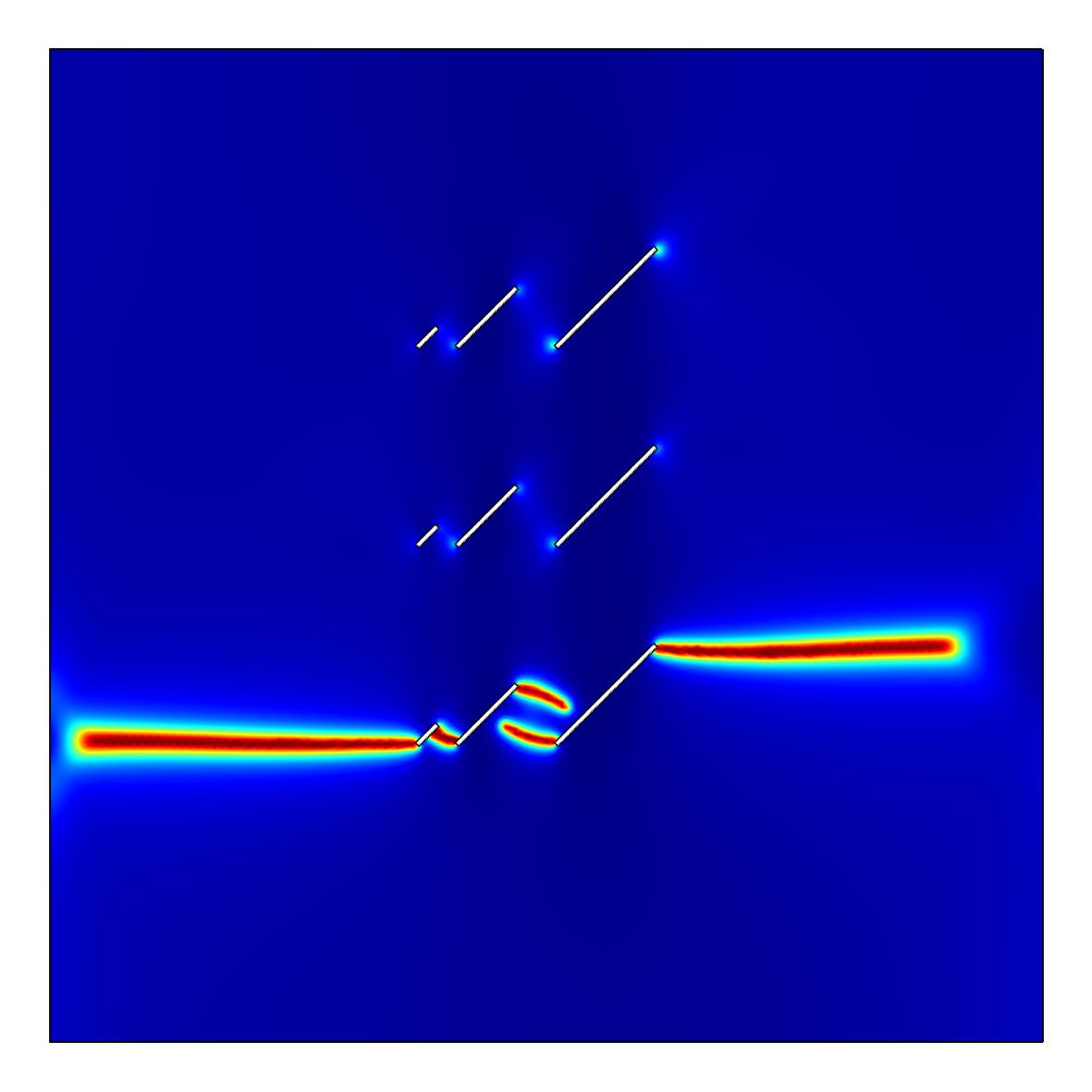}}
	\subfigure[]{\includegraphics[width = 6cm]{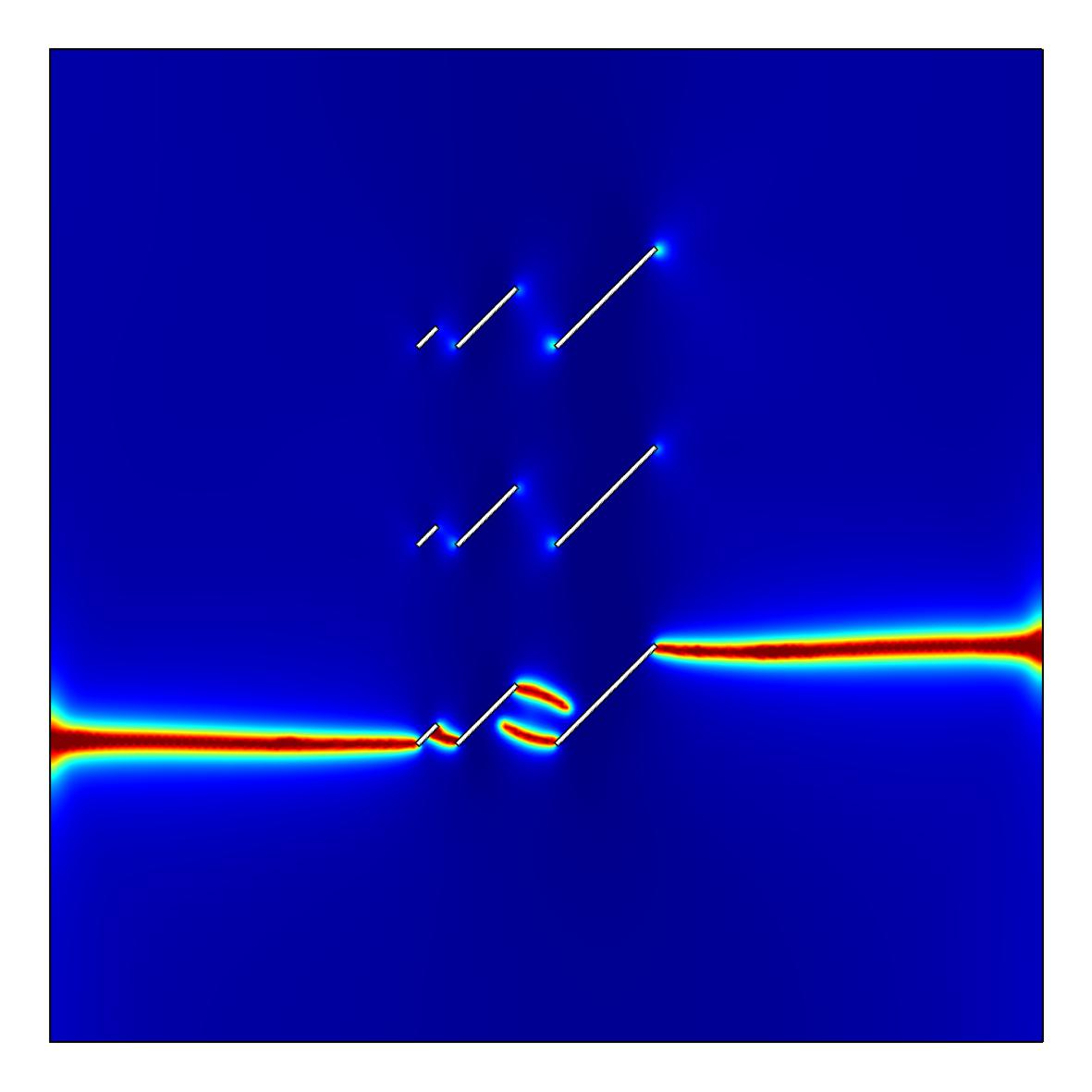}}
	\caption{Propagation and coalescence of the nine pre-existing flaws in a square rock sample at a displacement of (a) $u = 3.02\times10^{-3}$ mm, (b) $u = 3.04\times10^{-3}$ mm, (c) $u = 3.06\times10^{-3}$ mm,  (d) $u=3.08\times10^{-3}$ mm, (e) $u=3.12\times10^{-3}$ mm, and (f) $u=3.14\times10^{-3}$ mm}
	\label{Propagation and coalescence of the nine pre-existing flaws in a square rock sample}
	\end{figure}

\begin{figure}[htbp]
	\centering
	\includegraphics[width = 10cm]{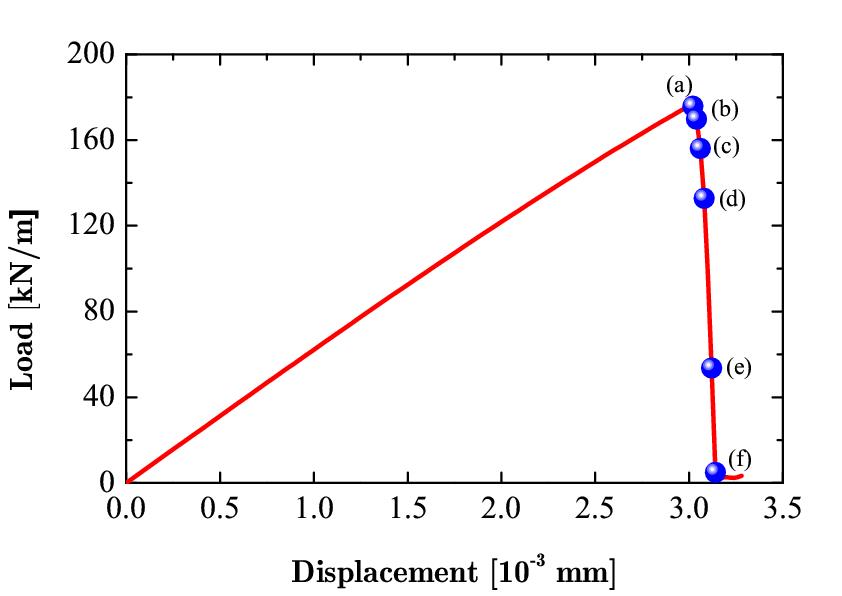}
	\caption{Load-displacement curves for the square rock sample with nine pre-existing flaws}
	\label{Load-displacement curves for the square rock sample with nine pre-existing flaws}
	\end{figure}

Figure \ref{Propagation and coalescence of the nine pre-existing flaws in a square rock sample}(a)-(f) shows the propagation and coalescence process of the nine pre-existing flaws, and the reaction force on the upper boundary of the rock sample is depicted in Fig. \ref{Load-displacement curves for the square rock sample with nine pre-existing flaws}. As the displacement $u$ increases, the phase field around the tips of the flaws and the load both increase. The first tensile cracks initiate from the left tips of the flaws $\textcircled{8}$ and $\textcircled{9}$ when the displacement $u$ reaches to  $3.02\times10^{-3}$ mm. At this time, the load achieves the maximum value. As the displacement increases to $3.04\times10^{-3}$ mm, the crack from the left tip of the flaw $\textcircled{8}$ propagates and links up the flaw $\textcircled{7}$. The crack from the left tip of the flaw $\textcircled{9}$ continues to propagate while new cracks initiate from the left tip of the flaw $\textcircled{7}$ and the right tips of the flaws $\textcircled{8}$ and $\textcircled{9}$. The load then has a drop after the peak pint (a). When $u$ reaches to $3.06\times10^{-3}$ mm and $3.08\times10^{-3}$ mm, the cracks initiating from the left tip of the flaw $\textcircled{9}$ and the right tip of the flaw $\textcircled{8}$ continue to propagate at a decreasing rate and at a small angle with the horizontal. However, the cracks from the left tip of the flaw $\textcircled{7}$ and the right tip of the flaw $\textcircled{9}$ propagate at relatively large velocity nearly along the horizontal direction. The load decreases sharply as the applied displacement increases. 

Figure \ref{Propagation and coalescence of the nine pre-existing flaws in a square rock sample}(e) shows that the crack from the left tip of the flaw $\textcircled{7}$ propagates close to the left boundary of the rock sample when $u=3.12\times10^{-3}$ mm. The crack initiating from the right tip of the flaw $\textcircled{9}$ propagates at a relatively small rate because of a larger distance from the boundary where the applied displacement is applied. When $u=3.14\times10^{-3}$ mm, the cracks from the left tip of  the flaw $\textcircled{7}$ and the right tip of the flaw $\textcircled{9}$ both reaches the left and right boundaries of the rock sample, indicating the failure of rock sample. The rock loses its load-bearing capacity and the load drops to near 0 in  Fig. \ref{Load-displacement curves for the square rock sample with nine pre-existing flaws}. In addition, Fig. \ref{Propagation and coalescence of the nine pre-existing flaws in a square rock sample} shows no cracks initiation from the tips of the flaws $\textcircled{1}$-$\textcircled{6}$. The reason is the stress shielding and amplification effects due to the interaction of the flaws \citet{zhou2015numerical}.

\subsection{Propagation and coalescence of twenty parallel flaws}

\begin{figure}
	\centering
	\subfigure[Doubly periodic rectangular array]{\includegraphics[width = 6cm]{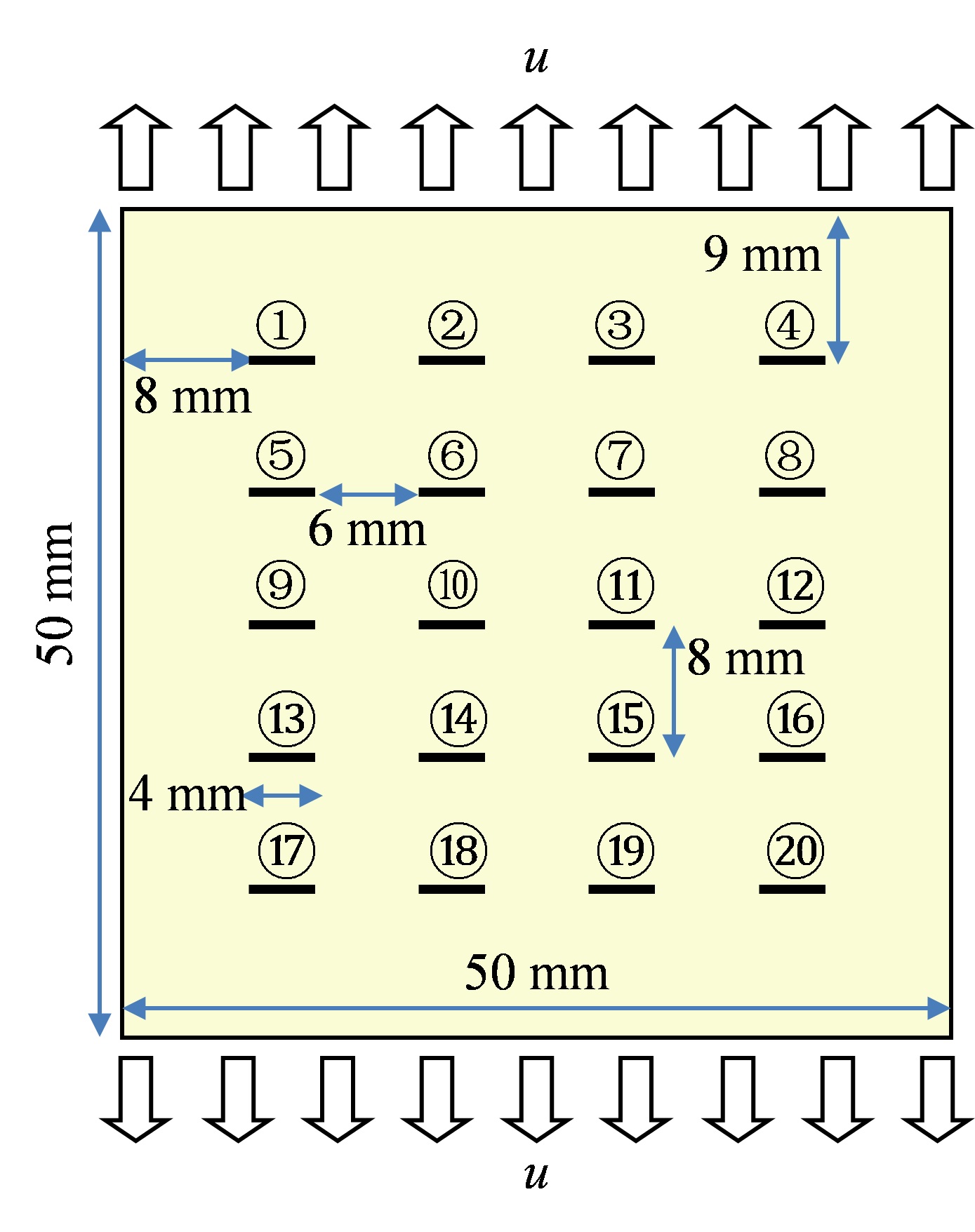}}
	\subfigure[Diamond-shaped array]{\includegraphics[width = 6cm]{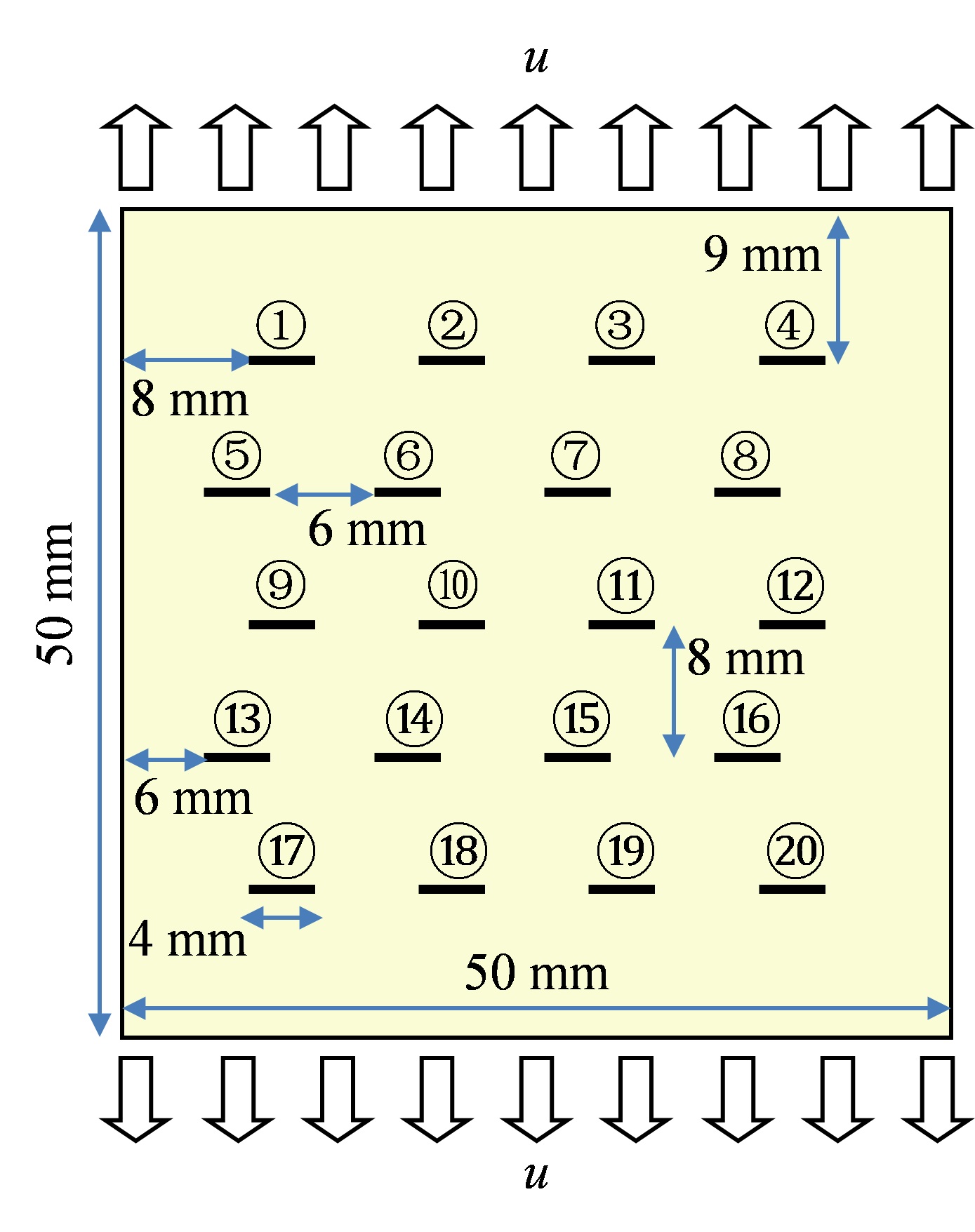}}
	\caption{Arrangement and geometry of the twenty pre-existing flaws and the boundary condition}
	\label{Arrangement and geometry of the twenty pre-existing flaws and the boundary condition}
	\end{figure}

A 2D square rock sample with twenty parallel pre-existing flaws subjected to tension is tested. All flaws have the same length, spacing, and inclination angle of $0^{\circ}$. We consider the flaws in doubly periodic rectangular and diamond-shaped arrays, respectively. The arrangement and geometry of the flaws are depicted in Fig. \ref{Arrangement and geometry of the twenty pre-existing flaws and the boundary condition}. The rock sample are 50 mm $\times$ 50 mm. These parameters are adopted: the rock density $\rho=2450$ kg/m$^3$, the Young's modulus $E=30$ GPa, the Poisson's ratio $\nu = 0.3$, $G_c = 100$ J/m$^2$, $k=1\times10^{-9}$, and the length scale $l_0=0.4$ mm. The rock sample is discretized by using uniform 8-node quadratic elements with the element size $h=0.2$ mm. We adopt the displacement increment $\Delta u  = 5\times10^{-6}$ mm for each time step.

\begin{figure}[htbp]
	\centering
	\subfigure[Doubly periodic rectangular array]{\includegraphics[width = 10cm]{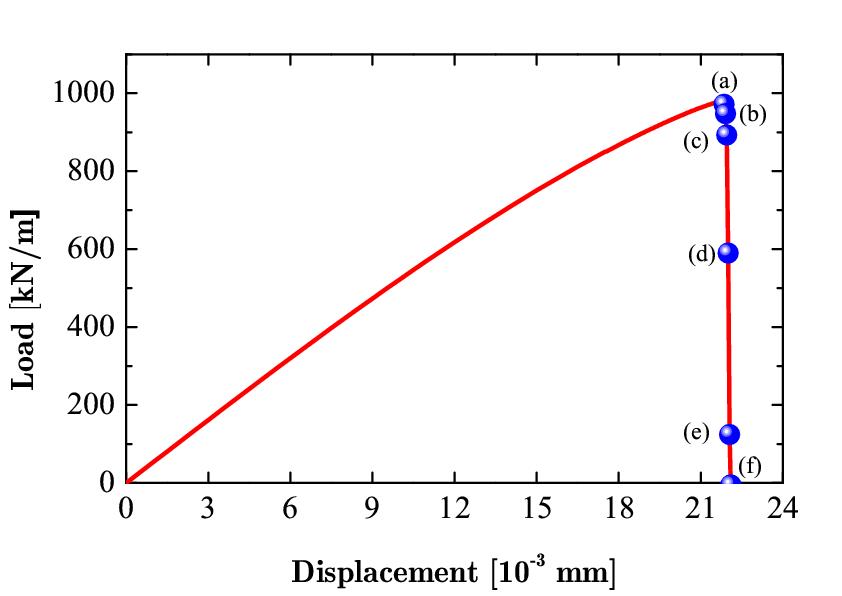}}\\
	\subfigure[Diamond-shaped array]{\includegraphics[width = 10cm]{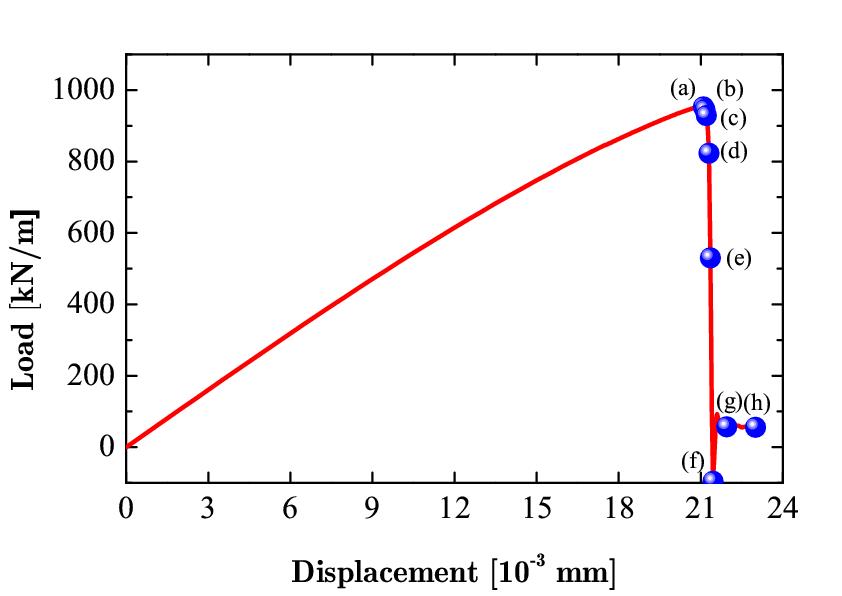}}
	\caption{Load-displacement curves for the square rock sample with twenty parallel pre-existing flaws}
	\label{Load-displacement curves for the square rock sample with twenty parallel nine pre-existing flaws}
	\end{figure}

\begin{figure}[htbp]
	\centering
	\subfigure[]{\includegraphics[width = 5cm]{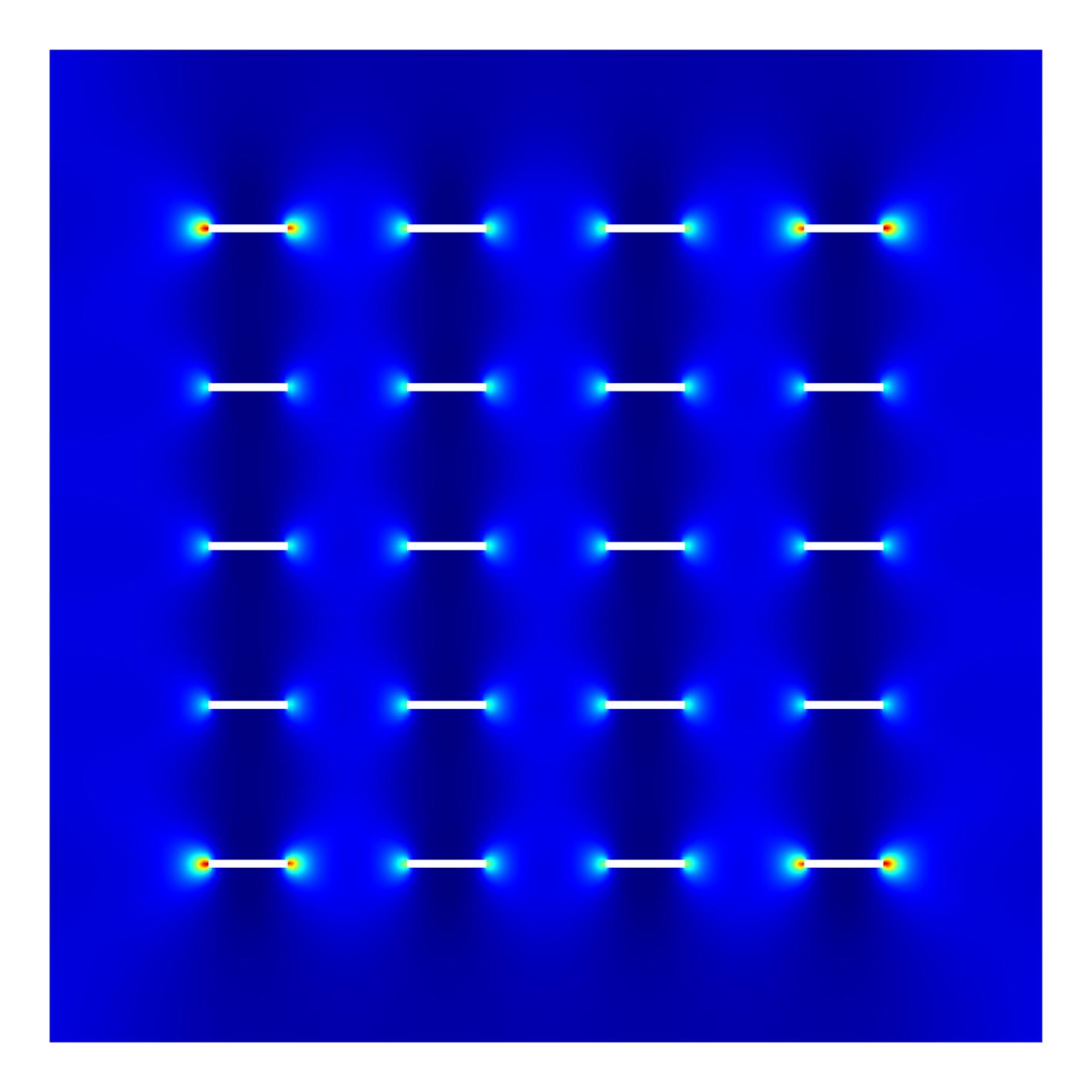}}
	\subfigure[]{\includegraphics[width = 5cm]{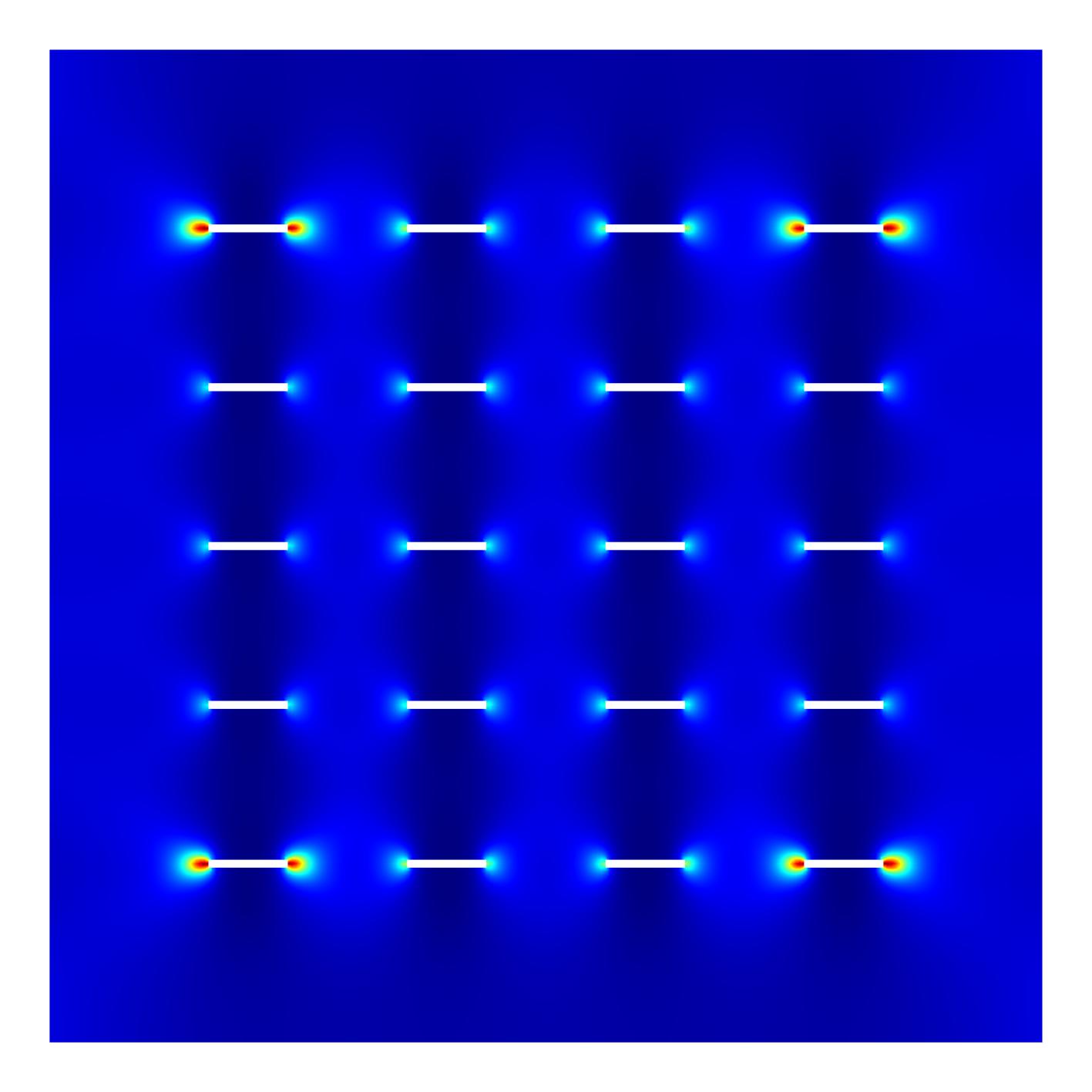}}	
	\subfigure[]{\includegraphics[width = 5cm]{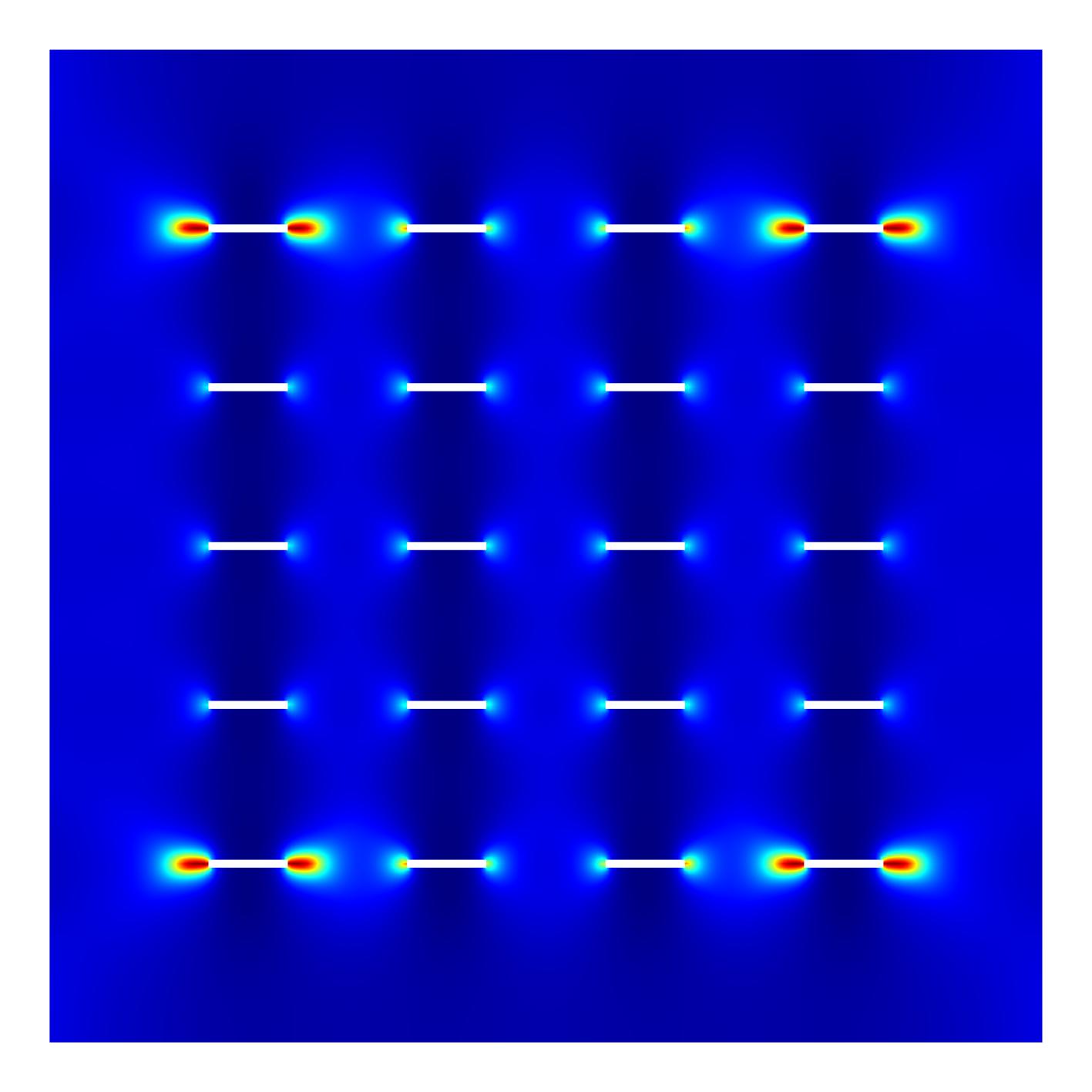}}

	\subfigure[]{\includegraphics[width = 5cm]{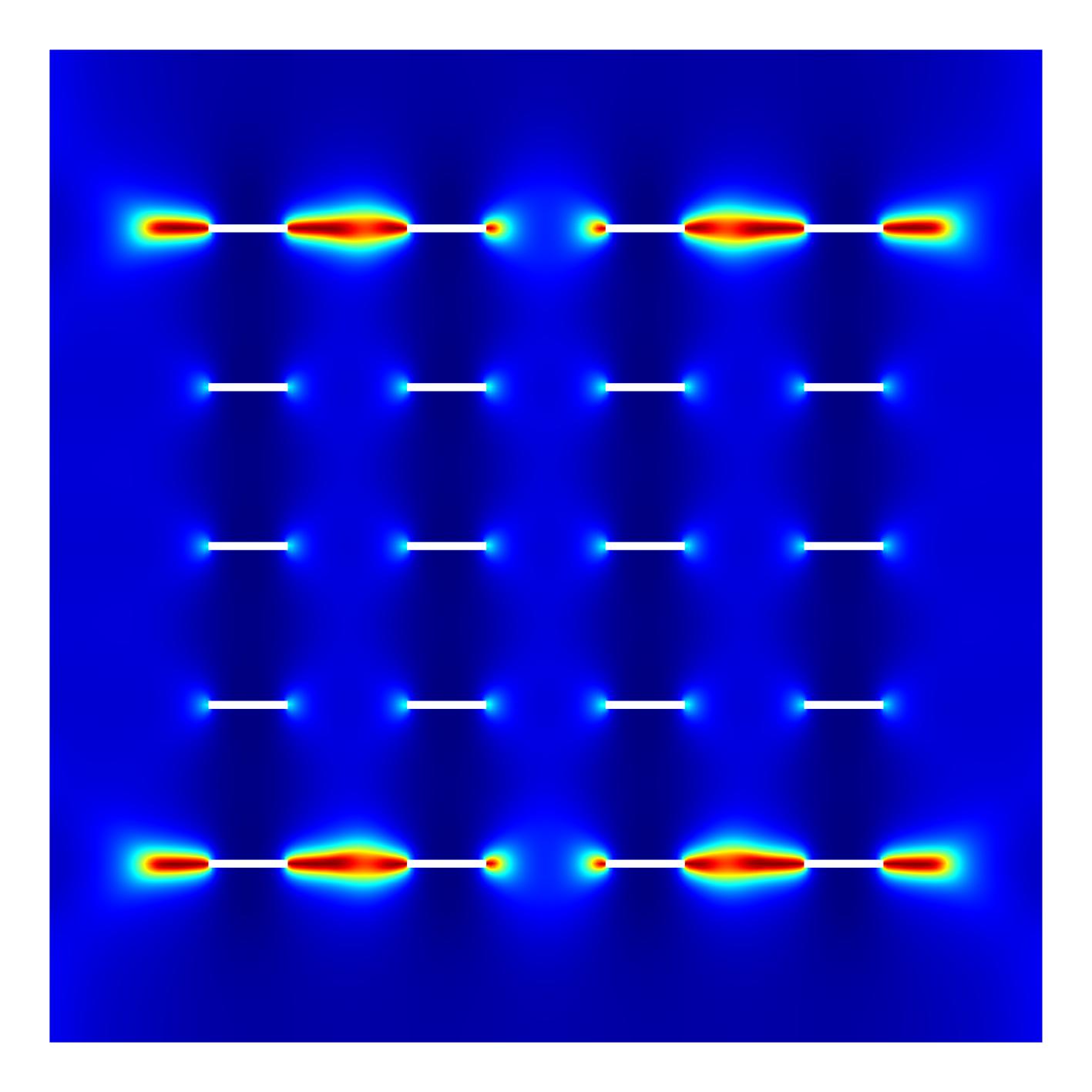}}
	\subfigure[]{\includegraphics[width = 5cm]{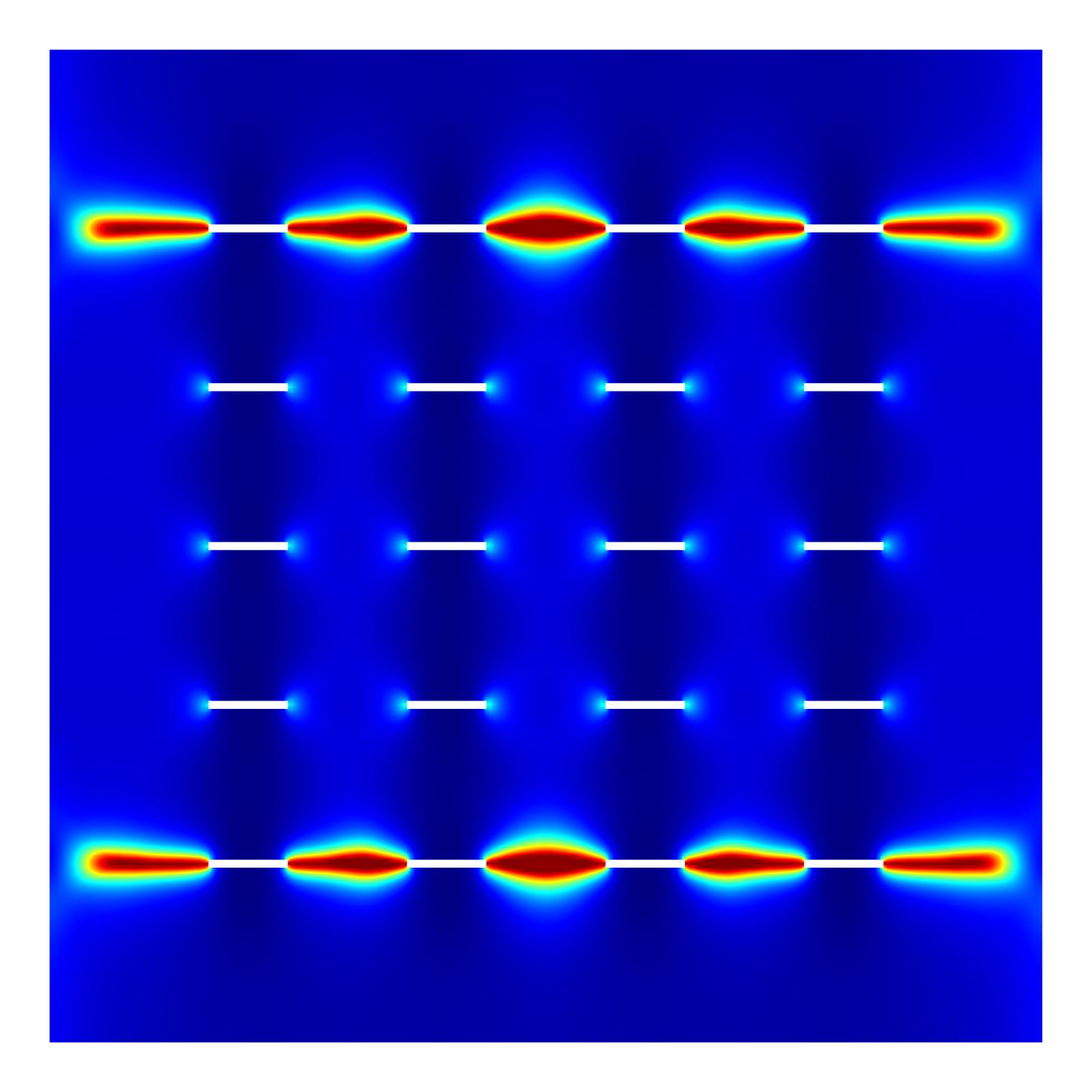}}
	\subfigure[]{\includegraphics[width = 5cm]{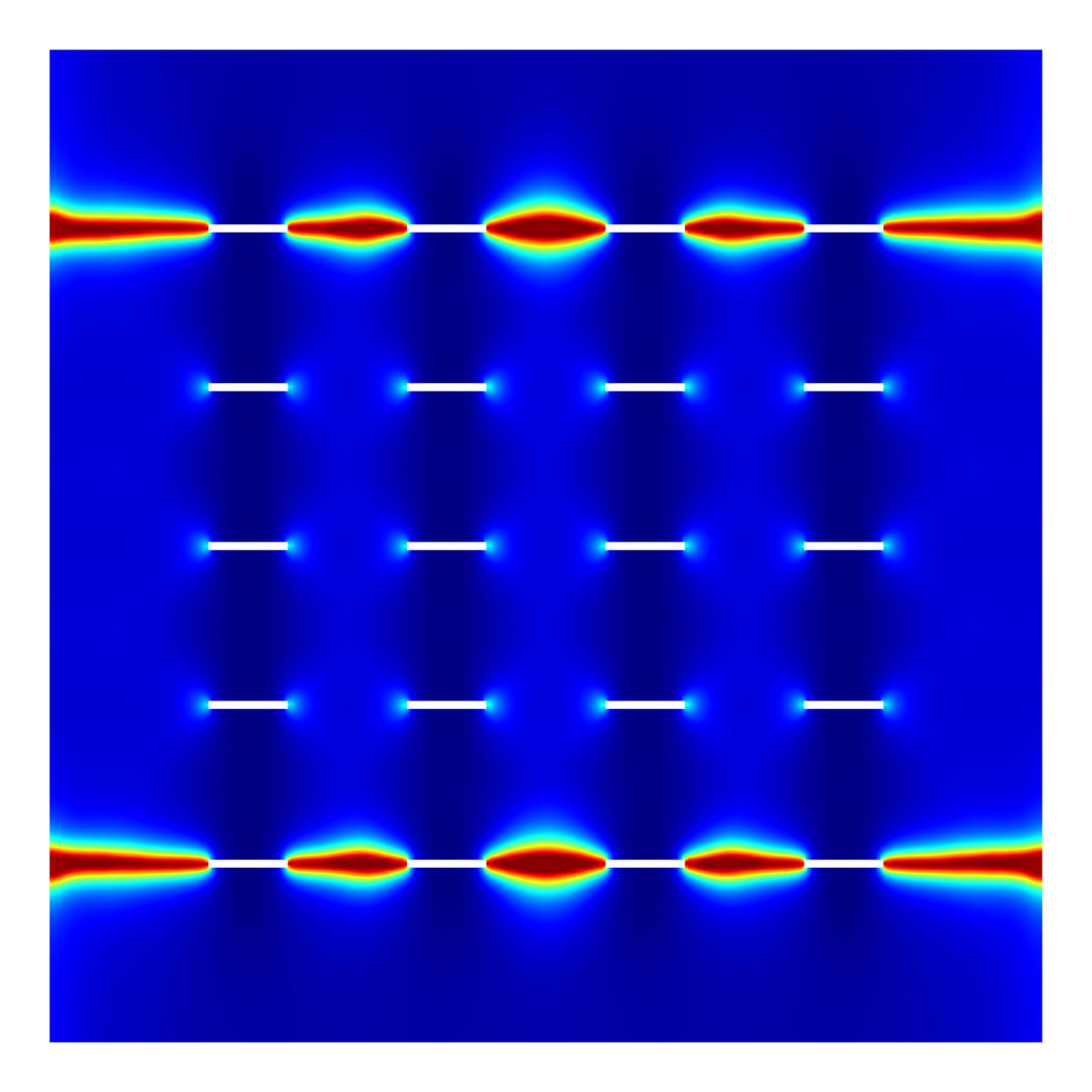}}
	\caption{Propagation and coalescence of the doubly periodic rectangular array of twenty pre-existing flaws at a displacement of (a) $u = 2.185\times10^{-2}$ mm, (b) $u = 2.19\times10^{-2}$ mm, (c) $u = 2.195\times10^{-2}$ mm,  (d) $u=2.2\times10^{-2}$ mm, (e) $u=2.205\times10^{-2}$ mm, and (f) $u=2.21\times10^{-2}$ mm}
	\label{Propagation and coalescence of the doubly periodic rectangular array of 20 pre-existing flaws in a square rock sample}
	\end{figure}

Figure \ref{Load-displacement curves for the square rock sample with twenty parallel nine pre-existing flaws} presents the load-displacement curves for the square rock sample with twenty parallel flaws. A sudden drop of the load is also observed after the maximum value is obtained. Figure \ref{Propagation and coalescence of the doubly periodic rectangular array of 20 pre-existing flaws in a square rock sample} presents the propagation and coalescence of the doubly periodic rectangular array of twenty pre-existing flaws. As the displacement $u$ increases, the phase field $\phi$ concentrates at the tip of each flaw and the load achieves the maximum when $u = 2.185\times10^{-2}$ mm. When the displacement $u$ reaches to $2.19\times10^{-2}$ mm, the first tensile cracks initiate from the left and right tips of the flaws $\textcircled{1}$, $\textcircled{5}$, $\textcircled{\footnotesize{17}}$, and $\textcircled{\footnotesize{20}}$. In addition, the load has a small drop from the peak. When $u=2.195\times10^{-2}$ mm, the first cracks continue to propagate perpendicular to the direction of the applied displacement and the load decreases. However, when $u=2.2\times10^{-2}$ mm, the cracks fully connect the flaws $\textcircled{1}$, $\textcircled{5}$, $\textcircled{\footnotesize{17}}$, and $\textcircled{\footnotesize{20}}$ with the flaws $\textcircled{2}$, $\textcircled{3}$, $\textcircled{\footnotesize{18}}$, and $\textcircled{\footnotesize{19}}$. Moreover, some new cracks initiate from the right tips of the flaws $\textcircled{2}$ and $\textcircled{\footnotesize{18}}$ as well as the left tips of the flaws $\textcircled{3}$ and $\textcircled{\footnotesize{19}}$. The load then drops to approximately $60\%$ of the maximum load. When  $u=2.205\times10^{-2}$ mm, the cracks initiating from the right tips of the flaws $\textcircled{2}$ and $\textcircled{\footnotesize{18}}$ and the cracks from the left tips of the flaws $\textcircled{3}$ and $\textcircled{\footnotesize{19}}$ coalesce. Finally, the cracks from the left tips of the flaws $\textcircled{1}$ and $\textcircled{\footnotesize{17}}$ as well as the cracks from the right tips of the flaws $\textcircled{4}$ and $\textcircled{\footnotesize{20}}$ reach the left and right sides of the rock sample when $u=2.21\times10^{-2}$ mm, indicating that the rock loses the load-bearing capacity. In addition, no cracks initiate from the tips of the flaws $\textcircled{5}$ - $\textcircled{\footnotesize{16}}$.

\begin{figure}[htbp]
	\centering
	\subfigure[]{\includegraphics[width = 5cm]{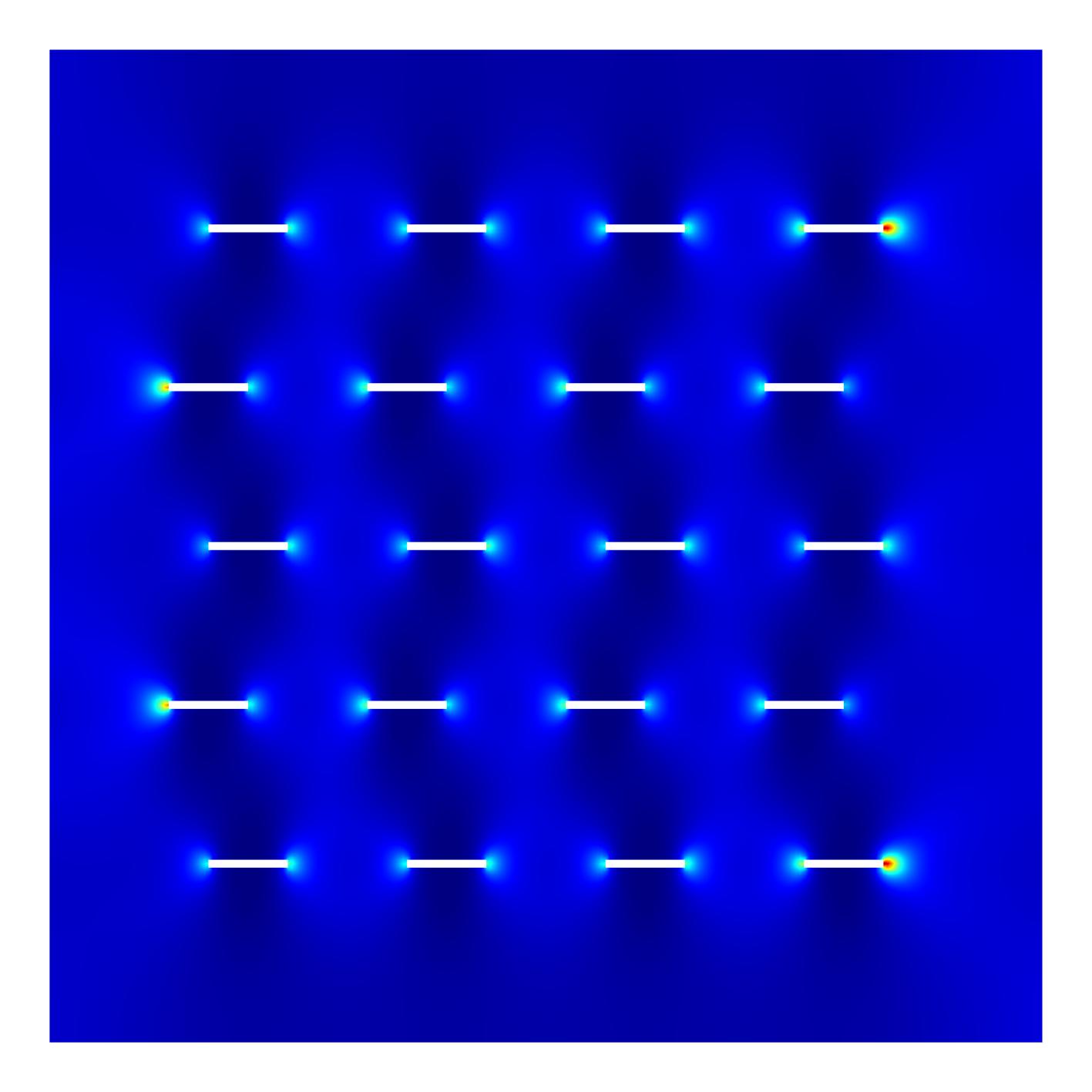}}
	\subfigure[]{\includegraphics[width = 5cm]{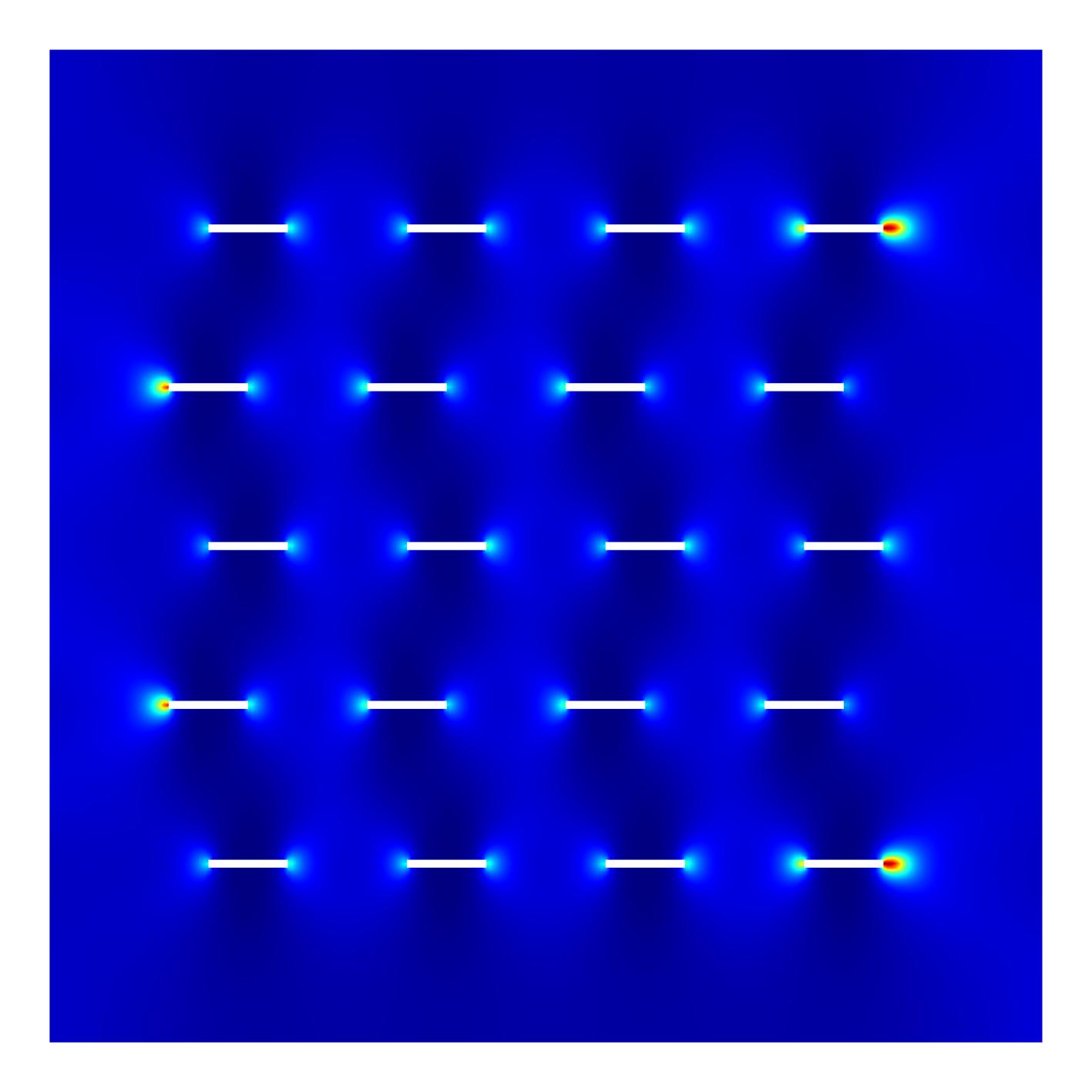}}	
	\subfigure[]{\includegraphics[width = 5cm]{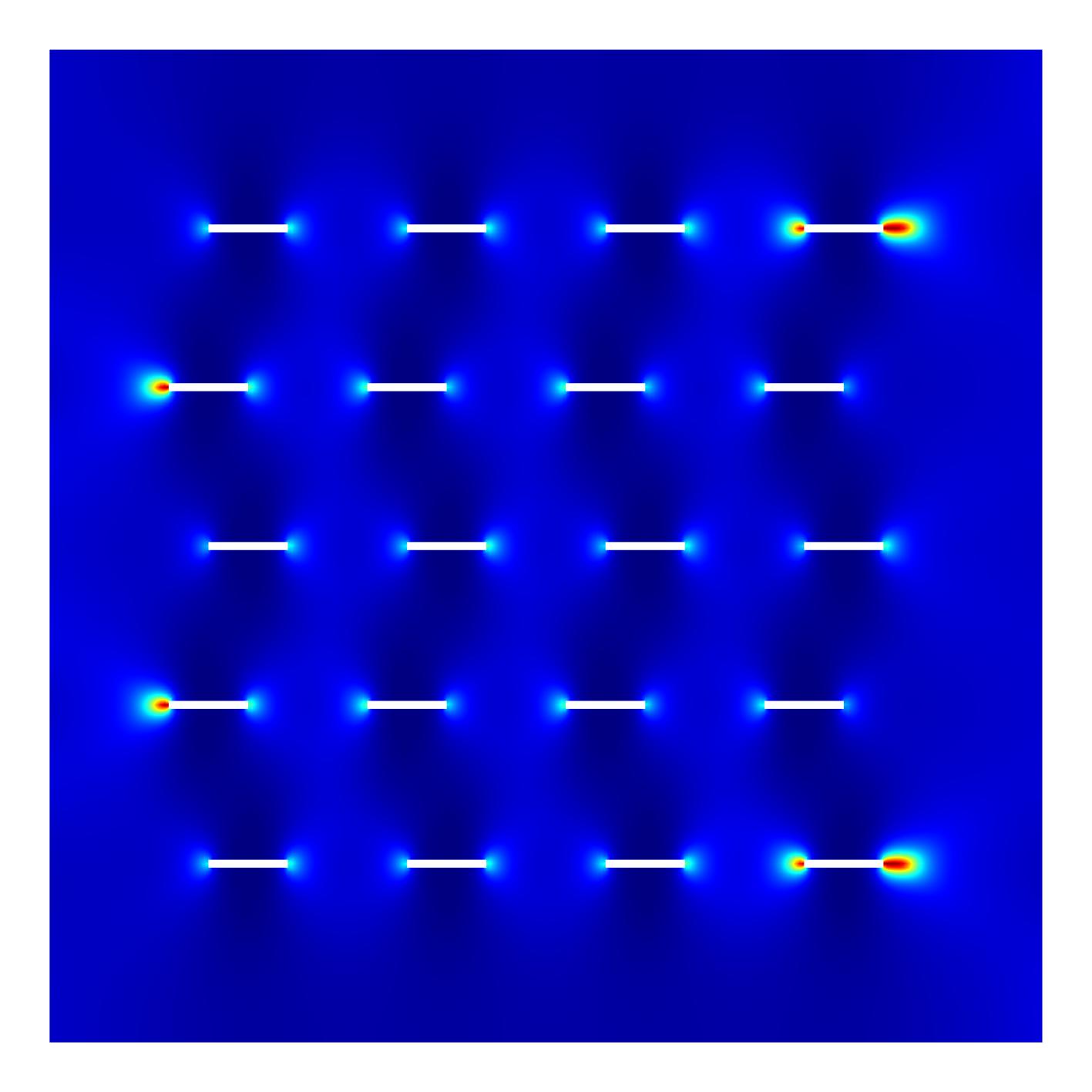}}

	\subfigure[]{\includegraphics[width = 5cm]{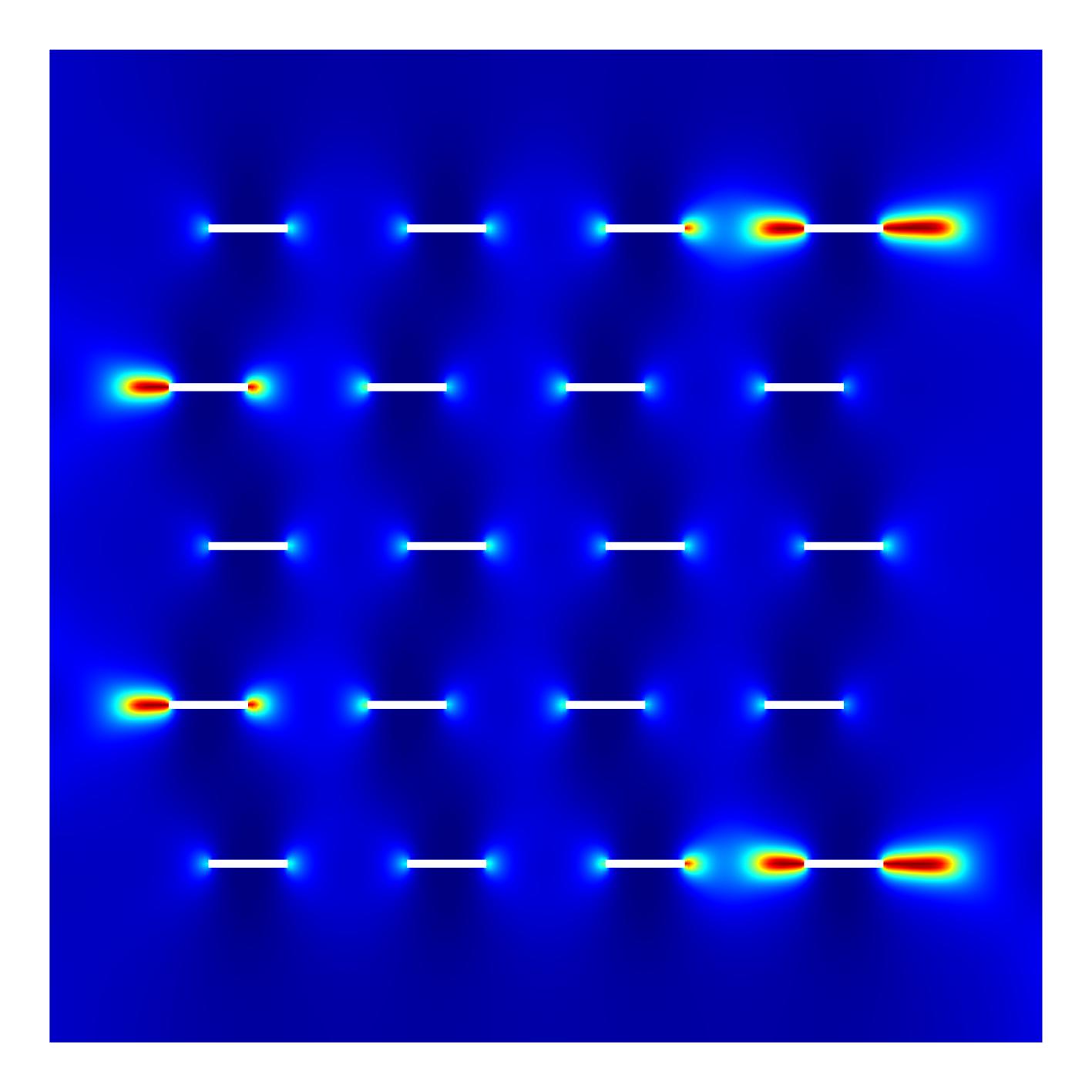}}
	\subfigure[]{\includegraphics[width = 5cm]{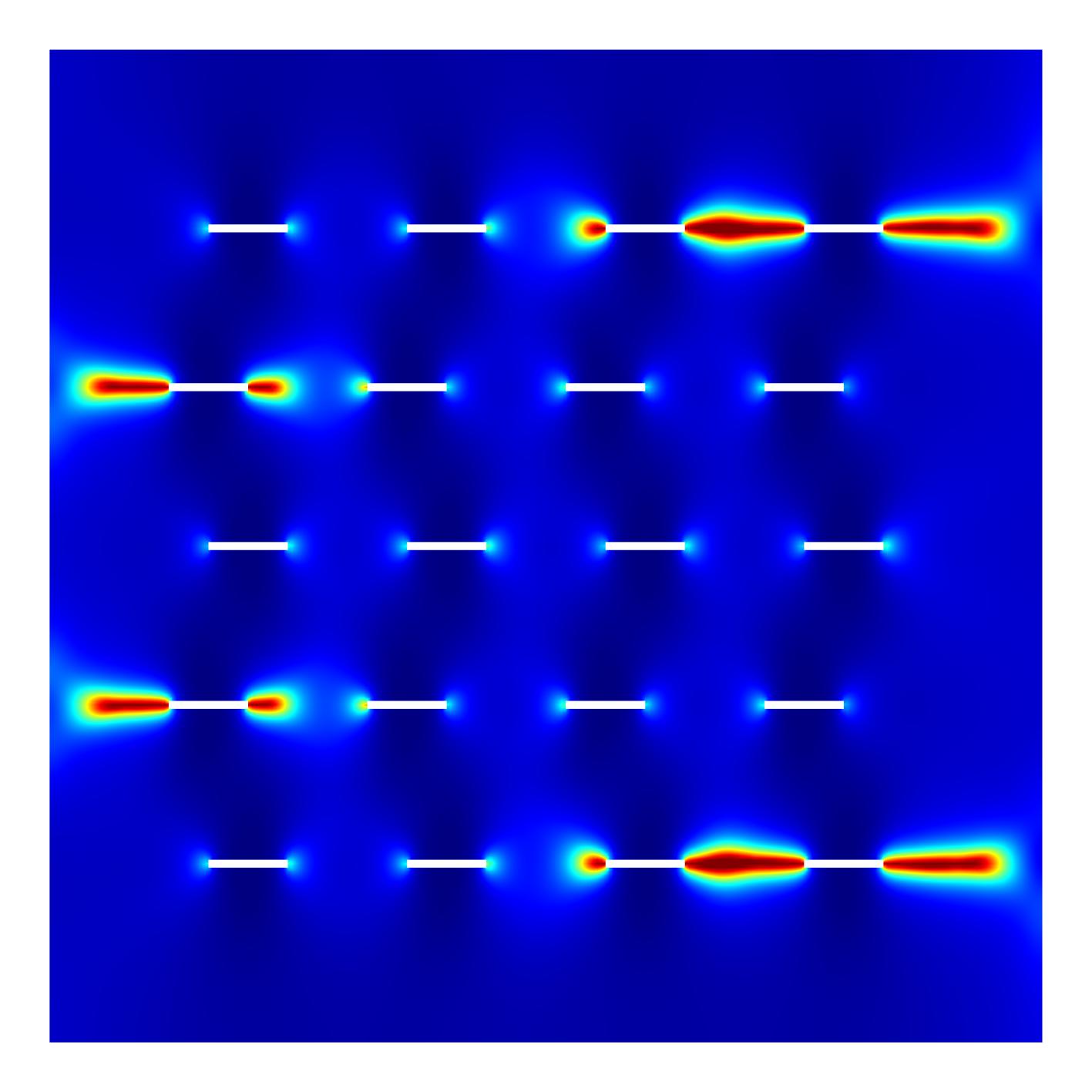}}
	\subfigure[]{\includegraphics[width = 5cm]{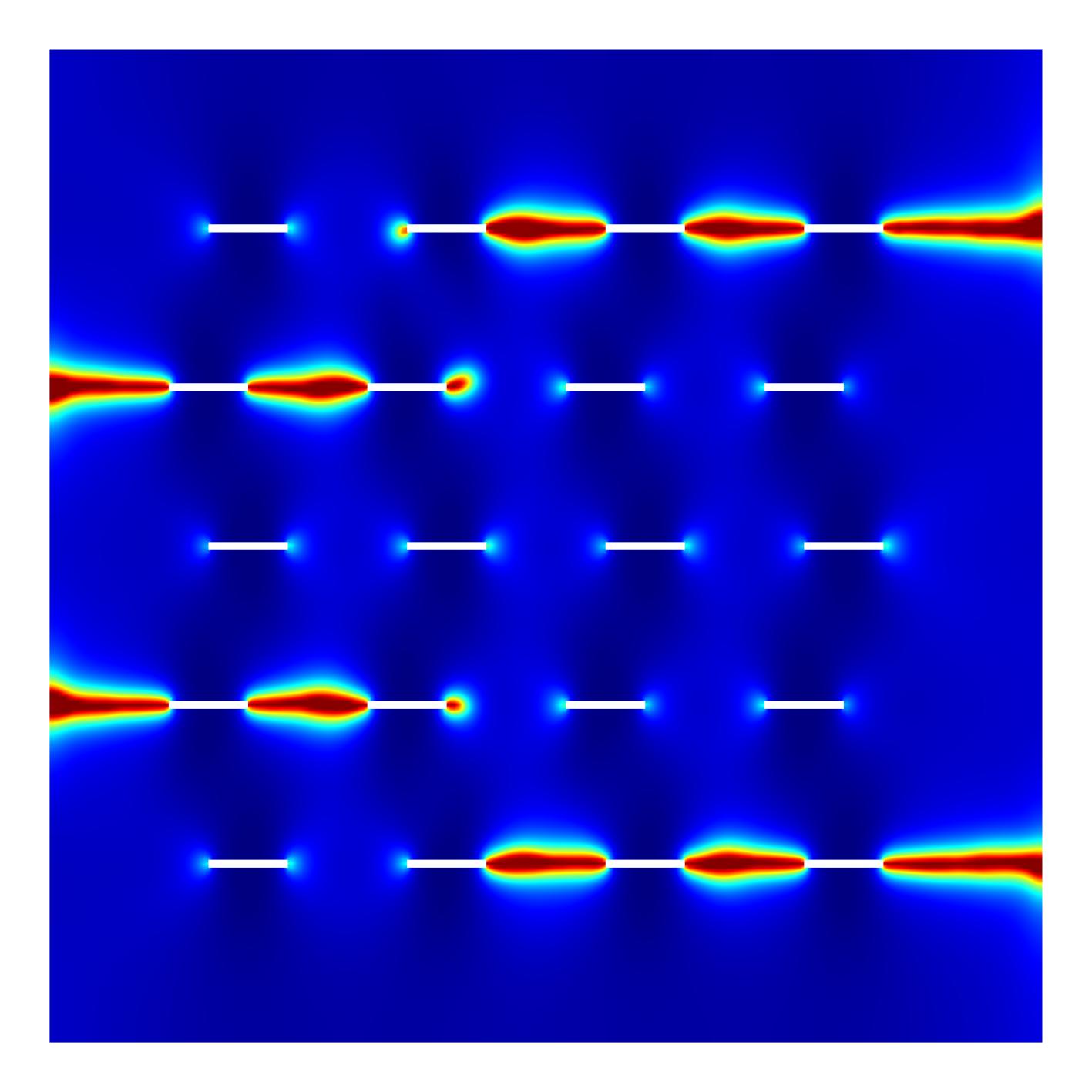}}

	\subfigure[]{\includegraphics[width = 5cm]{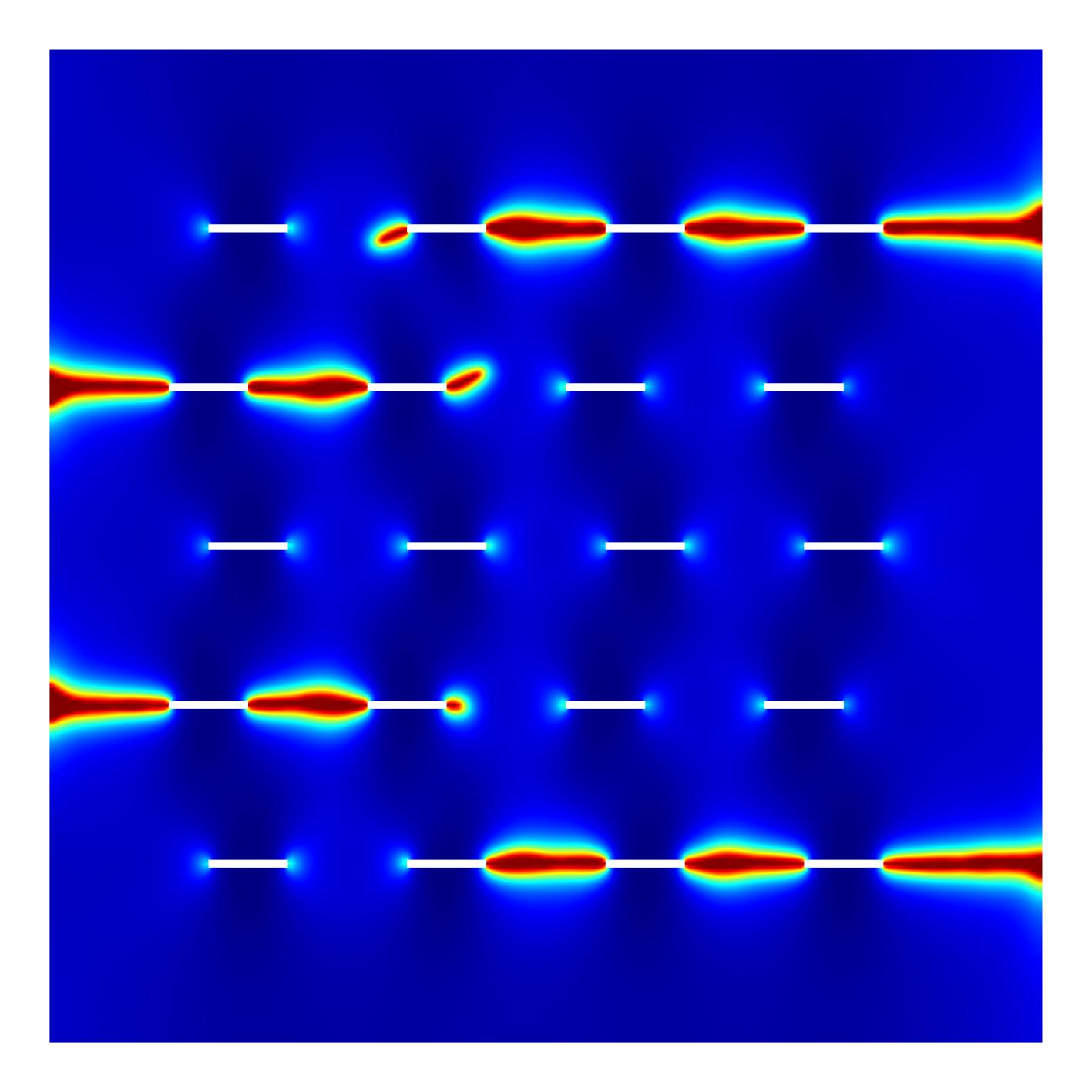}}
	\subfigure[]{\includegraphics[width = 5cm]{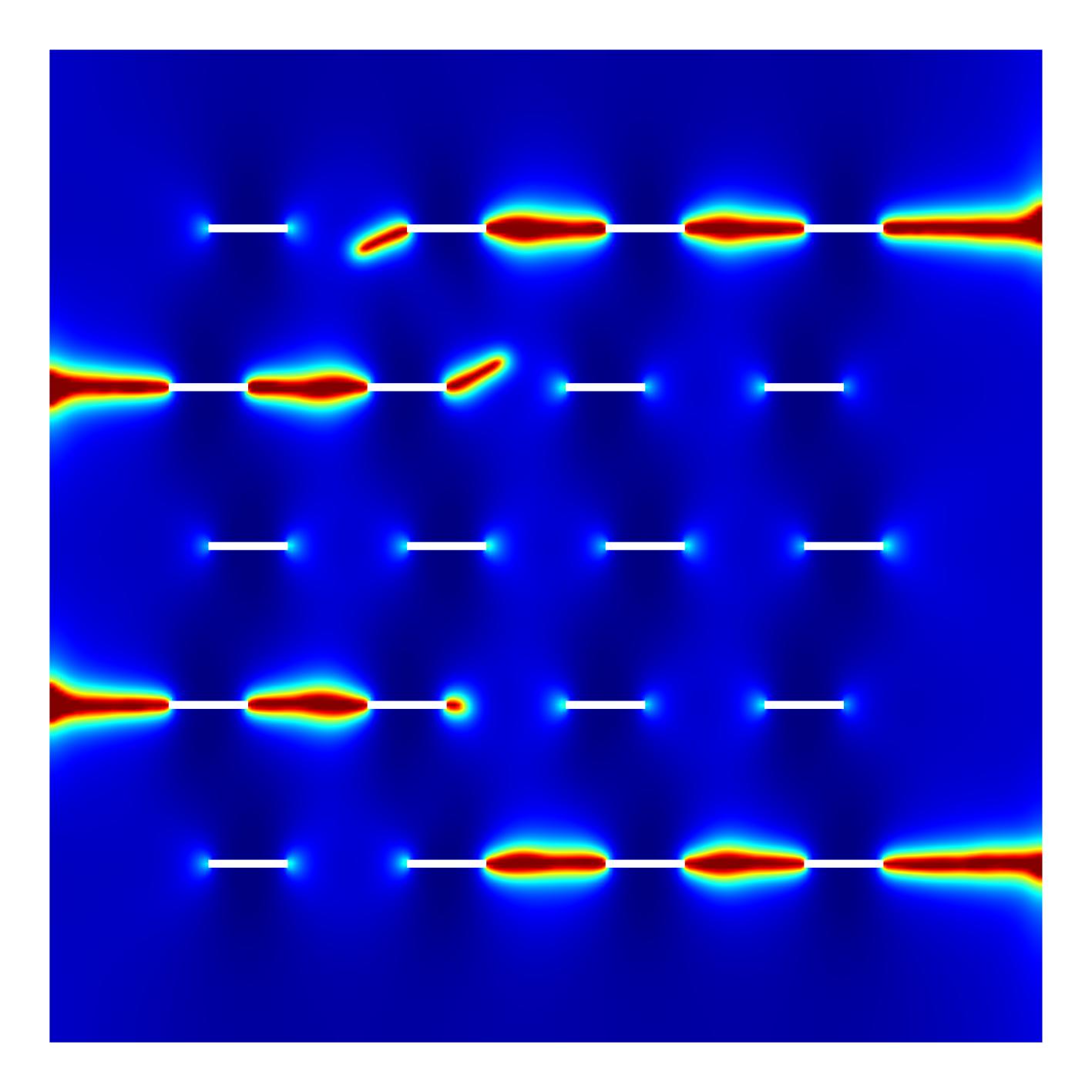}}

	\caption{Propagation and coalescence of the diamond-shaped array of twenty pre-existing flaws at a displacement of (a) $u = 2.11\times10^{-2}$ mm, (b) $u = 2.115\times10^{-2}$ mm, (c) $u = 2.120\times10^{-2}$ mm,  (d) $u=2.130\times10^{-2}$ mm, (e) $u=2.135\times10^{-2}$ mm, (f) $u=2.145\times10^{-2}$ mm, (f) $u=2.195\times10^{-2}$ mm, and (g) $u =2.3\times10^{-2}$ mm}
	\label{Propagation and coalescence of the diamond-shaped array of 20 pre-existing flaws in a square rock sample}
	\end{figure}

Figure \ref{Propagation and coalescence of the diamond-shaped array of 20 pre-existing flaws in a square rock sample} presents the propagation and coalescence of the twenty pre-existing flaws in the diamond-shaped array. The crack patterns for the diamond-shaped array are different from those for the doubly periodic rectangular array. As the displacement $u$ increases, the phase field $\phi$ rapidly increases around the tips closest to the left and right boundaries of the rock sample. When the displacement $u$ reaches to $2.11\times10^{-2}$ mm, the load increases to the maximum. When $u=2.115\times10^{-2}$ mm, the tensile cracks first emanate from the left tips of the flaws $\textcircled{5}$ and $\textcircled{\footnotesize{13}}$ as well as the right tips of the flaws $\textcircled{4}$ and $\textcircled{\footnotesize{20}}$. When $u=2.12\times10^{-2}$ mm, another two cracks initiate from the left tips of the flaws $\textcircled{4}$ and $\textcircled{\footnotesize{20}}$. The first four cracks continue to propagate along the horizontal direction, while the load at $u=2.12\times10^{-2}$ mm is close to that at $u=2.11\times10^{-2}$ mm and $u=2.115\times10^{-2}$ mm.

The flaws close to the top and bottom edges of the sample are placed perpendicular to the direction of loading. The arrangement of the internal flaws has little effect on the propagation of the edge flaws. Therefore, cracks propagating from those edge flaws will show typical features of Mode-I fracture. As shown in Fig. \ref{Propagation and coalescence of the diamond-shaped array of 20 pre-existing flaws in a square rock sample}, when $u=2.130\times10^{-2}$ mm, the originally initiating cracks propagate as expected, while new cracks emanate from the right tips of the flaws  $\textcircled{3}$, $\textcircled{5}$, $\textcircled{\footnotesize{13}}$, and $\textcircled{\footnotesize{19}}$. The load then starts to drop from the peak region. However, when $u=2.135\times10^{-2}$ mm, the cracks fully connect the flaws $\textcircled{3}$ and $\textcircled{\footnotesize{19}}$ with the flaws $\textcircled{4}$ and $\textcircled{\footnotesize{20}}$. In addition, two new cracks initiate from the left tips of the flaws $\textcircled{3}$ and $\textcircled{\footnotesize{19}}$. At this time, the load decreases to approximately half of the peak load. As the displacement $u$ increases to $2.145\times10^{-2}$ mm, the propagating cracks and the flaws $\textcircled{2}$, $\textcircled{3}$, $\textcircled{5}$, $\textcircled{6}$, $\textcircled{\footnotesize{13}}$, $\textcircled{\footnotesize{14}}$, $\textcircled{\footnotesize{18}}$, and $\textcircled{\footnotesize{19}}$ coalesce. At the same time, the cracks from the left tips of the flaws $\textcircled{5}$ and $\textcircled{\footnotesize{13}}$ as well as the cracks from the right tips of the flaws $\textcircled{4}$ and $\textcircled{\footnotesize{20}}$ reach the left and right sides of the rock sample. The load at $u=2.145$ mm is less than zero followed by new cracks emanating from the left tip of the flaw $\textcircled{2}$ and the right tip of the flaw $\textcircled{6}$.

When the displacement $u$ reaches to $2.195\times10^{-2}$ mm and $2.3\times10^{-2}$ mm, the crack initiating from the right tip of the flaw  $\textcircled{\footnotesize{14}}$ slowly propagates; however, the cracks from the left tip of the flaw $\textcircled{2}$ and the right tip of the flaw $\textcircled{6}$ continue to propagate obliquely at a relatively large velocity. The difference in the crack patterns from Figs. \ref{Propagation and coalescence of the doubly periodic rectangular array of 20 pre-existing flaws in a square rock sample} and \ref{Propagation and coalescence of the diamond-shaped array of 20 pre-existing flaws in a square rock sample} implies that initiation, propagation and coalescence of the cracks are  significantly affected by the arrangement of the flaws.

\subsection{Crack branching in a plate subjected to internal pressure}

\begin{figure}[htbp]
	\centering
	\includegraphics[width = 6cm]{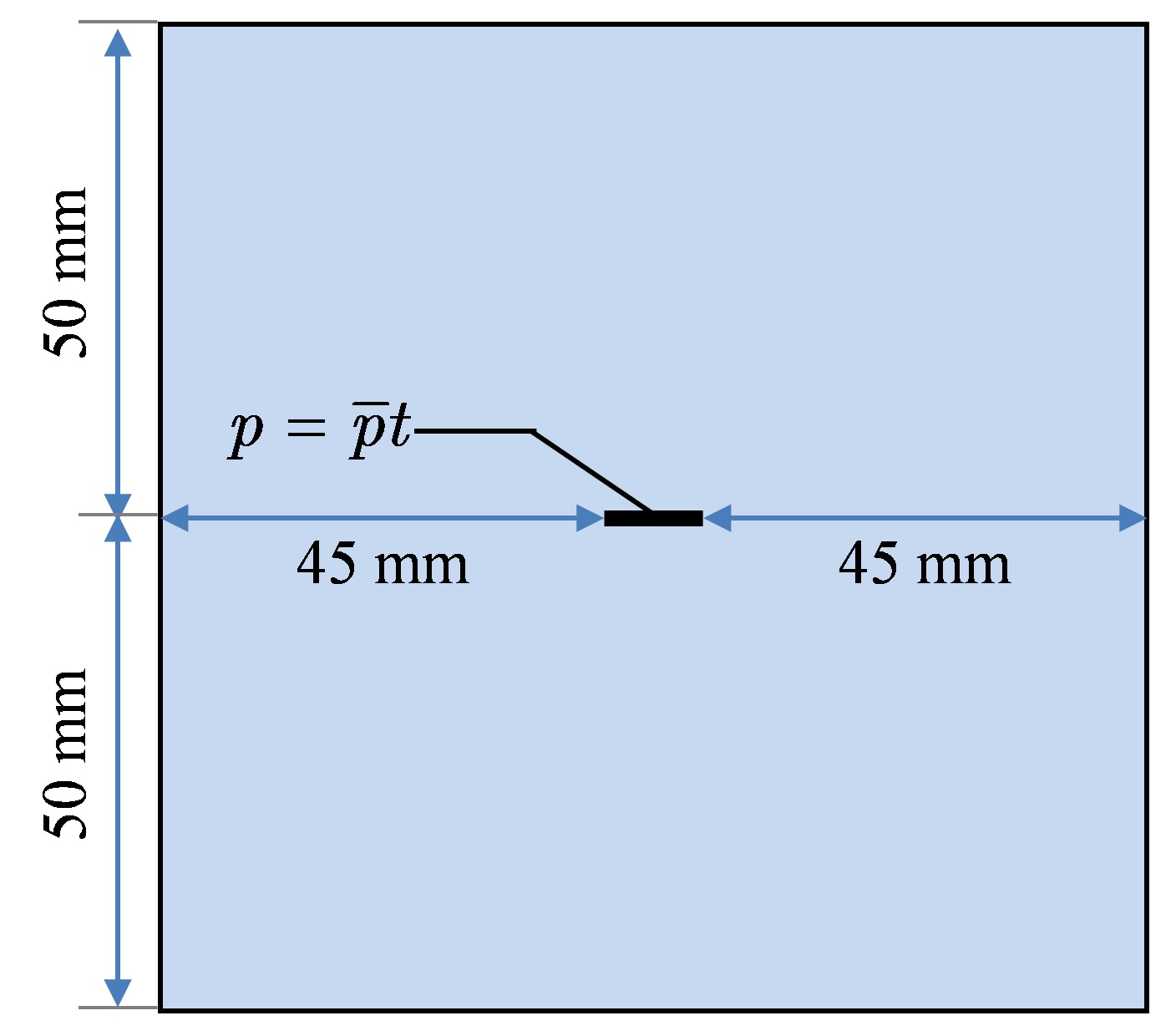}
	\caption{Geometry and boundary conditions of the plate subjected to internal pressure}
	\label{Geometry and boundary conditions of the plate subjected to internal pressure}
	\end{figure}

This example is a square plate subjected to internal pressure with geometry and boundary conditions shown in Fig. \ref{Geometry and boundary conditions of the plate subjected to internal pressure}. The internal pressure is applied on the upper and lower boundaries of the notch with $\bar p=1$ MPa/s, while the outer boundaries of the plate are traction-free. In this example, dynamic cracks are considered, and these parameters are adopted: the rock density $\rho=2450$ kg/m$^3$, the Poisson's ratio $\nu = 0.3$, $G_c = 1$ J/m$^2$, $k=1\times10^{-9}$, and the length scale $l_0=0.4$ mm. The plate is discretized by using uniform Q4 elements with the element size $h=l_0$, and we adopt the time step size $\Delta t  = 0.01$ $\mu$s.

\begin{figure}[htbp]
	\centering
	\subfigure[$m=1$]{\includegraphics[height = 4.5cm]{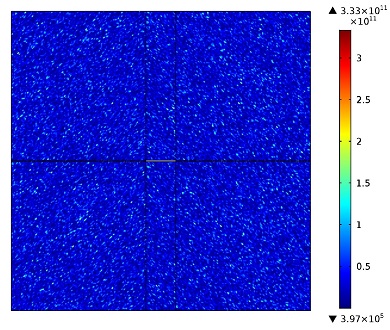}}
	\subfigure[$m=3$]{\includegraphics[height = 4.5cm]{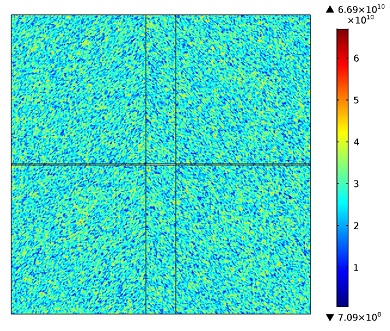}}	
	\subfigure[$m=5$]{\includegraphics[height = 4.5cm]{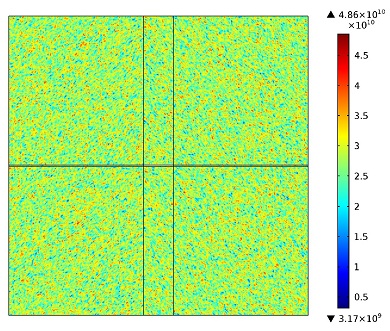}}\\
	\subfigure[$m=7$]{\includegraphics[height = 4.5cm]{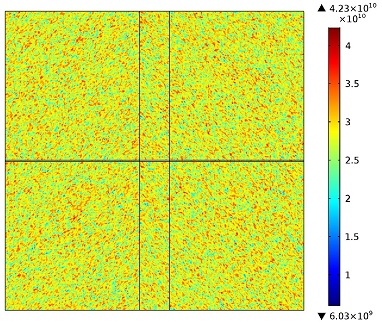}}
	\subfigure[$m=9$]{\includegraphics[height = 4.5cm]{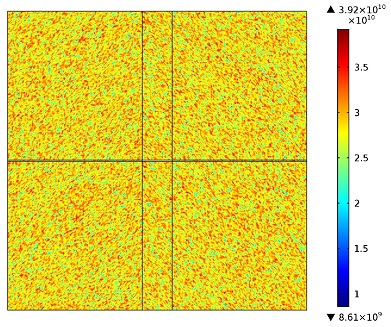}}
	\subfigure[Homogeneous]{\includegraphics[height = 4.5cm]{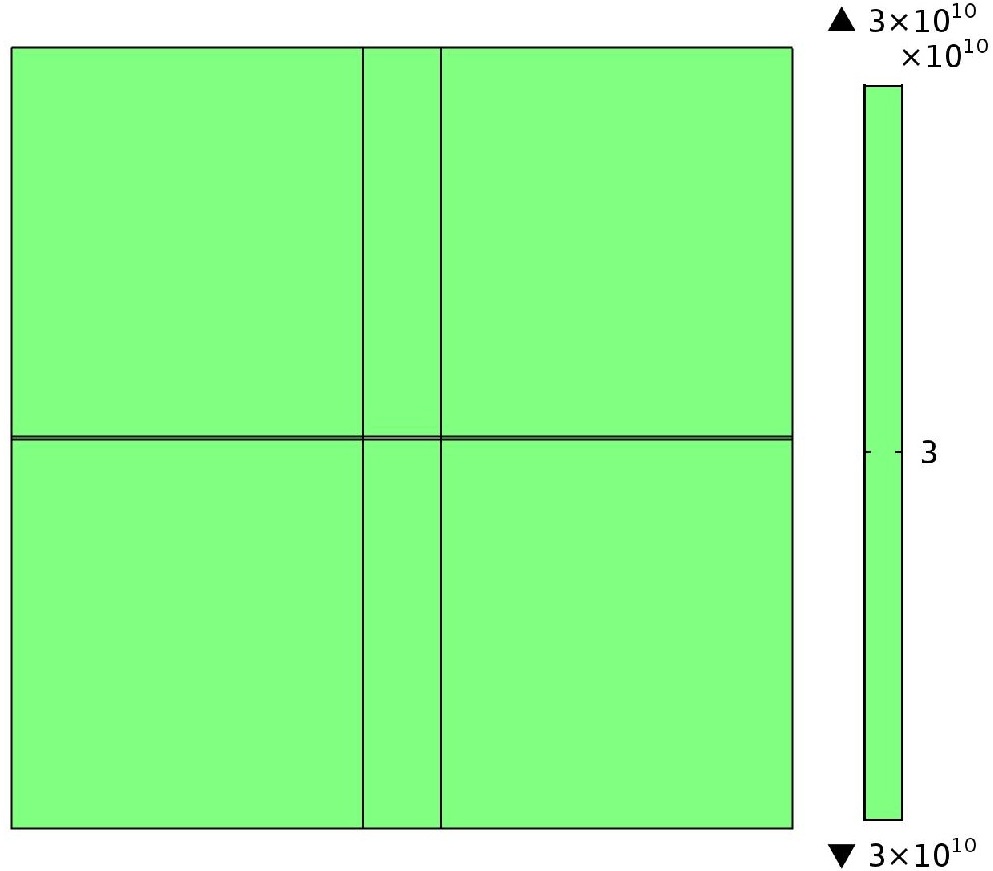}}

	\caption{Young's modulus of the plate}
	\label{Young's modulus of the plate}
	\end{figure}

We consider the plate as a heterogeneous material and apply a Weibull distribution to the Young's modulus:
	\begin{equation}
	\varphi(E)=\frac{m}{E_0}\left(\frac E{E_0}\right)^{m-1}\text{exp}\left(-\frac E{E_0}\right)^{m}
	\end{equation}

\noindent where $\varphi$ is the probability density function and coefficient $m$ determines the shape of $\varphi$. $m$ also reflects the homogeneity of the material. As $m$ increases, the material becomes more homogeneous and vice versa. In this paper, we consider $m=1$, 3, 5, 7, and 9 along with $m=\infty$ representing a homogeneous plate. We use $E_0=30$ GPa and Fig. \ref{Young's modulus of the plate} shows the distribution of Young's modulus for different $m$. 

\begin{figure}[htbp]
	\centering
		\begin{tabular}{ll}
		\centering
		&$t=20$ $\mu$s\hspace{1.5cm}$t=30$ $\mu$s\hspace{1.5cm}$t=50$ $\mu$s\hspace{1.5cm}$t=60$ $\mu$s\hspace{1.5cm}$t=70$ $\mu$s\\
		$m=1$ &\includegraphics[width = 3cm]{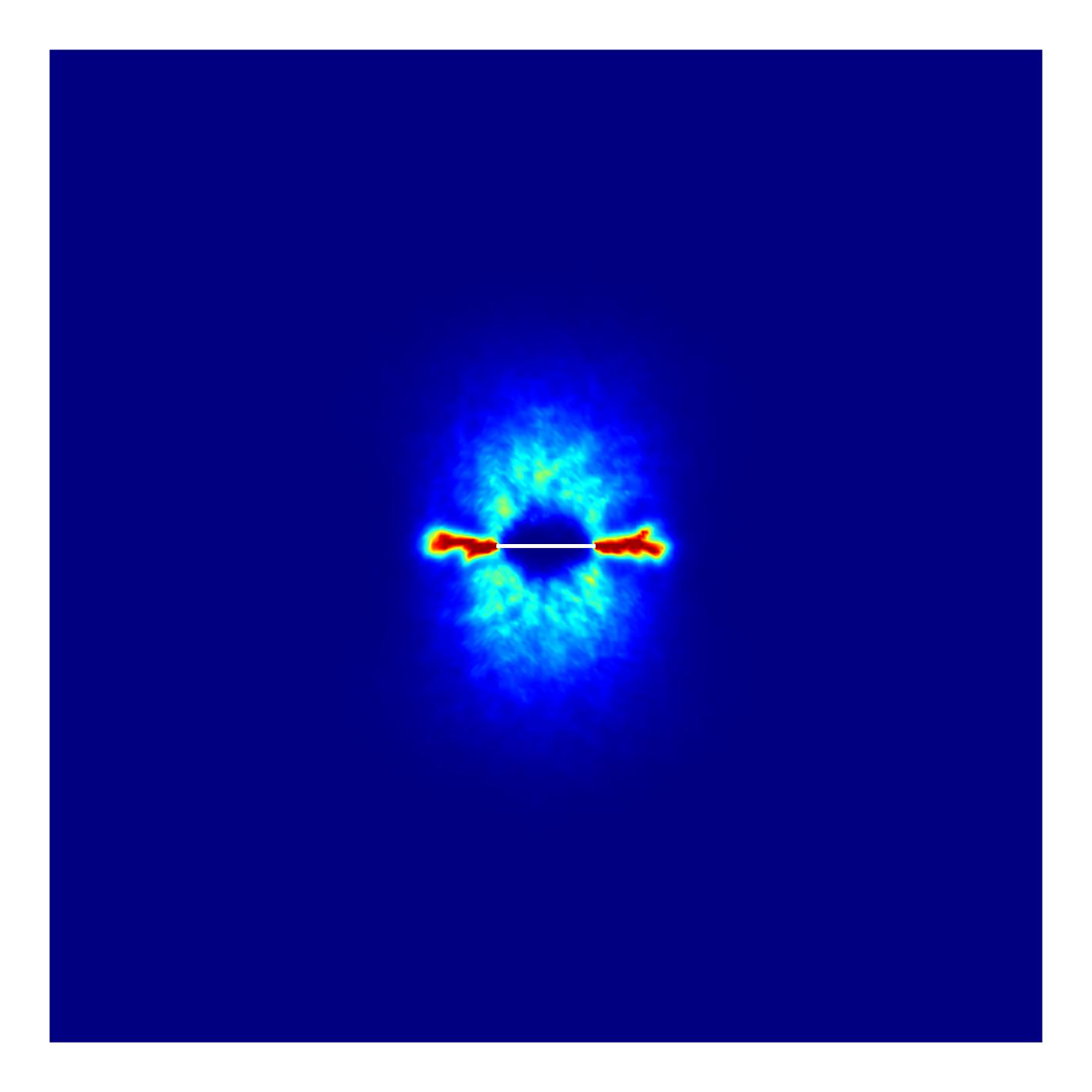}  \includegraphics[width = 3cm]{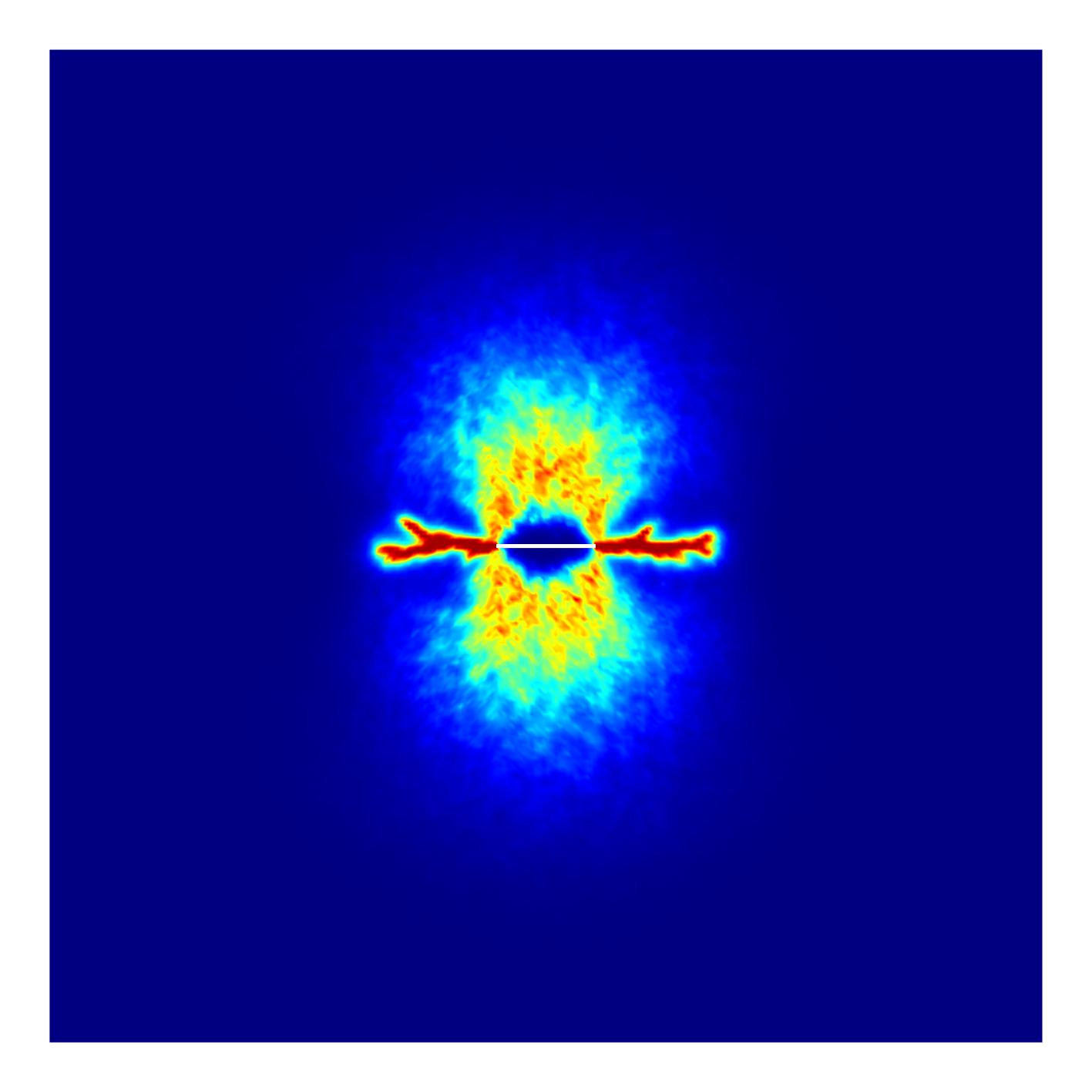}  \includegraphics[width = 3cm]{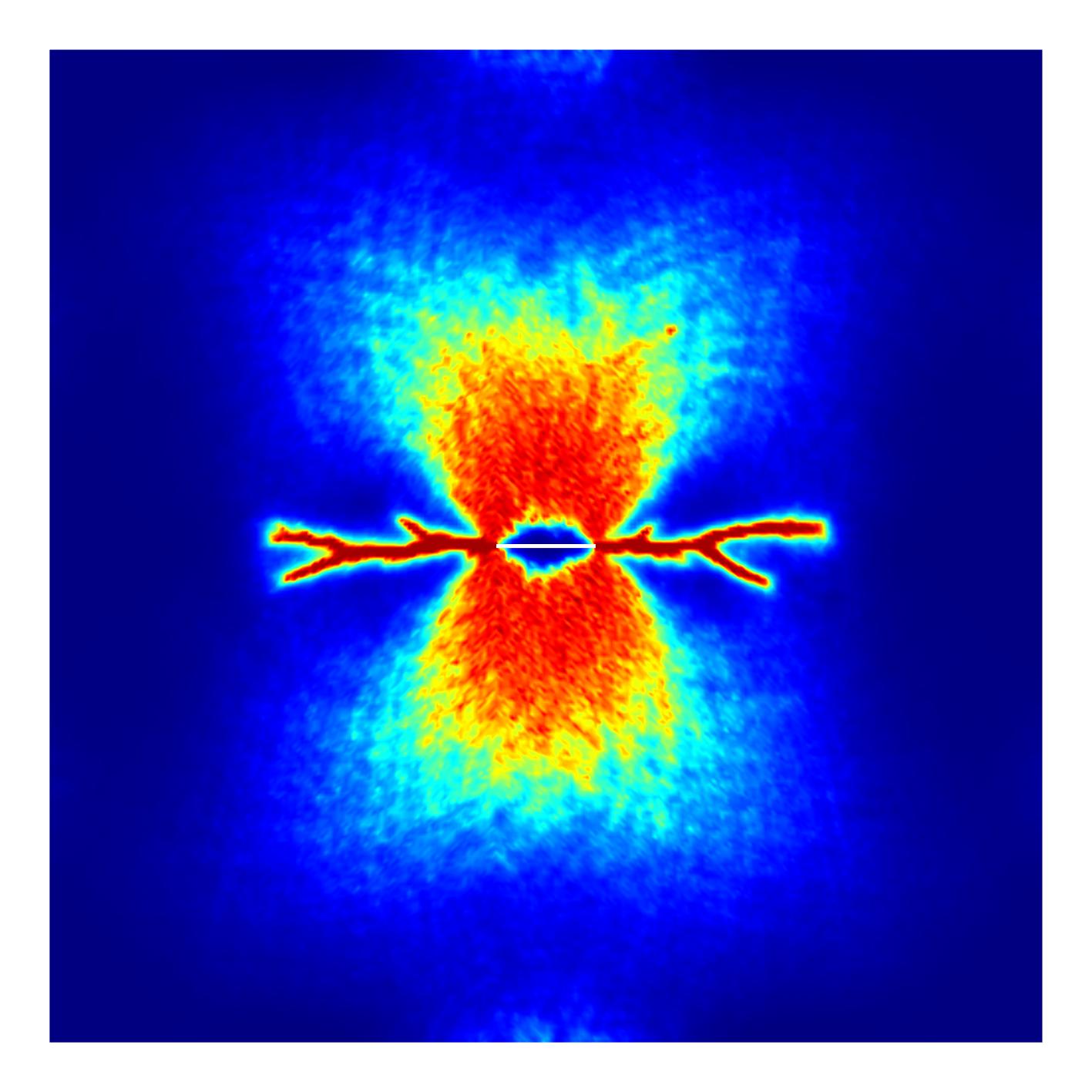} \includegraphics[width = 3cm]{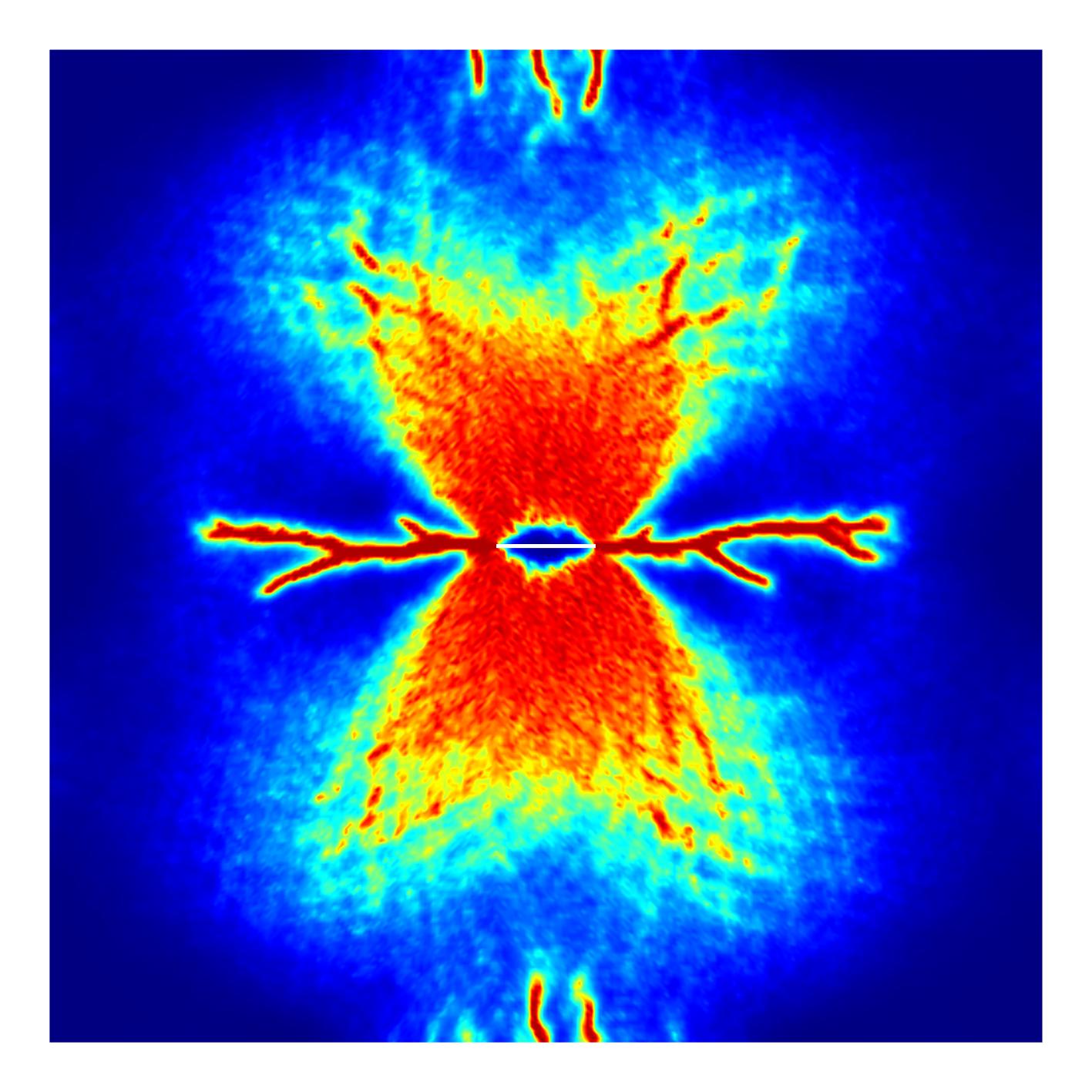}\includegraphics[width = 3cm]{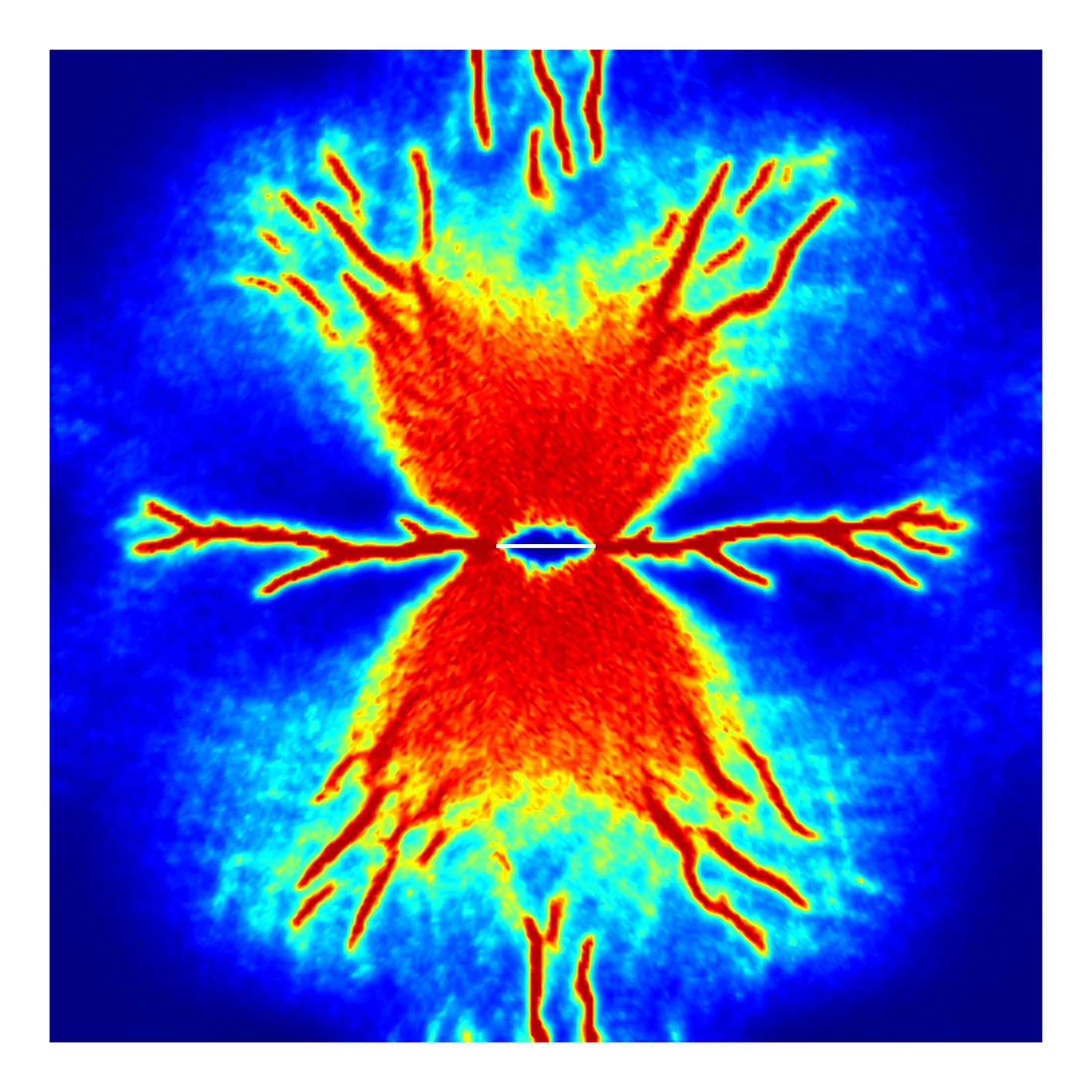}\\
		$m=3$ &\includegraphics[width = 3cm]{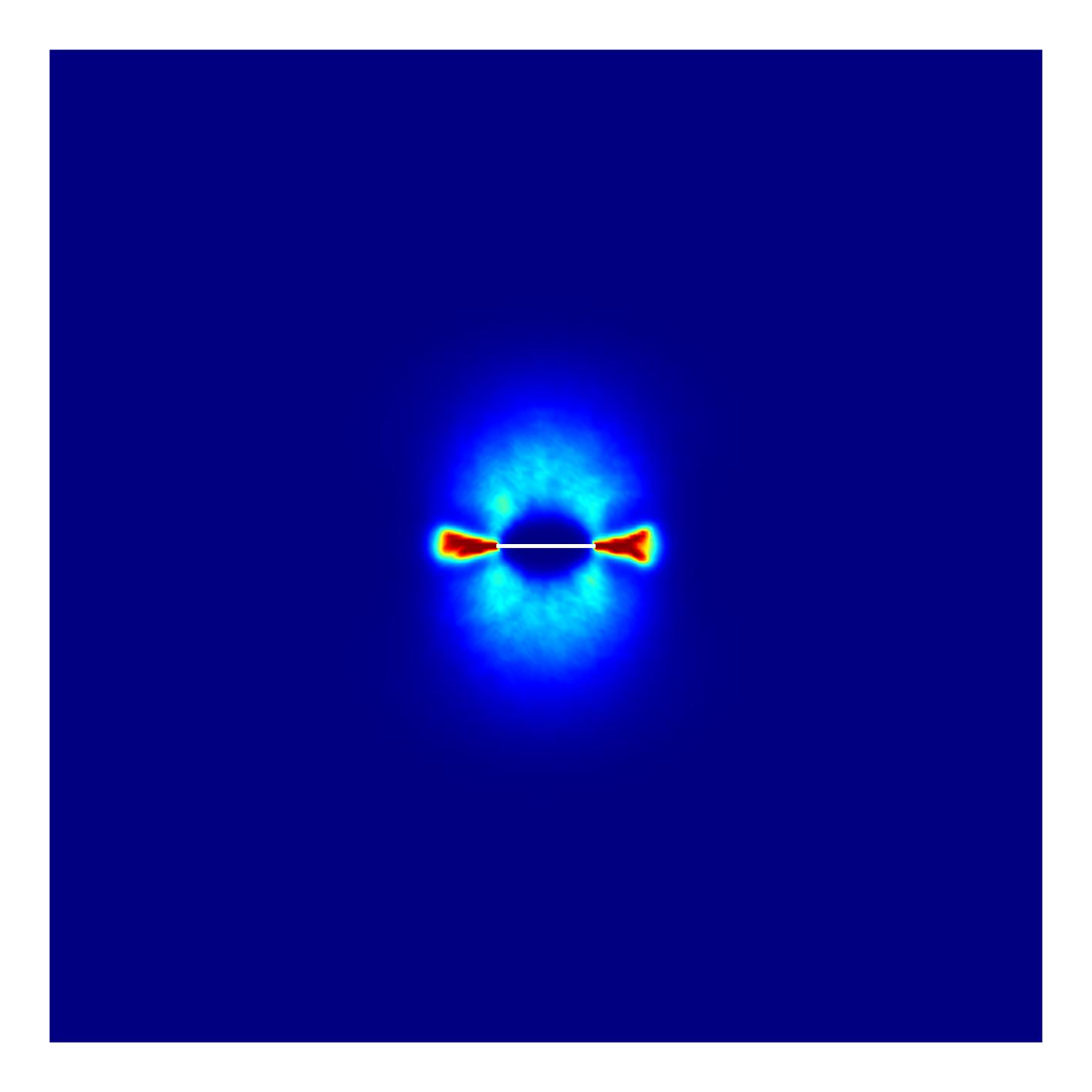}  \includegraphics[width = 3cm]{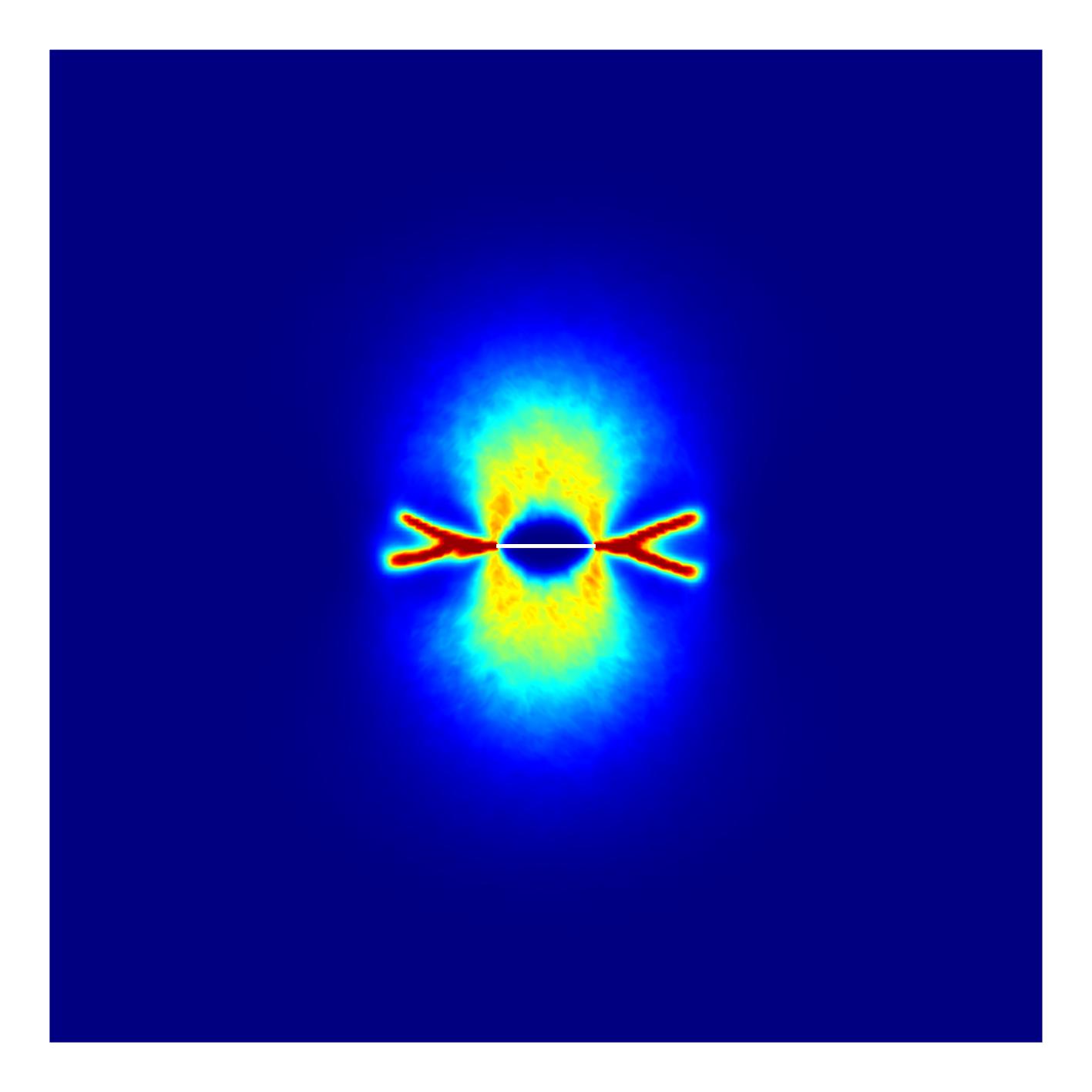}  \includegraphics[width = 3cm]{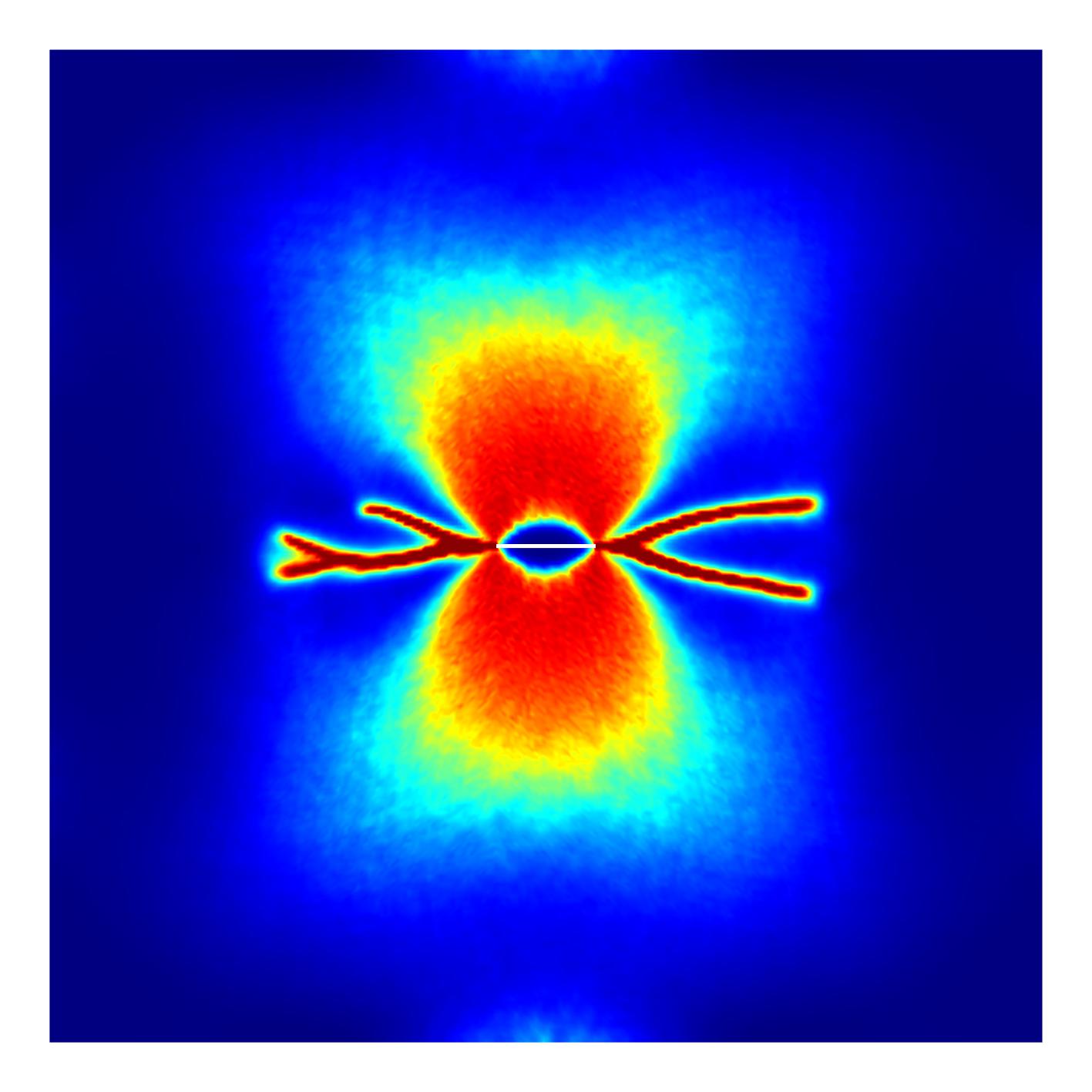} \includegraphics[width = 3cm]{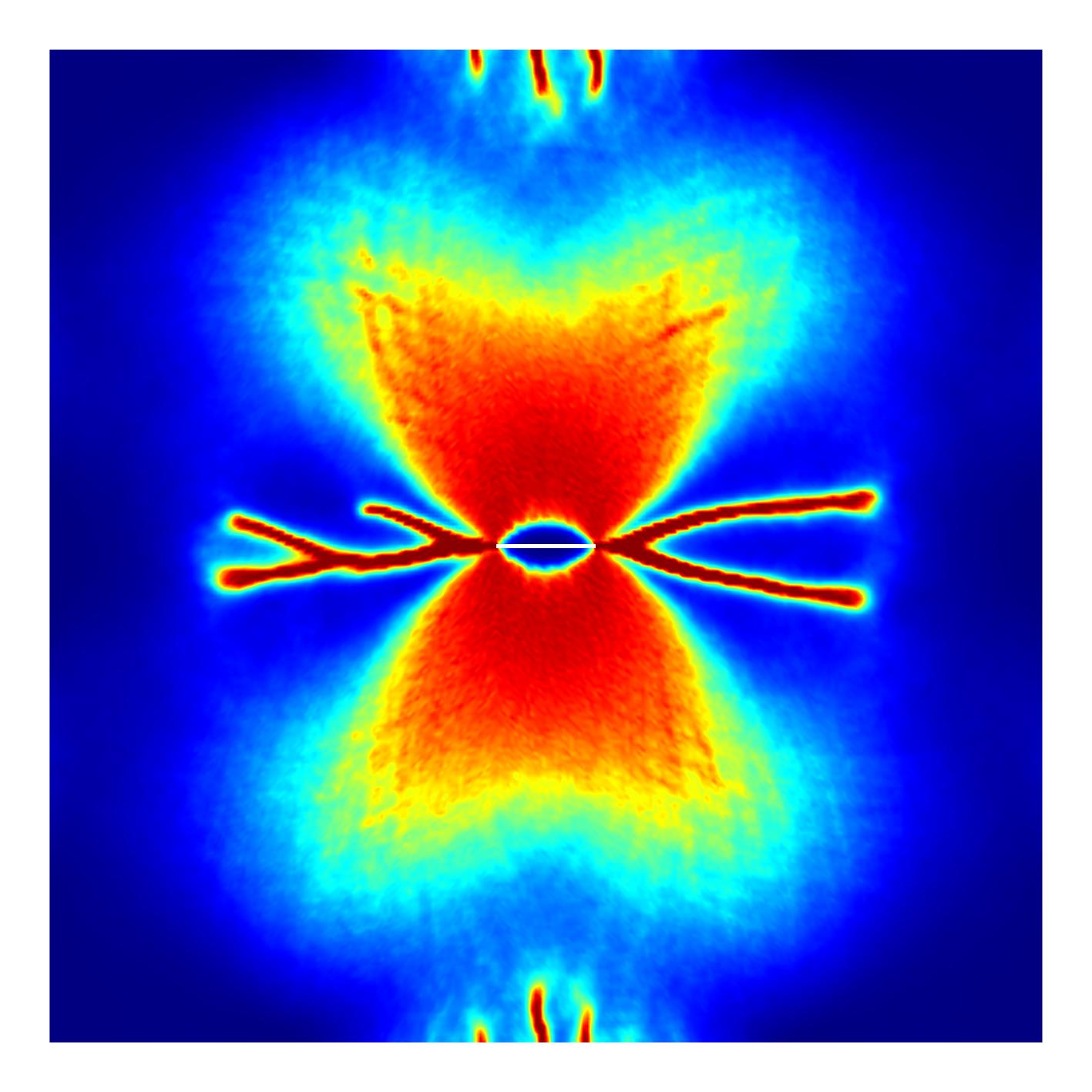}\includegraphics[width = 3cm]{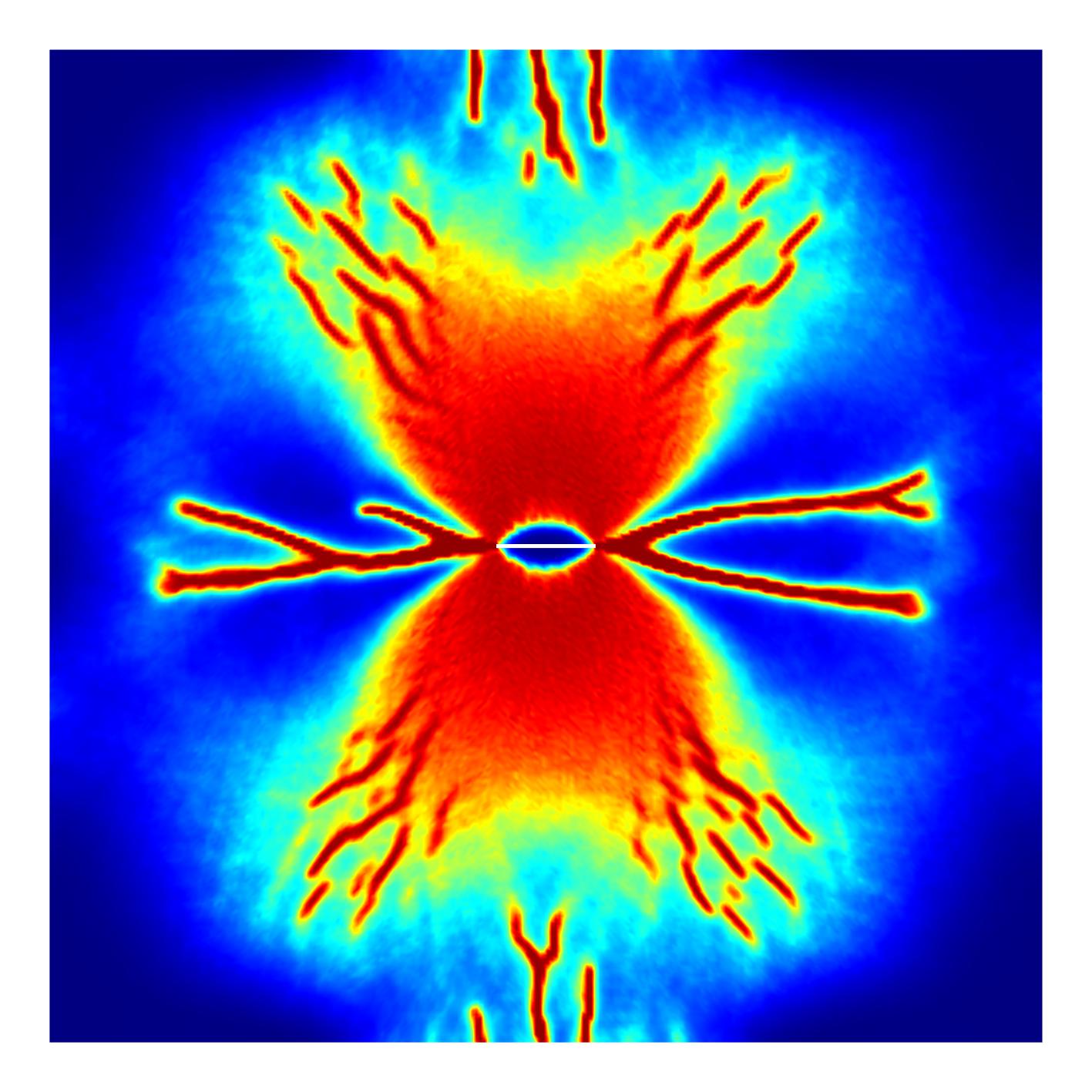}\\
		$m=5$ &\includegraphics[width = 3cm]{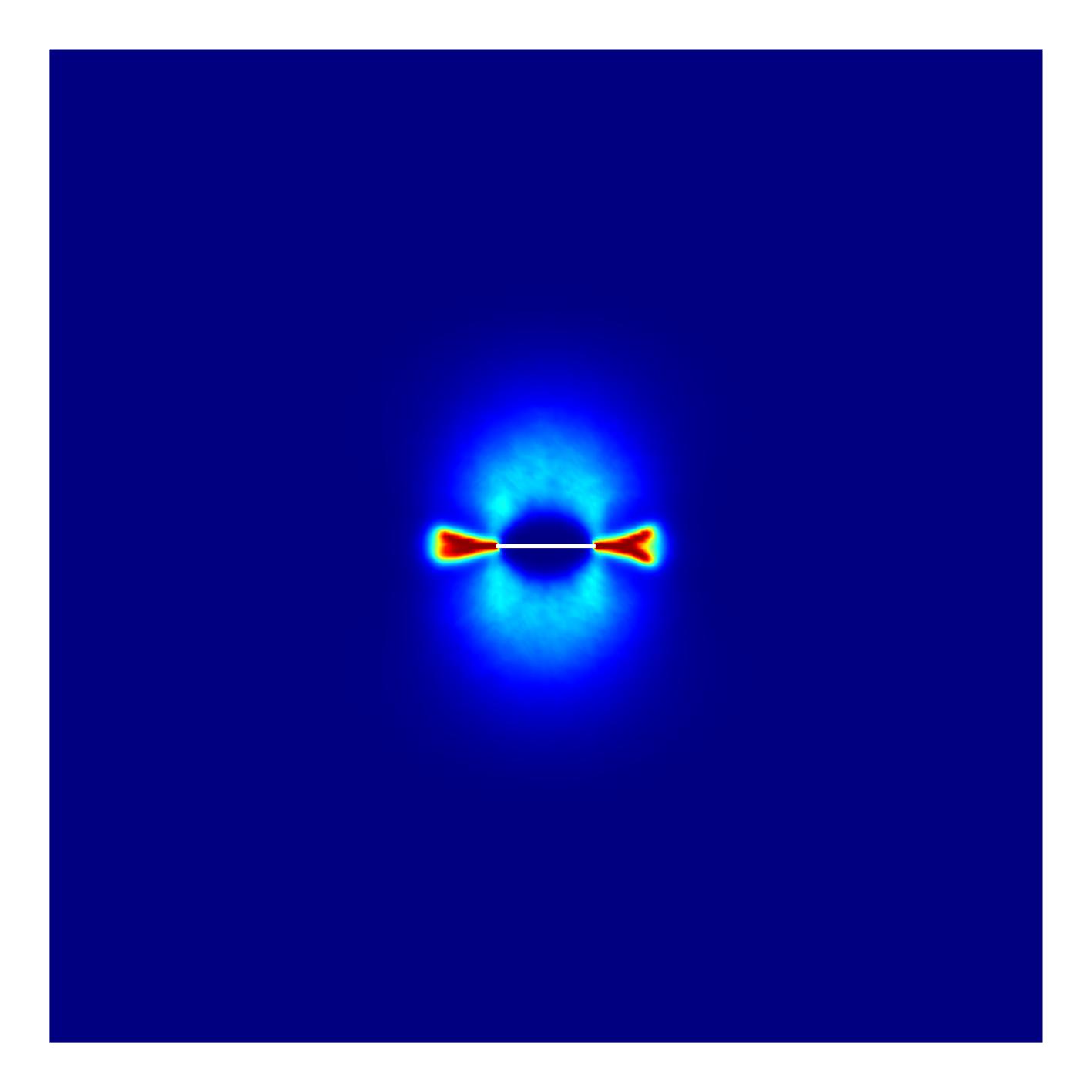}  \includegraphics[width = 3cm]{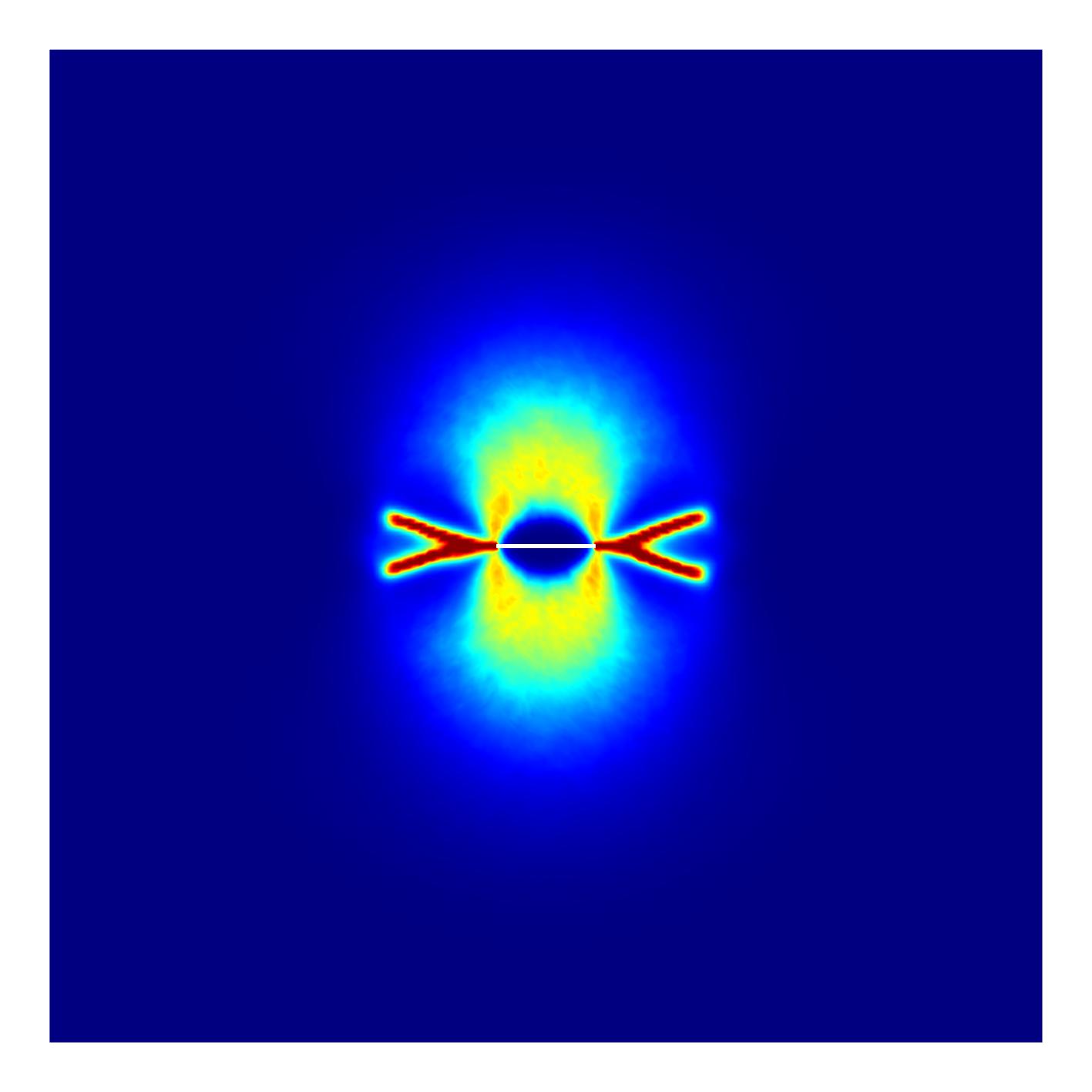}  \includegraphics[width = 3cm]{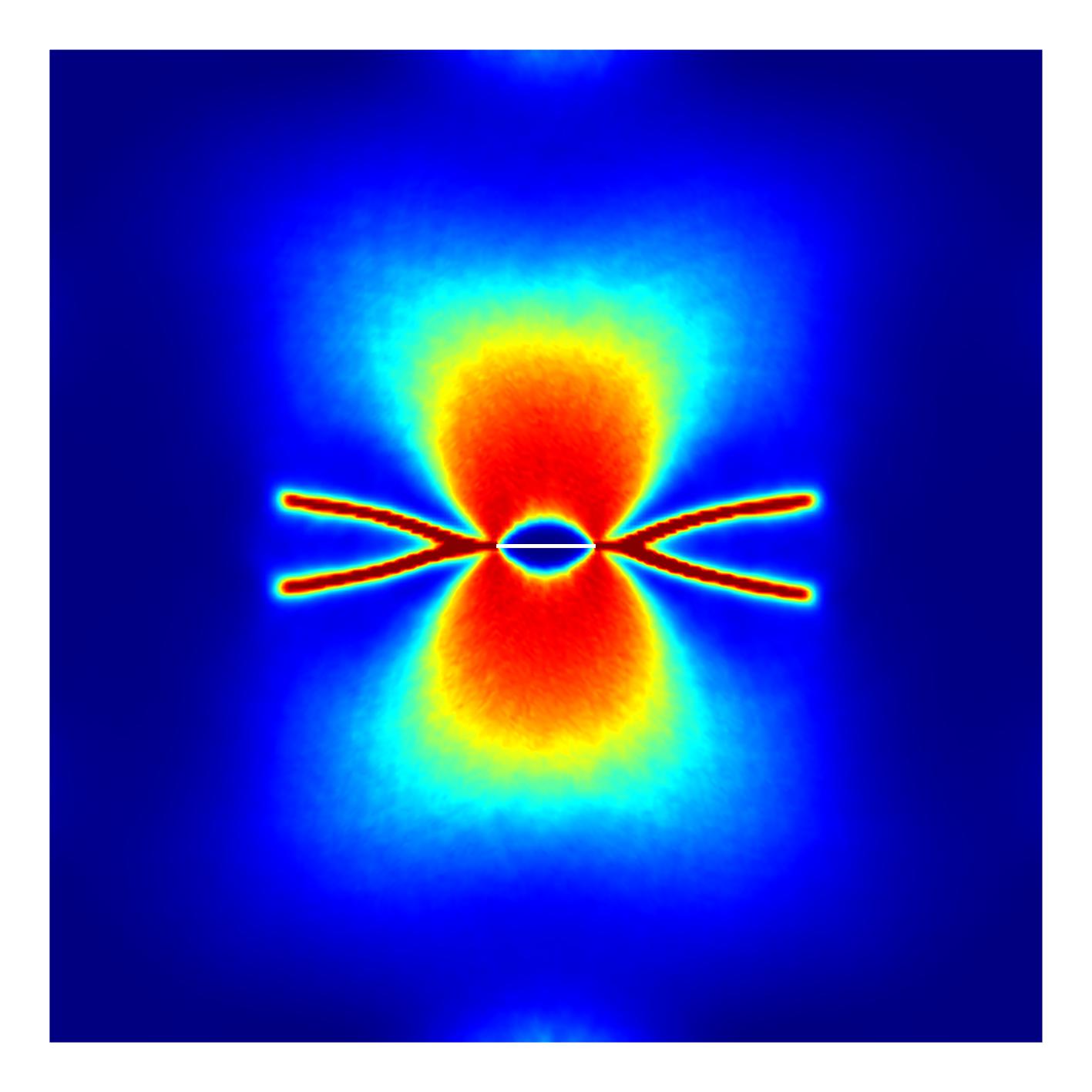} \includegraphics[width = 3cm]{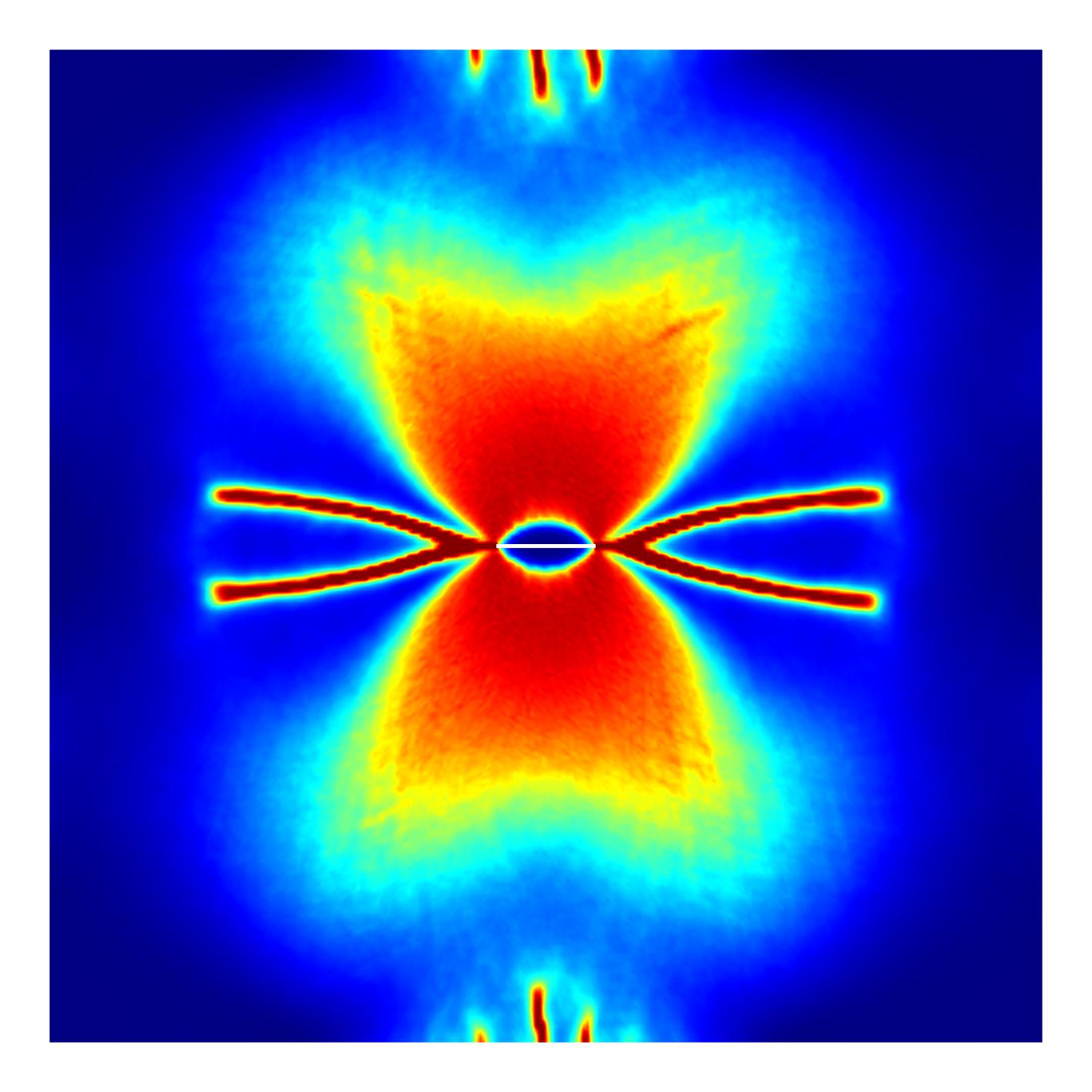}\includegraphics[width = 3cm]{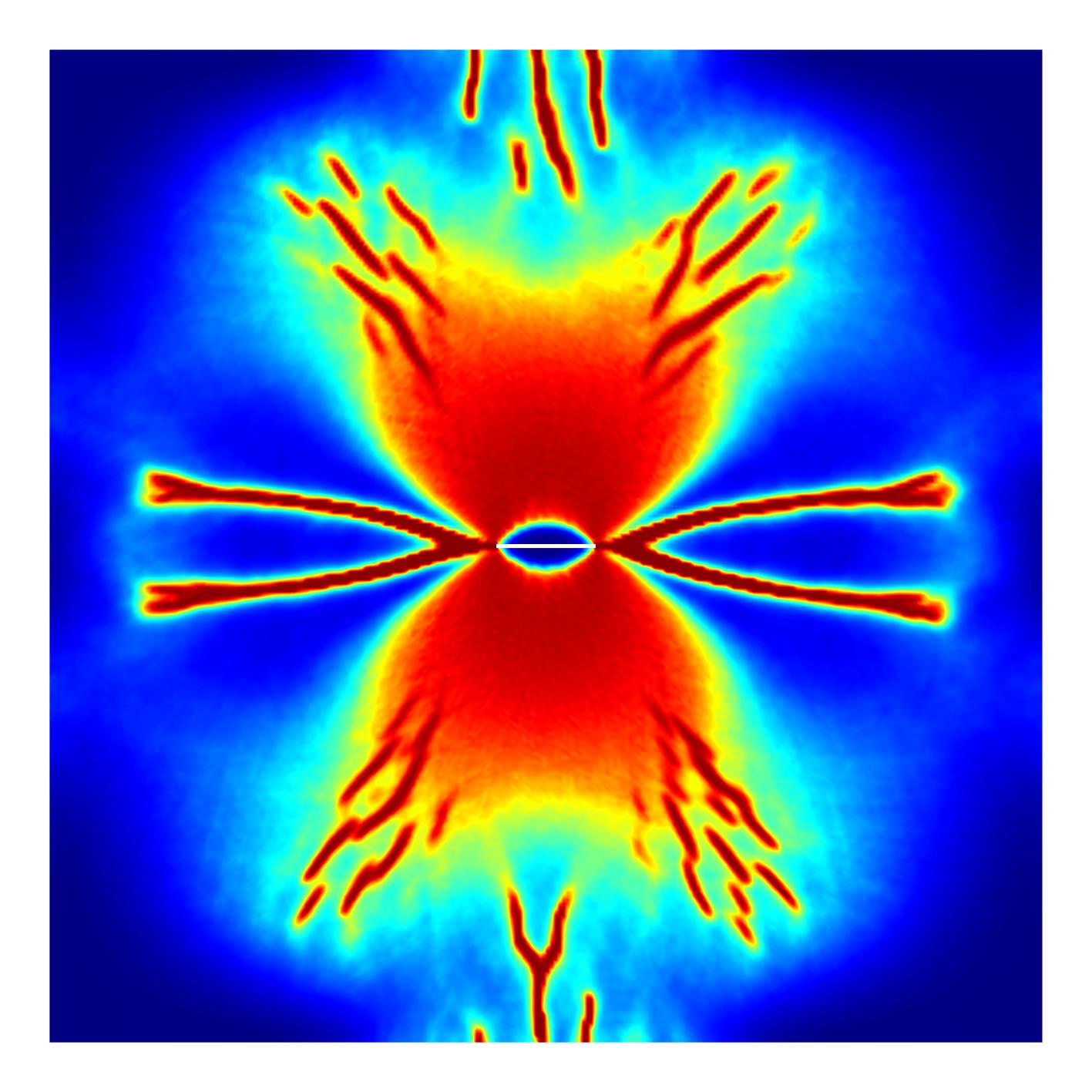}\\
		$m=7$ &\includegraphics[width = 3cm]{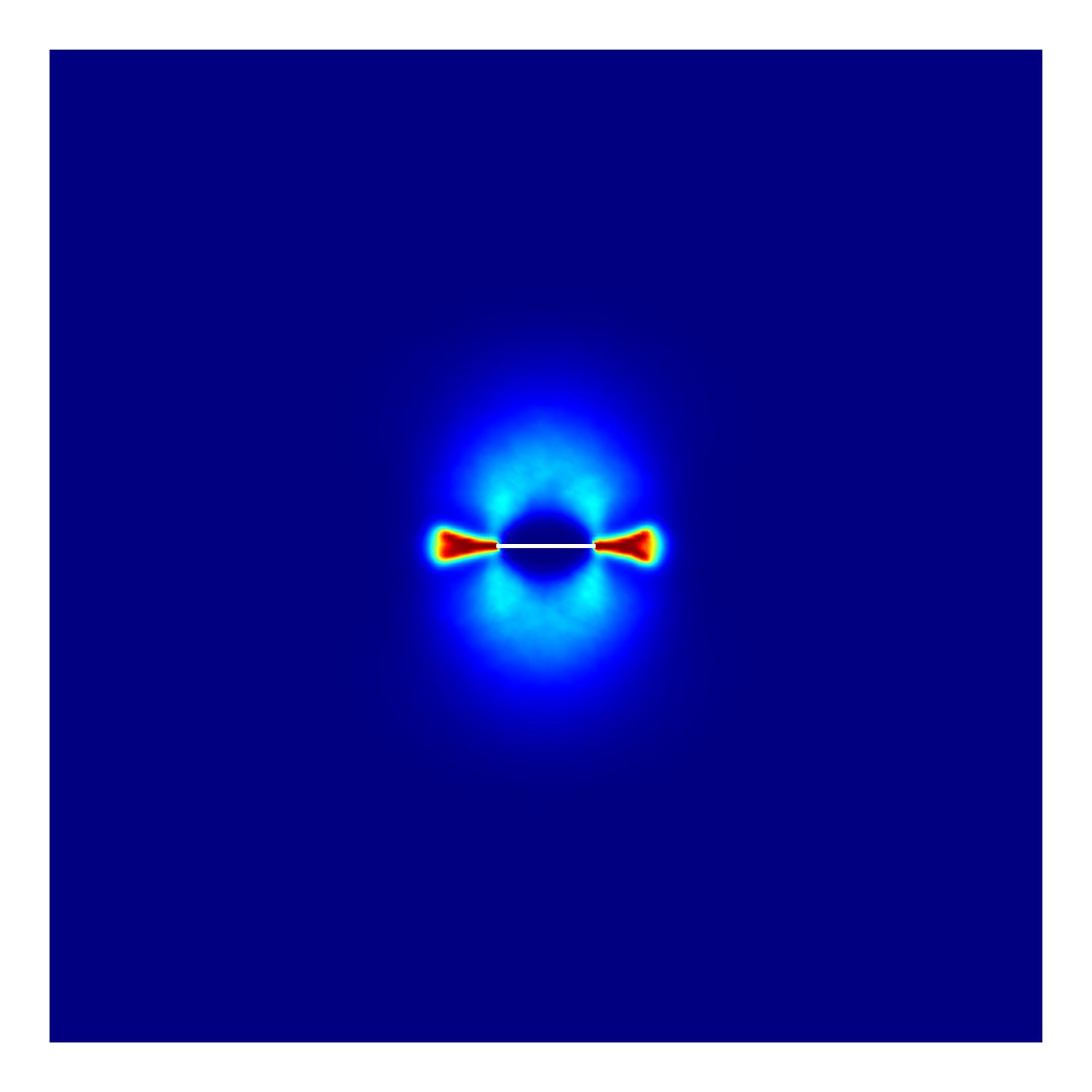}  \includegraphics[width = 3cm]{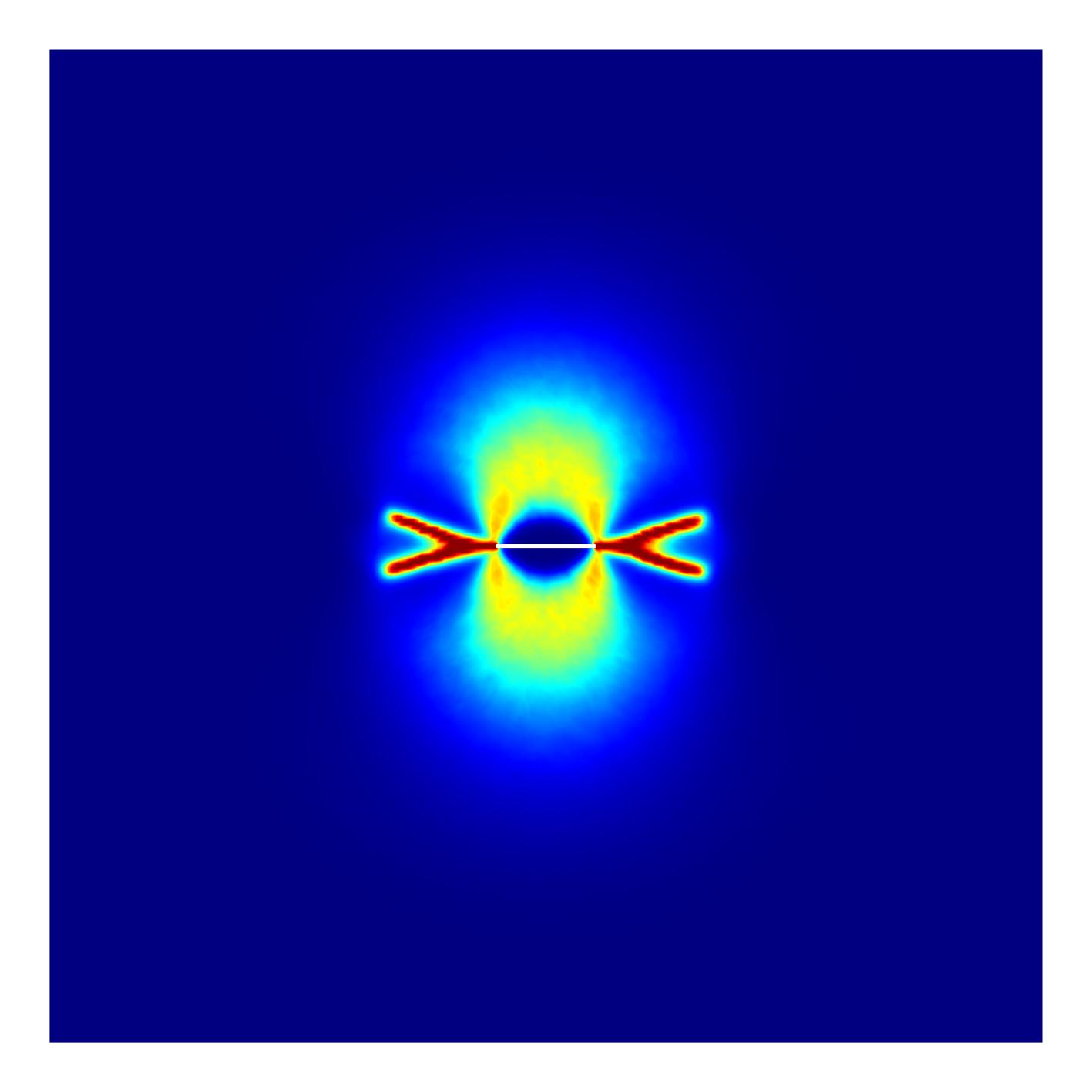}  \includegraphics[width = 3cm]{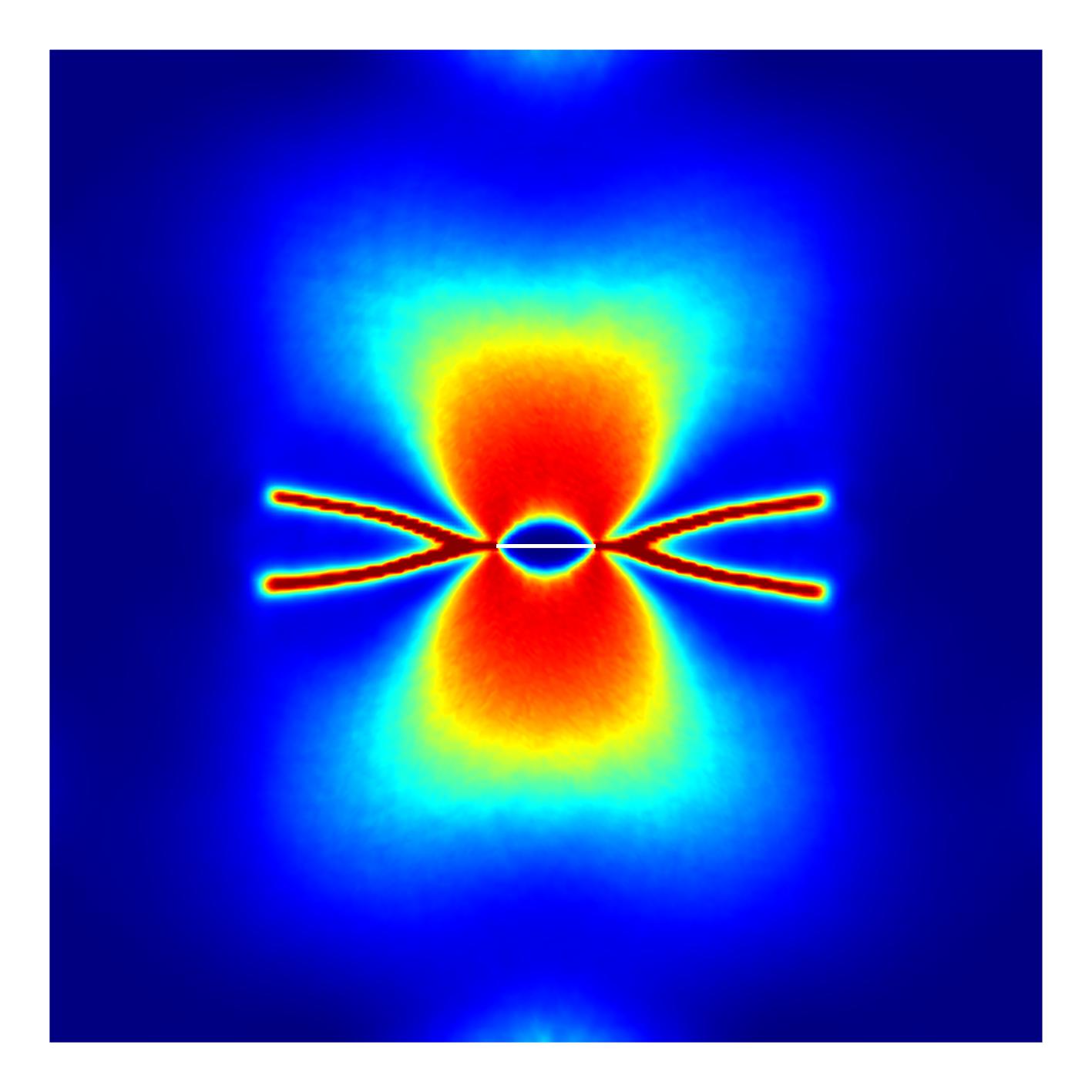} \includegraphics[width = 3cm]{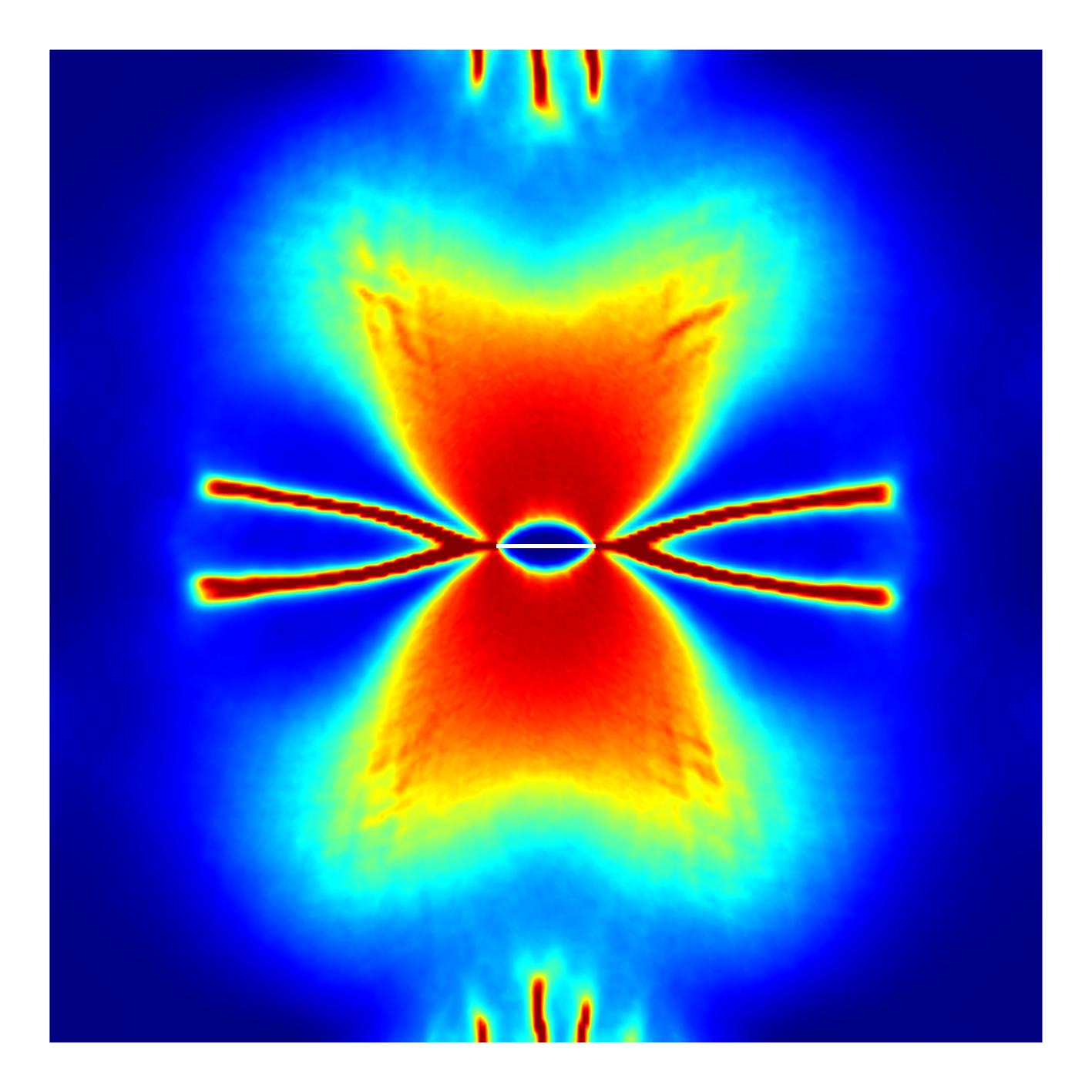}\includegraphics[width = 3cm]{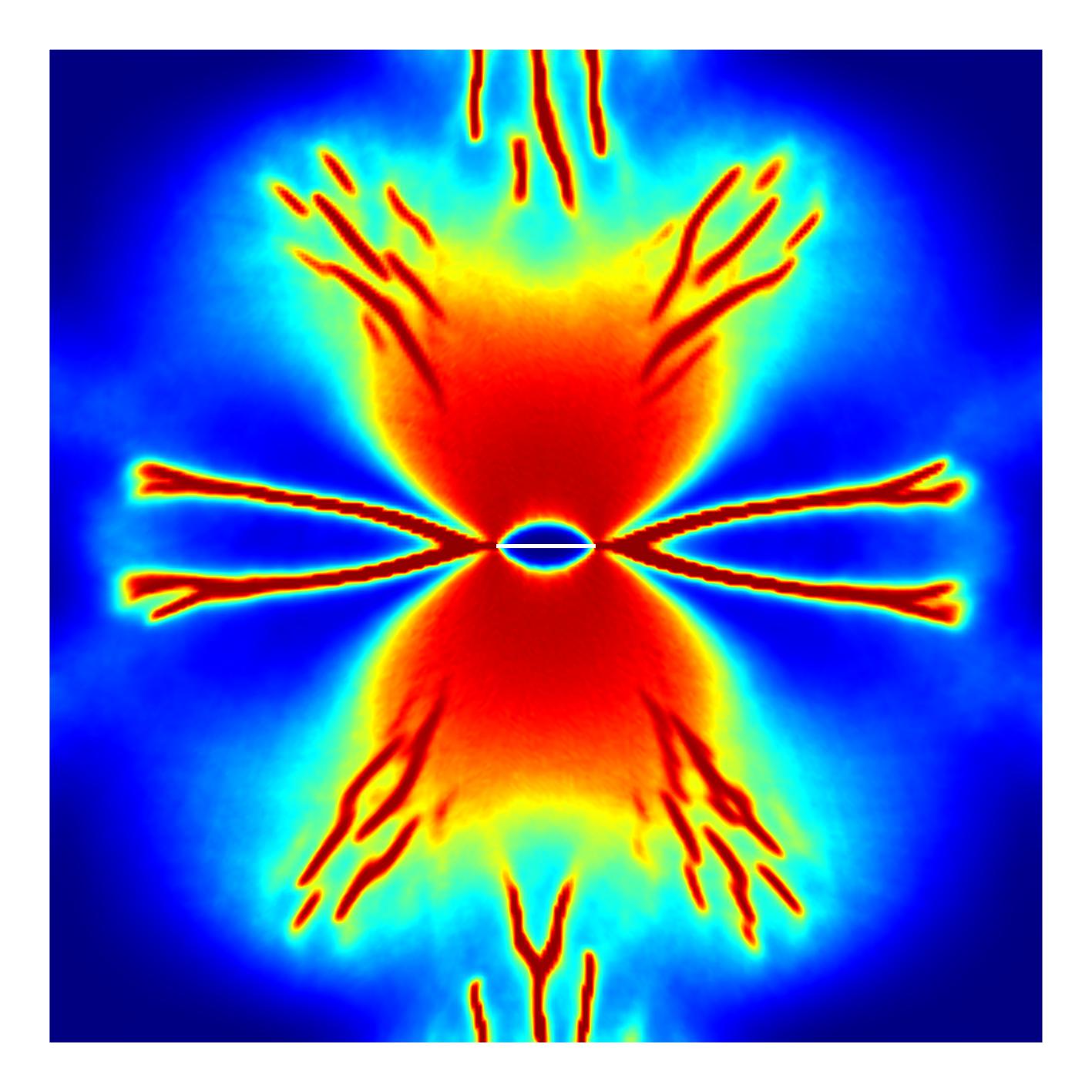}\\
		$m=9$ &\includegraphics[width = 3cm]{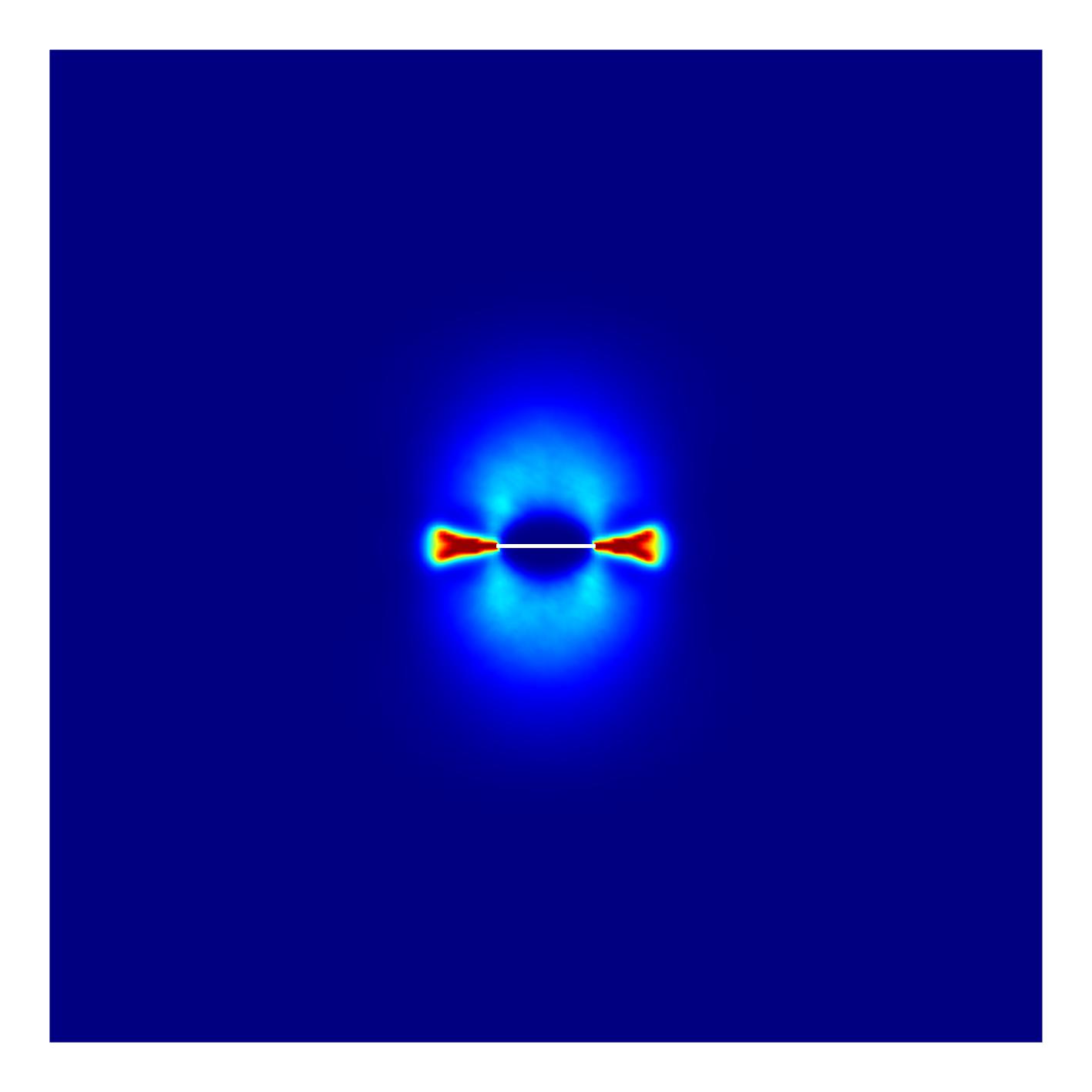}  \includegraphics[width = 3cm]{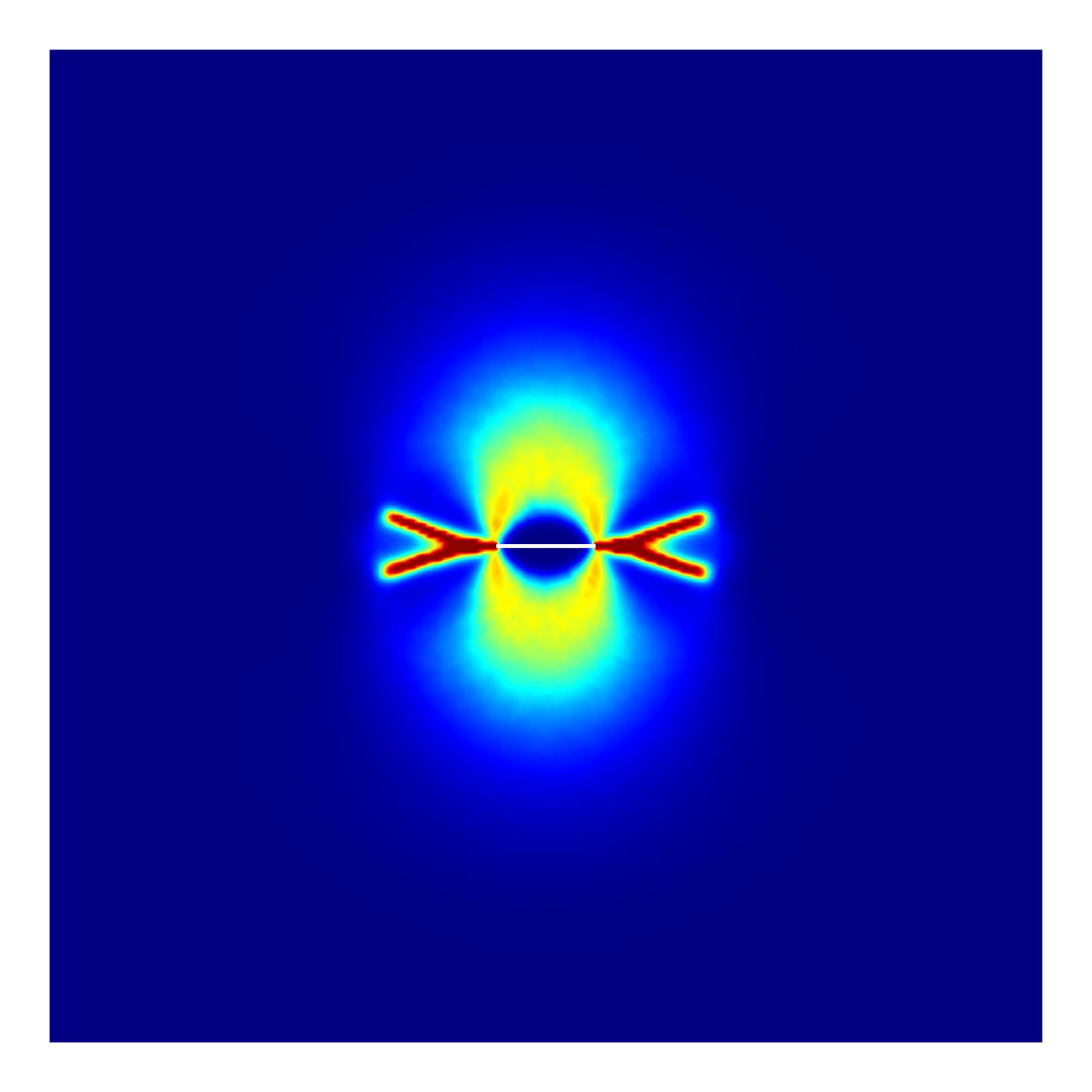}  \includegraphics[width = 3cm]{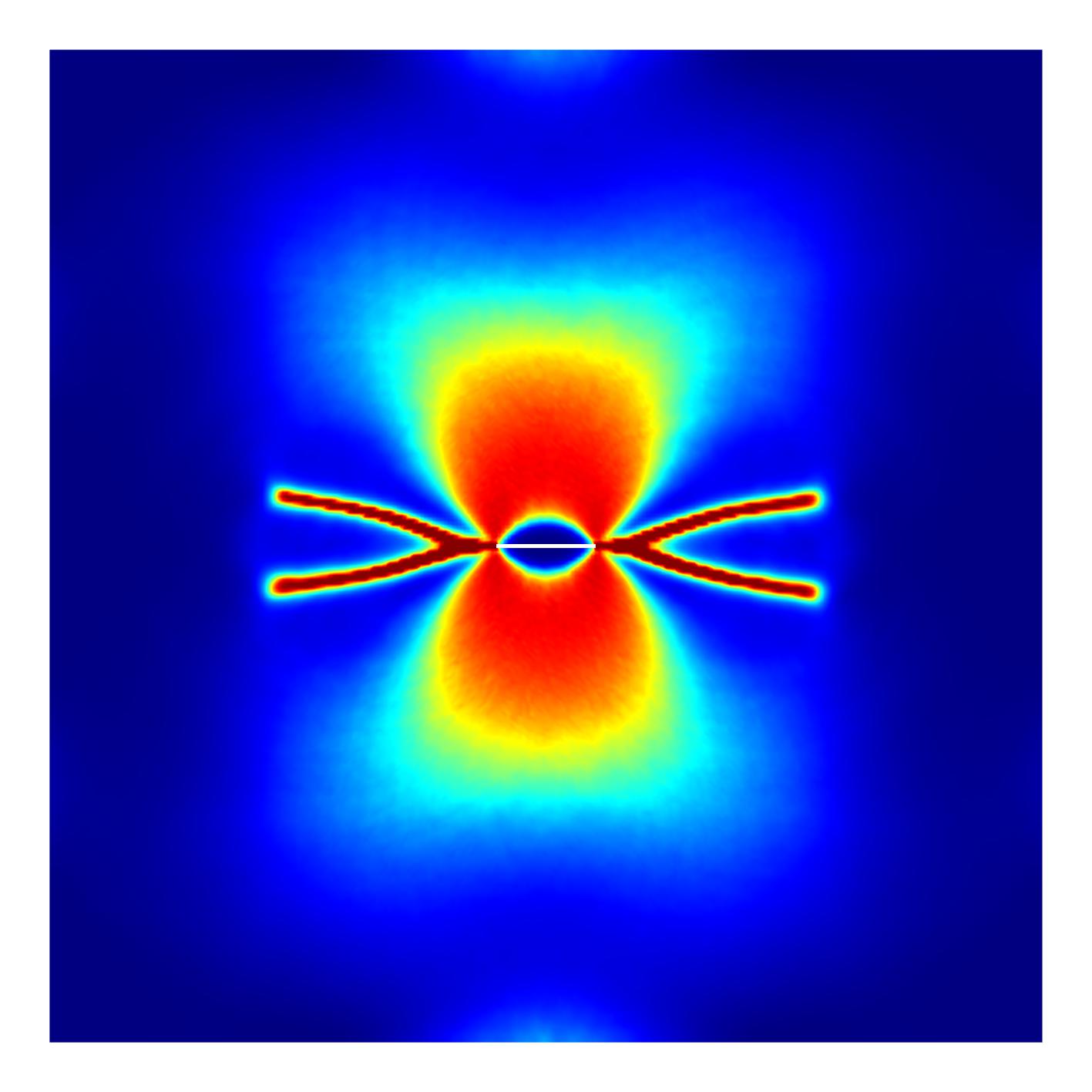} \includegraphics[width = 3cm]{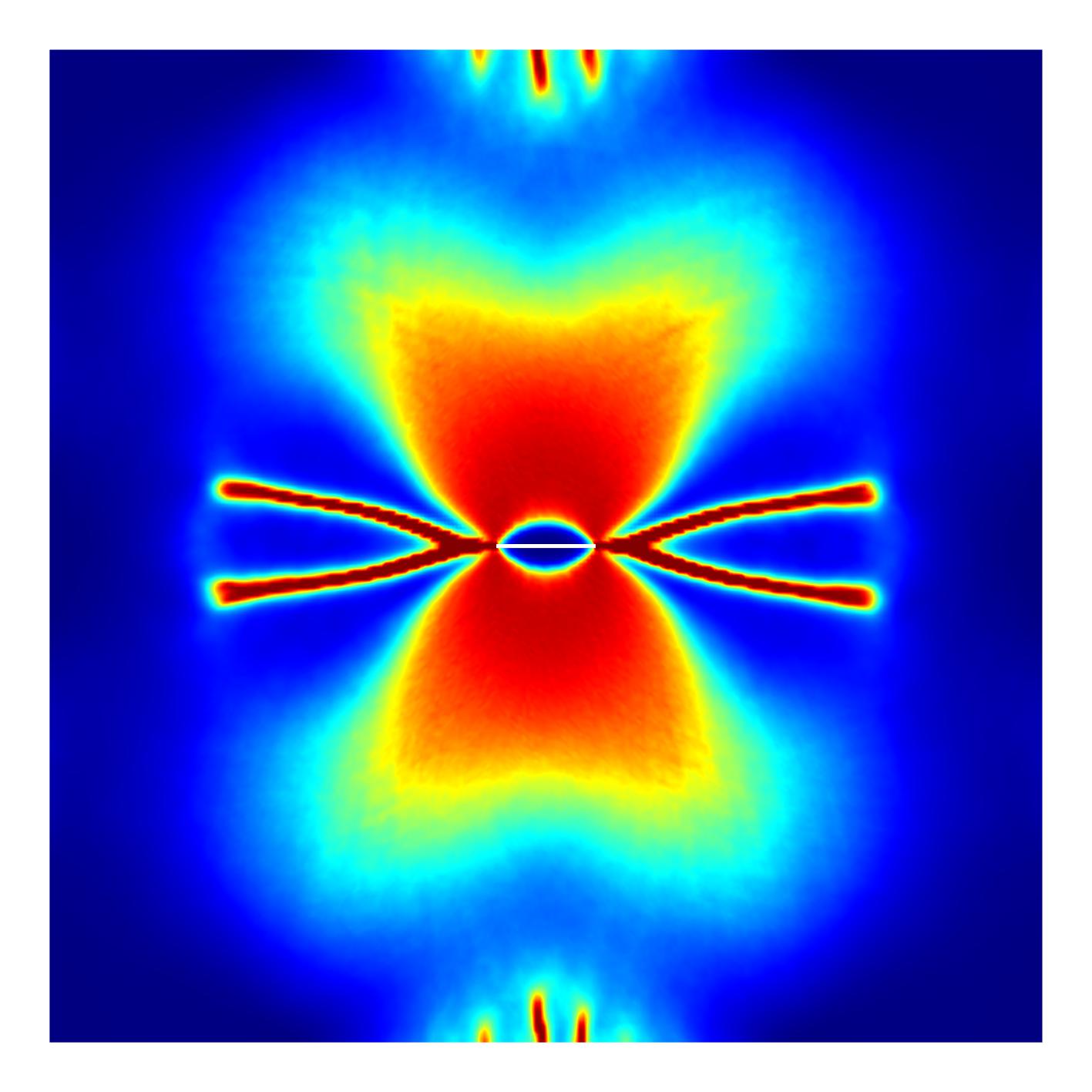}\includegraphics[width = 3cm]{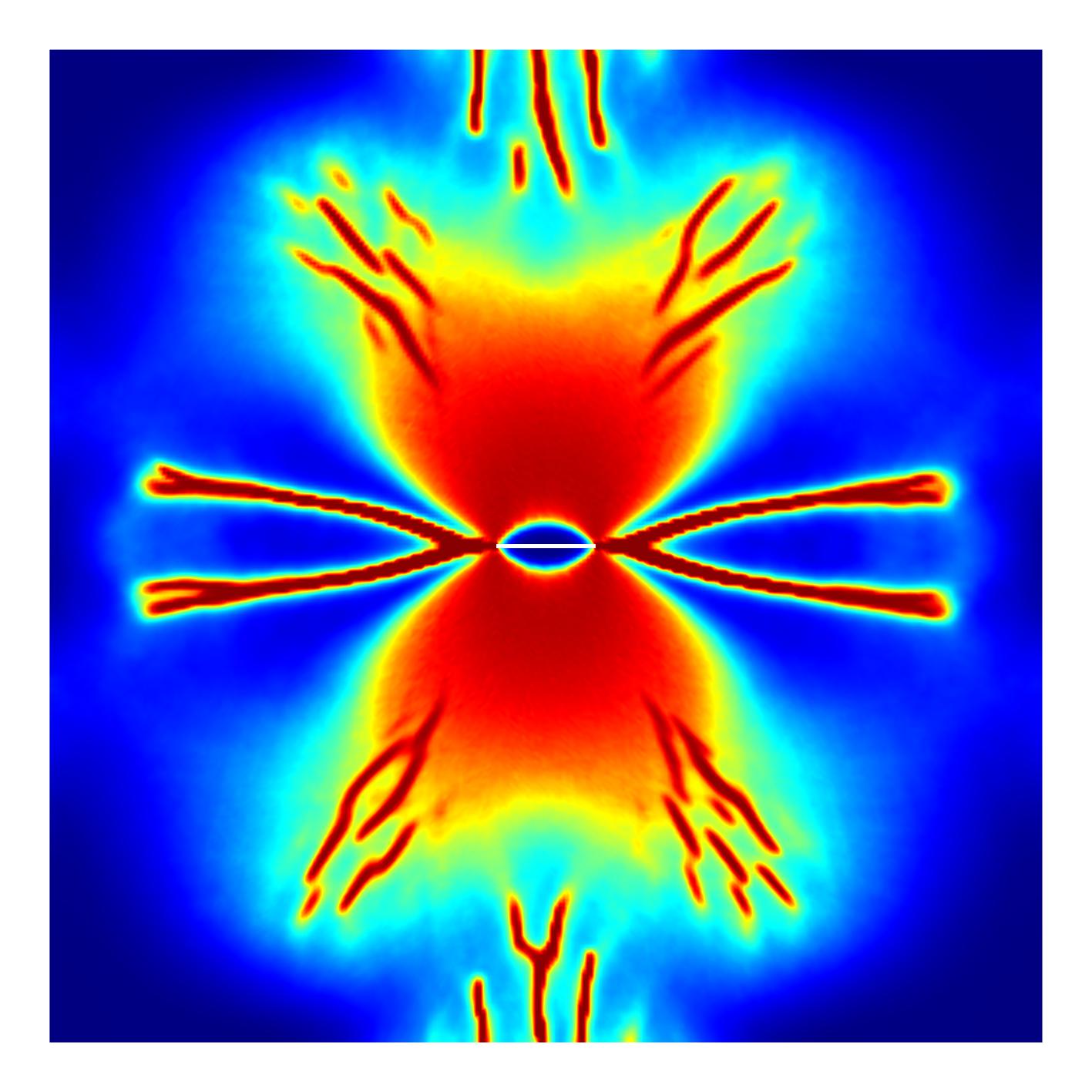}\\
		$m=inf$ &\includegraphics[width = 3cm]{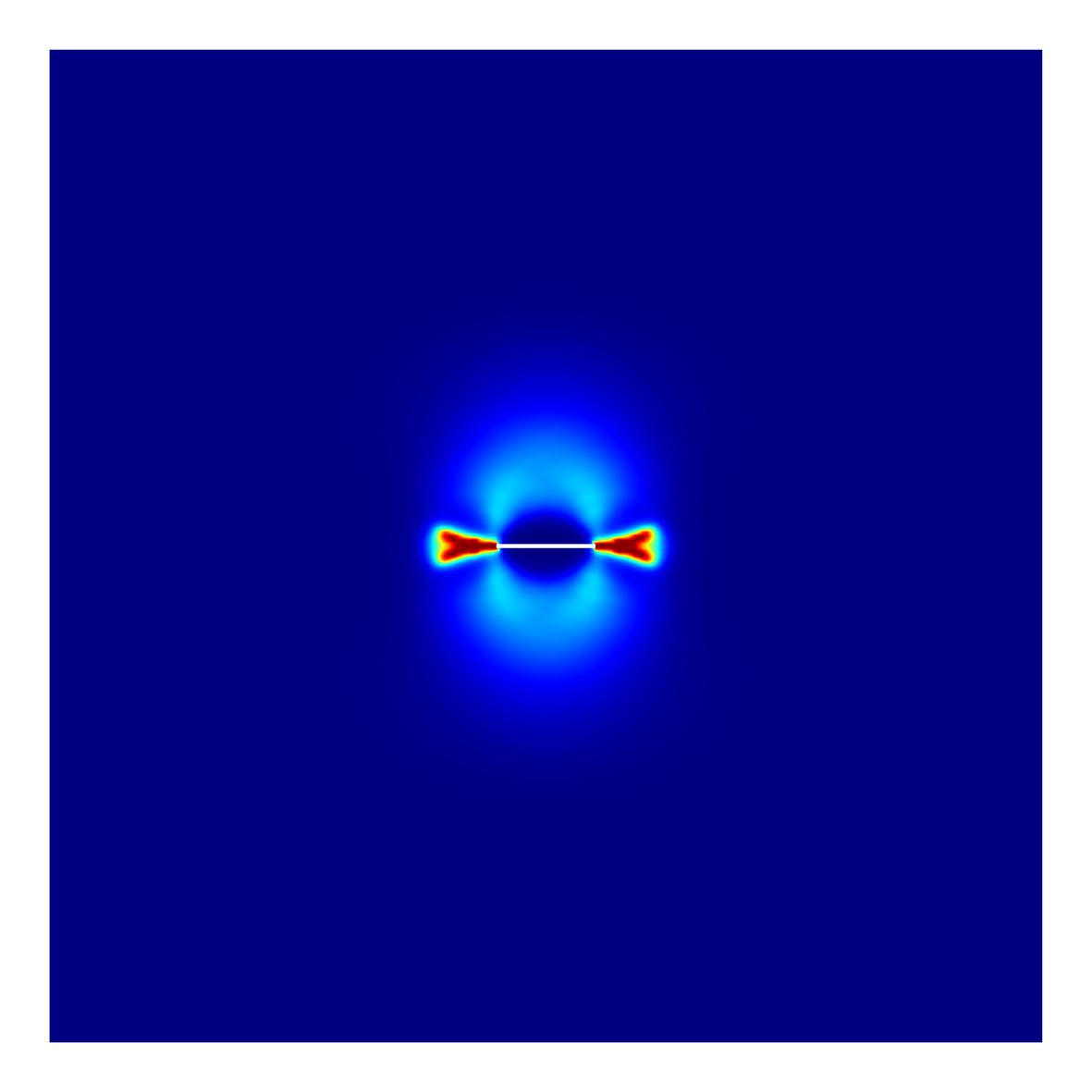}  \includegraphics[width = 3cm]{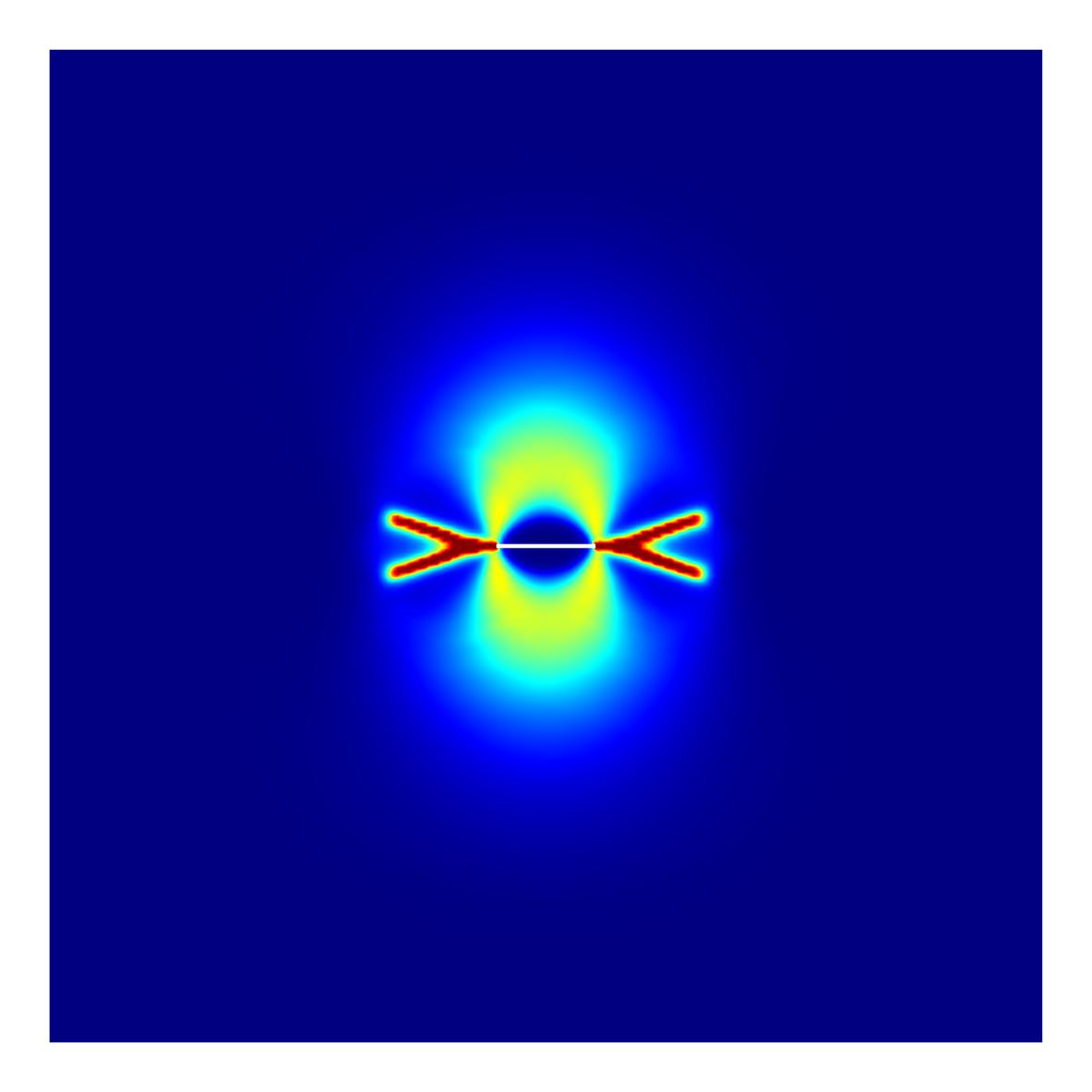}  \includegraphics[width = 3cm]{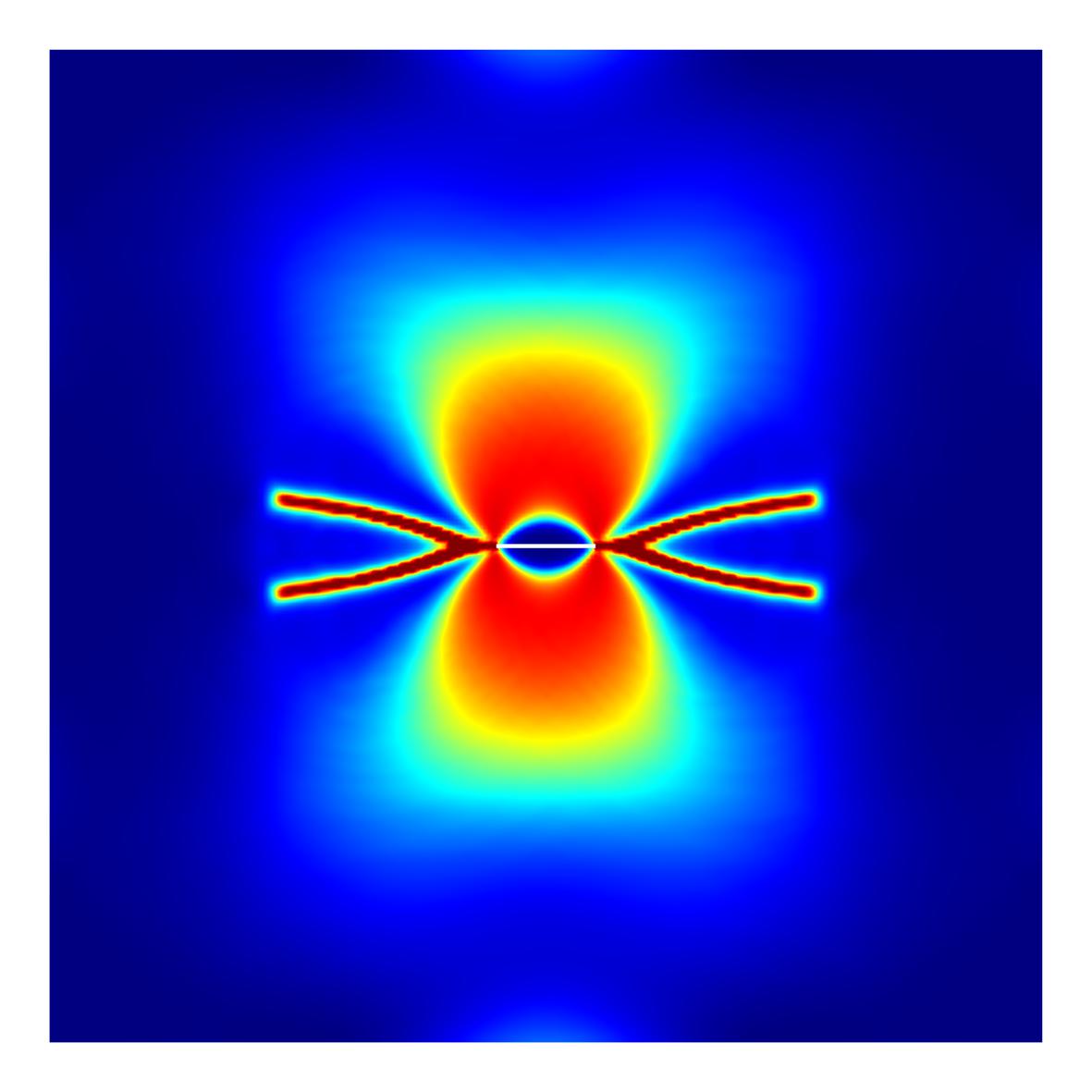} \includegraphics[width = 3cm]{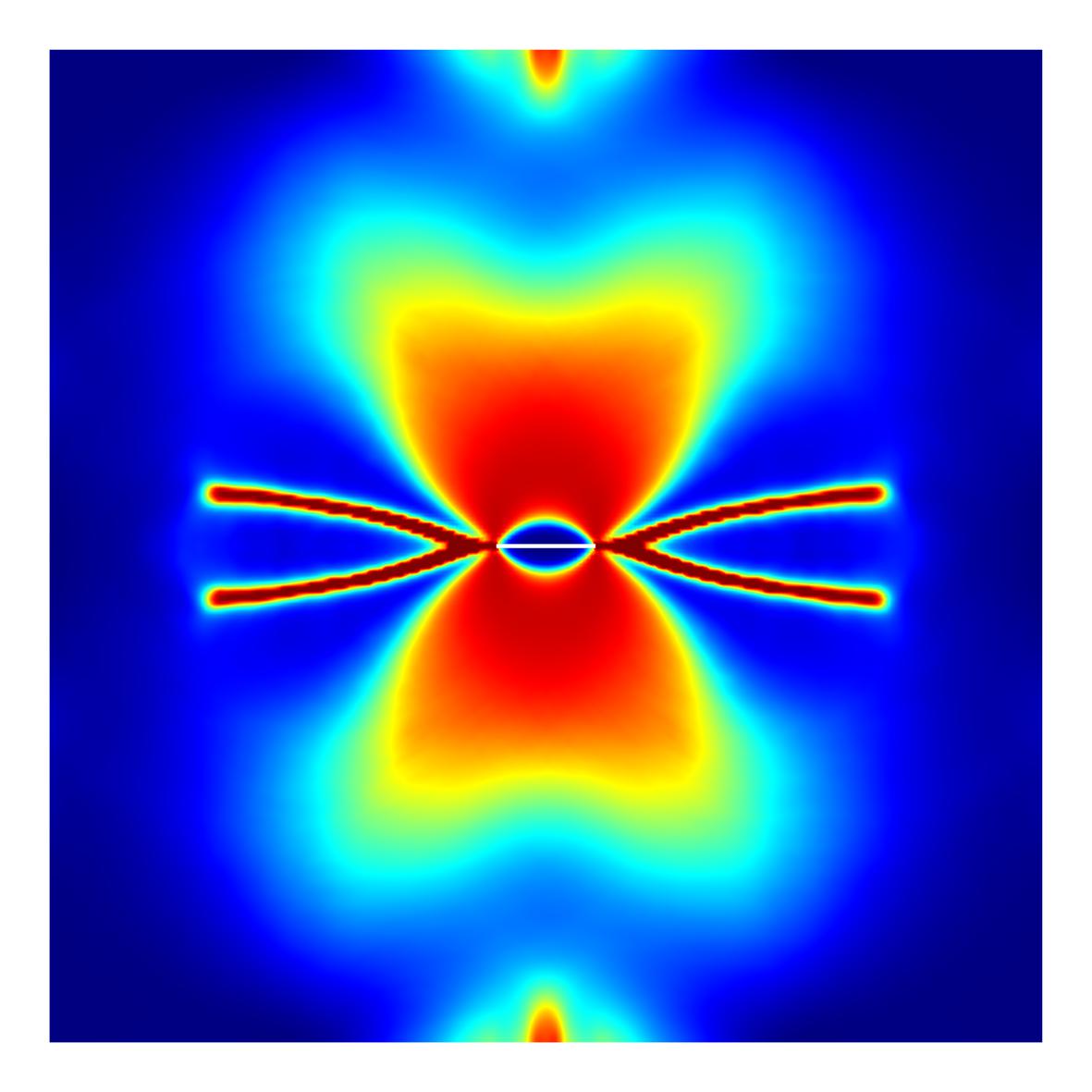}\includegraphics[width = 3cm]{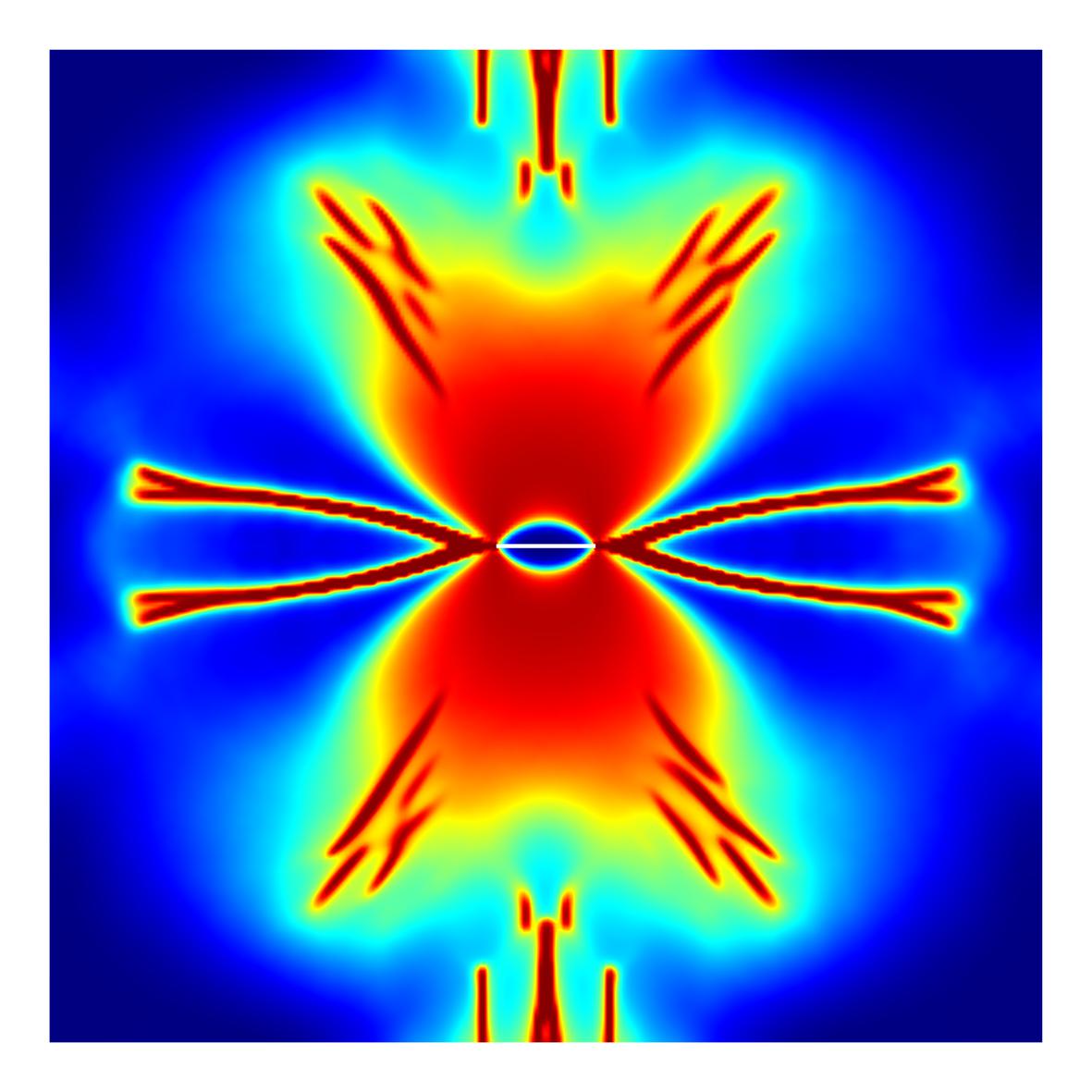}\\
	
		\end{tabular}
	\caption{Crack patterns of the plate subjected to internal pressure}
	\label{Crack patterns of the plate subjected to internal pressure}
	\end{figure}

Figure \ref{Crack patterns of the plate subjected to internal pressure} presents the crack patterns of the plate subjected to internal pressure for different $m$. Complex crack patterns, such as many crack branching, are observed. These observations are different from the previous examples. When the time $t$ reaches 20 $\mu$s, cracks initiate from the left and right tips of the notch and then start to branch. When the time $t$ increases to 30 $\mu$s, the branching cracks propagate. When $t=$ 50$\mu$s, large ``damage" region are observed around the notch where $\phi$ is large. At $t=60$ $\mu$s, some cracks initiate from the upper and lower boundaries of the plate, while the cracks from the notch keep propagating. Finally, when $t=70$ $\mu$s, many new cracks occur inside the plate and the cracks from the upper and lower boundaries of the plate start to branch. For a smaller $m$, more crack branching is observed during the crack propagation. Therefore, a smaller $m$  produces more complex crack patterns. 

\subsection{3D Petersson beam}\label{3D Petersson beam}

\begin{figure}[htbp]
	\centering
	\includegraphics[width = 12cm]{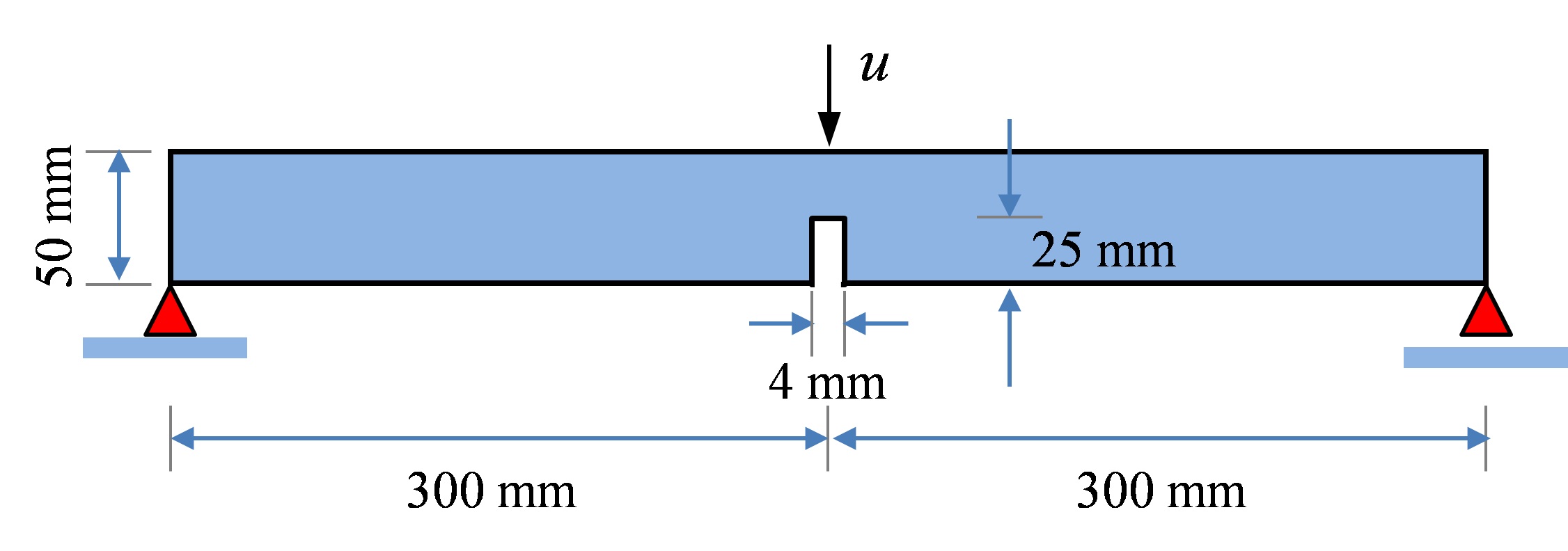}
	\caption{Geometry and boundary condition of Petersson beam}
	\label{Geometry and boundary condition of Petersson beam}
	\end{figure}

In this example, the phase field model is applied to a single edge notched beam subjected to three-point bending (the so-called Petersson beam). A 3D simulation is conducted in this example. The geometry and boundary conditions are depicted in Fig. \ref{Geometry and boundary condition of Petersson beam} according to \citet{petersson1981crack}. The thickness of the beam is 50 mm. The following mechanical properties of the beam are chosen \citet{petersson1981crack}: Young's modulus $E$ = 27 GPa, Poisson's ratio $\nu$ = 0.21, critical energy release rate $G_c$ = 56 J/m$^2$ and length scale $l_0$ = 1 mm. We choose the maximum element size $h$ = 5 mm in most of the beam but $h$ = 1 mm in the region where the crack is expected to propagate. A displacement increment of $\Delta u = 2.5\times10^{-4}$ mm is used.

\begin{figure}[htbp]
	\centering
	\includegraphics[width = 10cm]{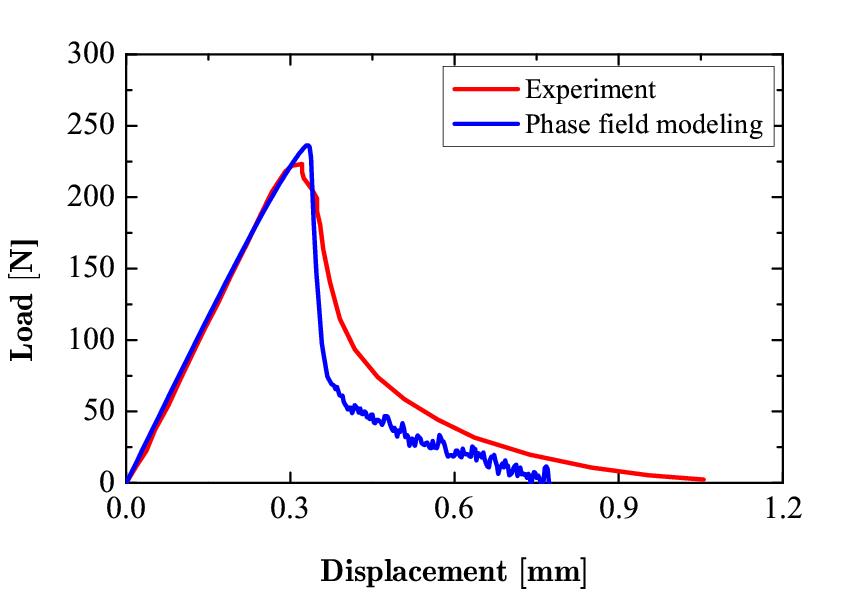}
	\caption{Load-displacement curves of 3D petersson beam}
	\label{Load-displacement curves of 3D petersson beam}
	\end{figure}

\begin{figure}[htbp]
	\centering
	\subfigure[]{\includegraphics[width = 8cm]{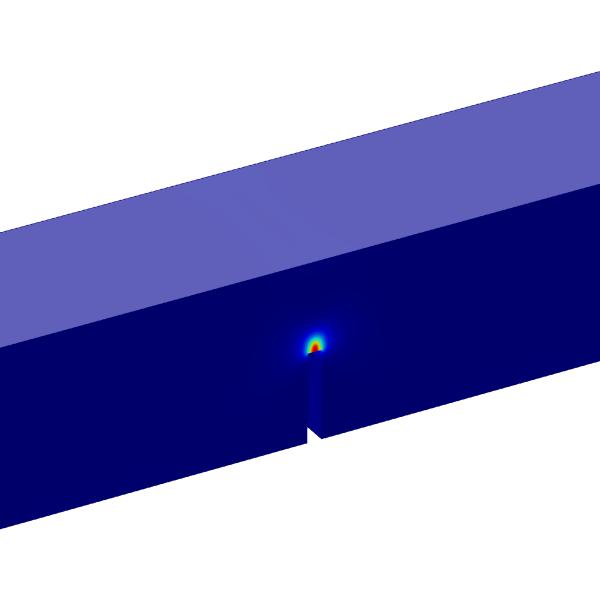}}
	\subfigure[]{\includegraphics[width = 8cm]{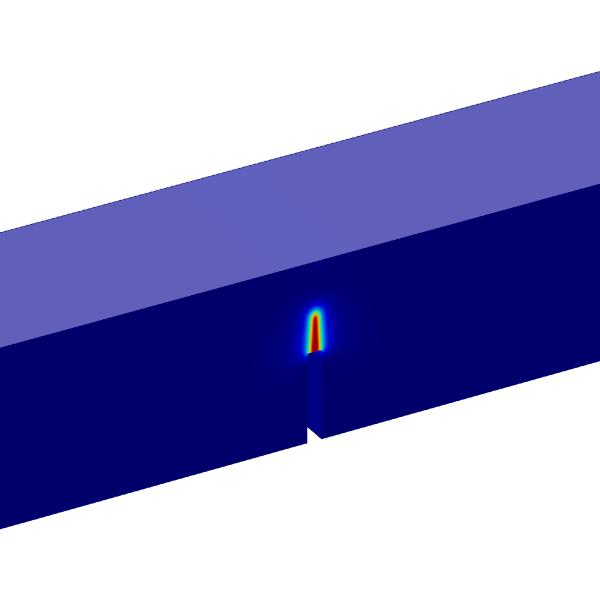}}\\
	\subfigure[]{\includegraphics[width = 8cm]{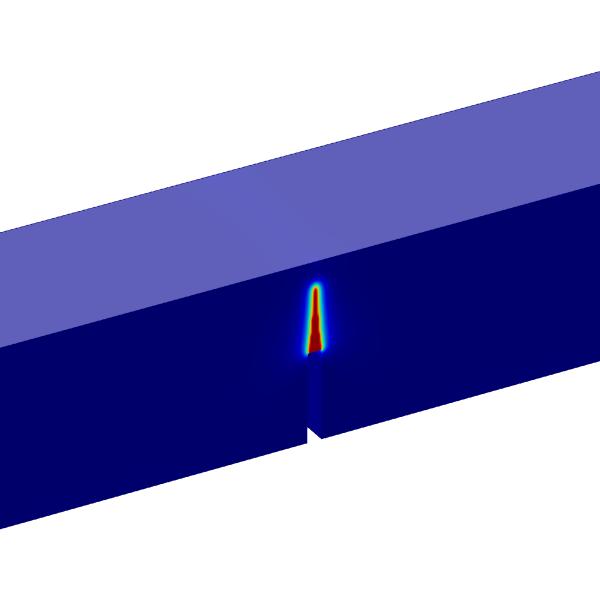}}
	\caption{Crack propagation in the 3D petersson beam at a displacement of (a) $u = 0.34$ mm, (b) $u = 0.355$ mm, and (c) $u = 0.46$ mm}
	\label{Crack propagation in the 3D petersson beam}
	\end{figure}

Figure \ref{Load-displacement curves of 3D petersson beam} compares the load-displacement curves obtained of the phase field model with the experimental test \citep{petersson1981crack}. The results obtained by the phase field model are in good agreement with the experimental test. Figure \ref{Crack propagation in the 3D petersson beam} presents the crack propagation in the beam at the displacements $u = 0.34$ mm, $0.355$ mm, and $0.46$ mm. 

\subsection{3D NSCB tests}\label{3D NSCB tests}

In the last example, we simulate the crack propagation in a 3D NSCB specimen. The geometry and boundary conditions are similar to the 2D tests except a thickness of 16 mm in 3D. The same parameters are used as those in the 2D tests and $G_c=7.6$ J/m$^2$. We refine the elements with the maximum size $h=4.5\times10^{-4}$ m in the region where the crack is expected to propagate, while in the rest region $h=1.8\times10^{-3}$ m. In addition, the displacement increment $\Delta u = 5\times10^{-7}$ mm is applied.

\begin{figure}[htbp]
	\centering
	\subfigure[]{\includegraphics[width = 6cm]{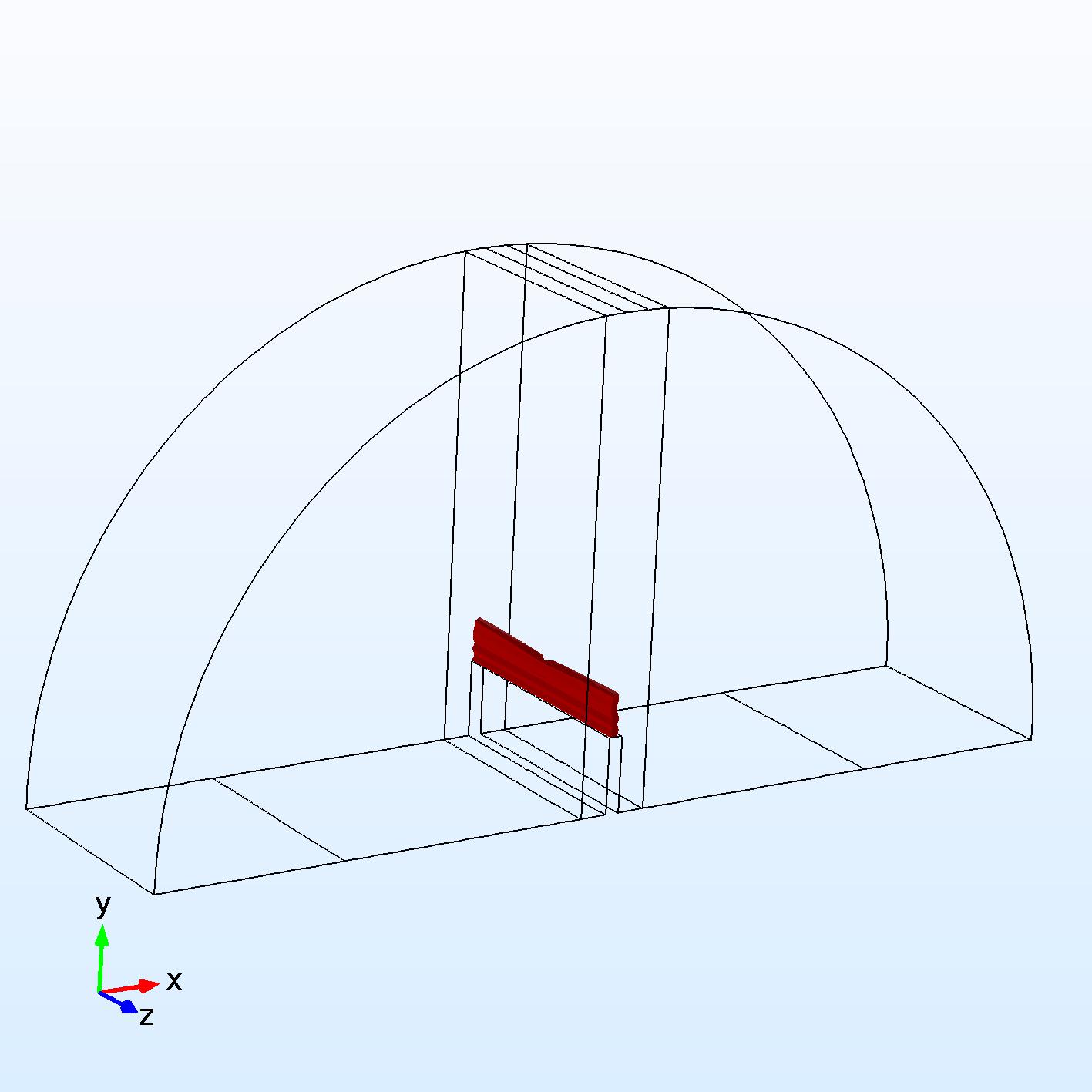}}
	\subfigure[]{\includegraphics[width = 6cm]{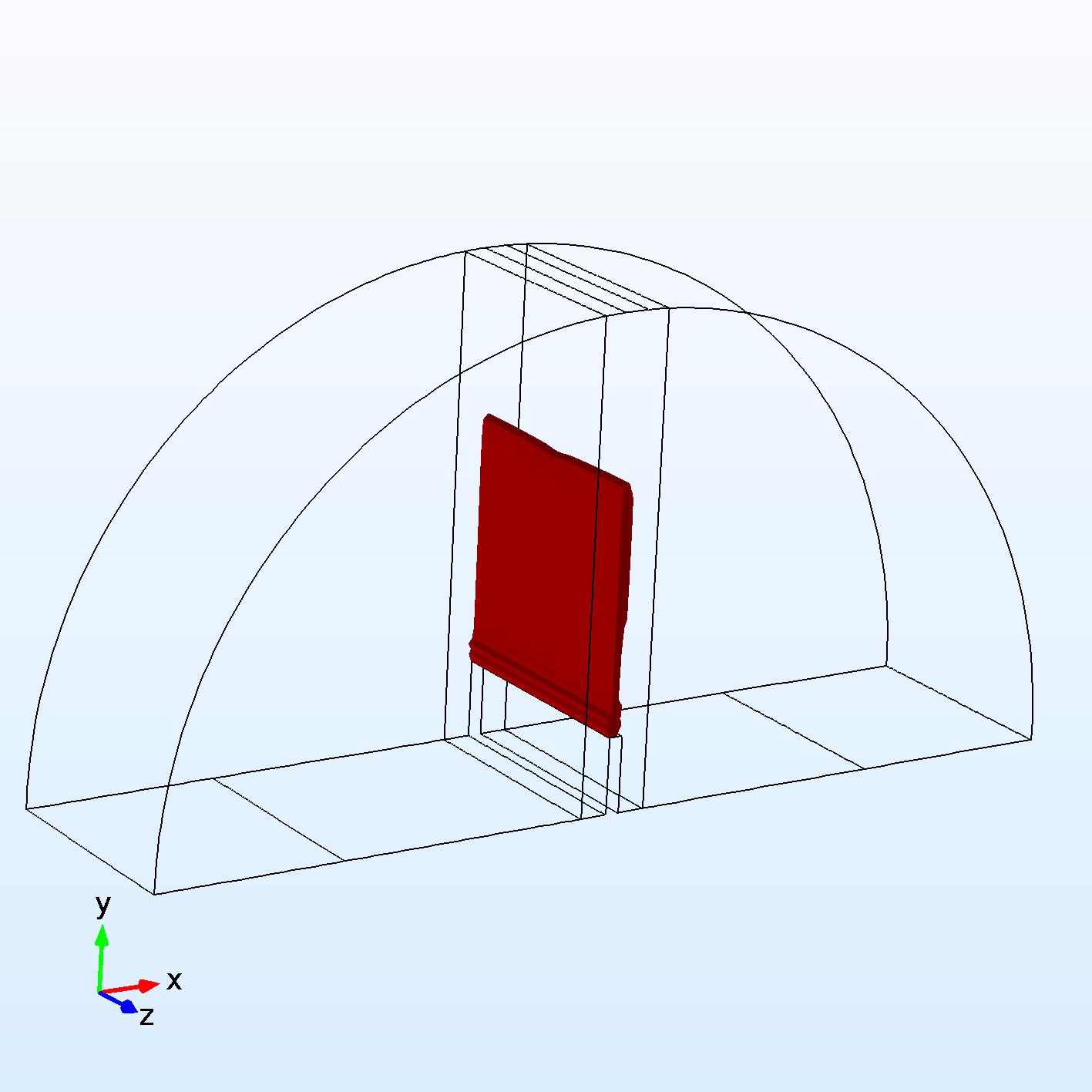}}
	
	\subfigure[]{\includegraphics[width = 6cm]{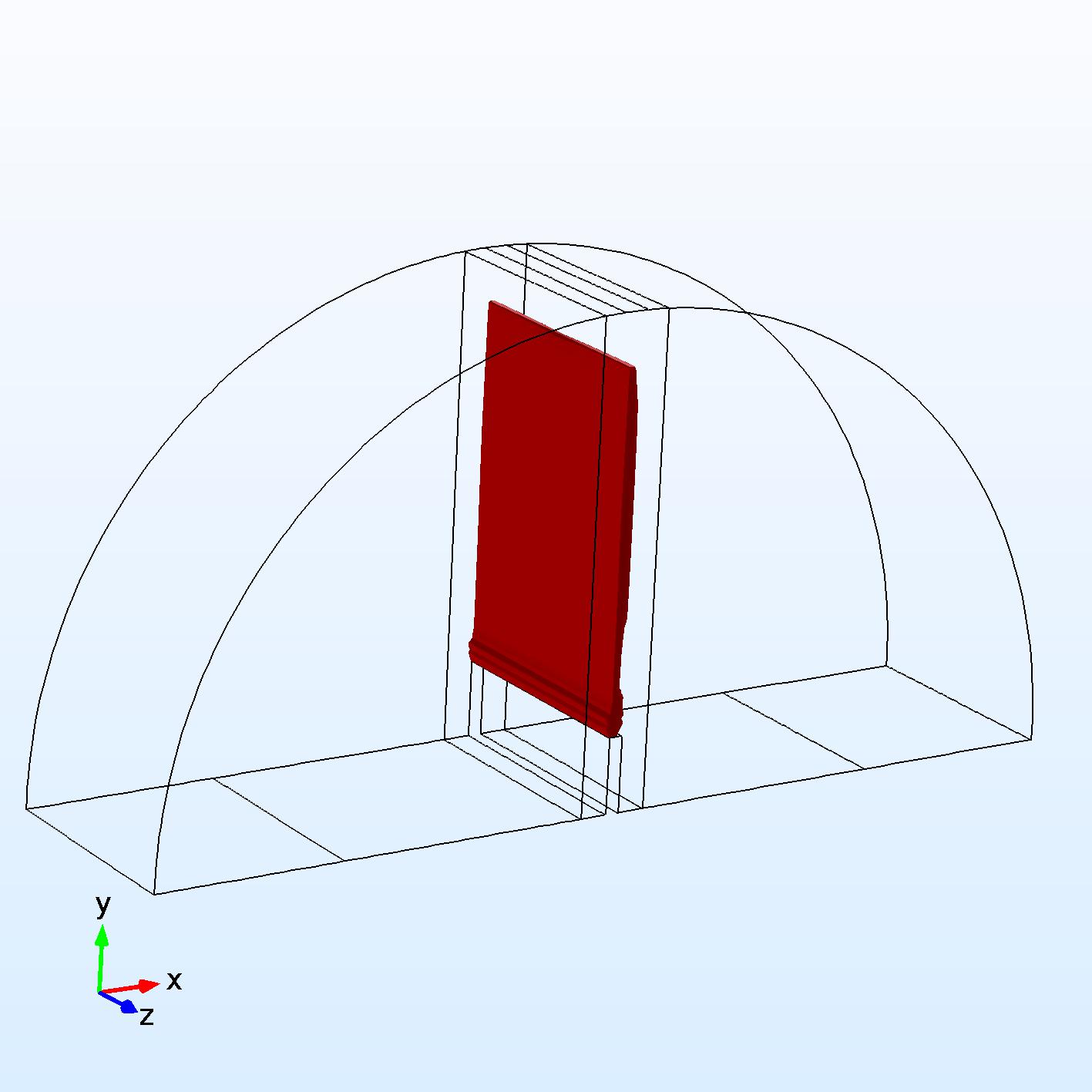}}
	\caption{Crack propagation of the 3D NSCB test at a displacement of (a) $u = 6.67\times10^{-3}$ mm, (b) $u = 6.71\times10^{-3}$ mm, and (c) $u = 6.745\times10^{-3}$ mm for $G_c=7.6$ J/m$^2$}
	\label{Crack propagation of the 3D NSCB test}
	\end{figure}

Figure \ref{Crack propagation of the 3D NSCB test} presents the crack propagation at the displacements $u = 6.67\times10^{-3}$ mm, $6.71\times10^{-3}$ mm, and $6.745\times10^{-3}$ mm. Only the domain with $\phi>0.95$ are displayed for the crack shape. The crack patterns are the same as those in the 2D simulation, and also in good agreement with the results of the experimental tests \citep{gao2015application}. The example of 3D NSCB test shows the ability and practicability of the phase field method in modeling crack propagation of rocks in 3D.

\section {Conclusions}

The phase field theory for fracture is applied to study the crack propagation, branching and coalescence in rocks. Implementation details of the phase field modeling in COMSOL are presented with the consideration of cracks only due to tension. The numerical simulations of the 2D notched semi-circular bend (NSCB) tests and Brazil splitting tests are then performed. The presented results are in good agreement with those of the previous experimental tests. Subsequently, the crack propagation and coalescence in plates with multiple echelon flaws and twenty parallel flaws are studied. We also present the complex crack patterns in a plate subjected to internal pressure, the increase of which produces crack propagation and branching. Finally, the simulation of a 3D Petersson beam and a 3D NSCB test are performed to show the practicability of phase field modeling in 3D rocks.

All the numerical examples presented by this work show that the initiation, propagation, coalescence, and branching of cracks are autonomous, while the phase field modeling does not require external criterion for fracture and setting propagation path in advance. These observations highlight the advantages of the phase field method over other numerical methods in modeling complex crack propagation in rocks. Therefore, the phase field modeling approach will be useful and practicable for other crack problems in rock engineering in future research. In addition, the presented phase field model cannot predict the shear cracks when a rock reaches its shear strength. The reason is that the shear strength is not involved in the formulation of the phase field method and the crack propagation is only driven by the elastic energy. In this sense, a modified phase field model coupled with the shear model of rocks will be also attractive in the future.

\section*{Acknowledgement}
The financial support provided by the Sino-German (CSC-DAAD) Postdoc Scholarship Program 2016, the Natural Science Foundation of China (51474157), and RISE-project BESTOFRAC (734370) is gratefully acknowledged.
\bibliography{references}
\end{document}